%% file: PRD_lvbb_9.7fb.tex
\documentclass[aps,prd,showpacs,twocolumn,superscriptaddress,floatfix]{revtex4}  

\usepackage{graphicx}  
\usepackage{bm}        
\usepackage{amssymb}   

\usepackage{placeins}
\usepackage{epsfig}

\usepackage{color}

\newcommand{\MET}{\mbox{\ensuremath{\not\!\!E_T}}}
\newcommand{\deltaR}{\mbox{\ensuremath{\Delta\mathcal{R}}}}
\newcommand{\ppbar}{\mbox{\ensuremath{p\bar{p}}}}
\newcommand{\pbar}{\mbox{\ensuremath{\bar{p}}}}

\newcommand{\explimA} {4.7} 
\newcommand{\obslimA} {5.8} 

\newcommand{\diboexpsigMVA}{1.8}

\newcommand{\diboxsecMVA}  {{0.50 $\pm$ 0.34\,(stat.)~$\pm$ 0.36\,(syst.)}}
\newcommand{\diboSMxsec}   {{4.4~$\pm$~0.3~pb}}

\input{dictionary.tex}

\begin{document}

\hspace{5in}\mbox{FERMILAB-PUB-13-030-E}

\title{Search for the standard model Higgs boson in $\bm{\ell\nu}$+jets final states in 9.7~fb$\bm{^{-1}}$ of $\bm{p\bar{p}}$ collisions with the D0 detector}
\input author_list.tex           
\date{January 25, 2013}

\begin{abstract}
We present, in detail, a search for the standard model Higgs boson, $H$, in final states with a charged lepton (electron or muon), 
missing energy, and two or more jets in data corresponding to 9.7~fb$^{-1}$ of integrated luminosity collected at a center of mass energy  of $\sqrt{s}$ = 1.96~TeV
with the D0 detector at the Fermilab Tevatron $p\bar{p}$ Collider.
The search uses $b$-jet identification to categorize events for improved signal versus background separation and 
is sensitive to associated production of the $H$ with a $W$ boson, $WH\to\ell\nu b\bar{b}$; gluon fusion with the Higgs decaying to $W$ boson pairs, $H\to WW\to\ell\nu jj$; and associated production with a vector boson where the Higgs decays to $W$ boson pairs,
$VH\to VWW \to \ell\nu jjjj$ production (where $V = W$ or $Z$).  
We observe good agreement between data and expected background.
We test our method by measuring $WZ$ and $ZZ$ production with $Z\to b\bar{b}$ and find production rates consistent with the standard model prediction.
For a Higgs boson mass of 125~GeV, we set a 95\%\ C.L. upper limit on the production of a standard model Higgs boson of
\obslimA$\times\sigma_{\rm SM}$, where $\sigma_{\rm SM}$ is the standard model Higgs boson production cross section, while the expected limit is 
\explimA$\times\sigma_{\rm SM}$. We also interpret the data considering models with fourth generation fermions, or a fermiophobic Higgs boson.
\end{abstract}

\maketitle

\section{Introduction}

The Higgs boson is the massive physical state that emerges from electroweak symmetry breaking in the Higgs mechanism~\cite{Englert:1964et,Higgs:1964pj,Guralnik:1964eu}. 
This mechanism generates the masses of the weak gauge 
bosons and explains the fermion masses through their Yukawa couplings to the Higgs boson field. The mass of the Higgs boson ($M_H$) is a free 
parameter in the standard model (SM). Precision measurements of various SM electroweak parameters constrain $M_H$
to be less than $152$~GeV at the 95\% C.L.~\cite{Aaltonen:2012bp,Abazov:2012bv,bib:LEPEWWG}. Direct searches at the CERN $e^+e^-$ Collider 
(LEP)~\cite{Barate:2003sz} exclude $M_H < 114.4$~GeV at the 95\%\ C.L. The ATLAS and CMS 
Collaborations, using $pp$ collisions at the CERN LHC,
exclude masses from $110 < M_H < 600$~GeV, except for a narrow region between 122 and 
127~GeV~\cite{Atlas-jul2012,CMS-PAS-HIG-12-008}. Both experiments observe a resonance at a mass of $\approx 125$~GeV, 
primarily in the $\gamma\gamma$ and $ZZ$ final states,  with a significance greater than 5 standard deviations (s.d.)\ that is consistent with SM Higgs boson production~\cite{atlas-obs, cms-obs}. 
The CDF and D0 Collaborations at the Fermilab Tevatron Collider report a combined analysis that excludes
the region $147 < M_H < 179$~GeV~\cite{CDFandD0:2012aa} and shows evidence at the 3 s.d.\ level for a particle decaying 
to $b\bar{b}$, produced in association with a $W$ or $Z$ boson, consistent with SM $WH/ZH$ production~\cite{tevatron-bbbar}.
Demonstrating that the observed resonance is the SM Higgs boson requires also observing it at the predicted rate
in the $b\bar{b}$ final state, which is the dominant decay mode for masses below $M_H\lesssim135$~GeV.

The dominant production process for the Higgs boson at the Tevatron Collider is gluon fusion ($gg\to H$), followed by
the associated production of a Higgs boson with a vector boson ($VH$), then via vector boson fusion ($VVqq^{\prime}\to Hqq^{\prime}$). 
At masses below $M_H \approx 135$~GeV, the Higgs boson mainly decays to a pair of
$b$ quarks, while for larger masses, the dominant decay is to a pair of $W$ bosons.
Because the  $H\to b\bar{b}$ process is difficult to 
distinguish from background at hadron colliders, it is more effective to search for the Higgs boson produced in association with a vector boson for this decay channel.

This Article presents a search by the D0 collaboration for the SM Higgs boson using events containing
one isolated charged lepton ($\ell=e$ or $\mu$),
a significant imbalance in transverse energy ($\MET$), 
and two or more jets.
It includes a detailed description of the $WH \to \ell\nu b\bar{b}$ search, initially presented in Ref.~\cite{Abazov:2012wh97} and used as an input to the result presented in Ref.~\cite{tevatron-bbbar}, differing from and superseding that result due to an updated treatment of some systematic uncertainties as described in Sec.~\ref{sec:syst} below.
The complete analysis comprises searches for the production and decay channels: $WH\rightarrow\ell\nu b\bar{b}$, $H \rightarrow WW^{*} \rightarrow \ell\nu jj$ 
(where $j=u,d,s,c$), and $VH \rightarrow VWW^{*} \rightarrow \ell \nu jjjj$ (where $V=W$ or $Z$).
This search also considers contributions from $ZH$ production and from the decay $H\to ZZ$ when 
one of the charged leptons
from $Z\to\ell\ell$ decay is not identified in the detector.
We optimize the analysis by subdividing data into mutually exclusive subchannels
based on charged lepton flavor, jet multiplicity, and the number and quality of candidate $b$ quark jets. This search also extends the most recent D0 $WH \to \ell\nu b\bar{b}$ search~\cite{Abazov:2012wh97}
 by adding subchannels with looser $b$-quark jet identification requirements and subchannels with 
four or more jets. These additional subchannels are primarily sensitive to $H \rightarrow WW^{*} \rightarrow \ell\nu jj$ and 
$VH \rightarrow VWW^{*} \rightarrow \ell\nu jjjj$ production and extend the reach of our
search to $M_H=200$~GeV.  We present a measurement 
of $VZ$ production with $Z\to b \bar{b}$ as a cross check on our methodology in Sec.~\ref{sec:diboson}. In addition to our standard 
model interpretation, we consider interpretations of our result in models with a fourth generation of fermions, and models with a fermiophobic Higgs as described in Sec.~\ref{sec:BSM}.

Several other searches for $WH \to \ell\nu b\bar{b}$ production have been reported at a $p\bar{p}$ 
center-of-mass energy of $\sqrt{s}=1.96$~TeV, most recently 
by the CDF Collaboration~\cite{Aaltonen:2012wh}.
The results presented here supersede previous searches by the D0 Collaboration, presented in Refs.~\cite{Abazov:2005aa, Abazov:2007hk, Abazov:2008eb, Abazov:2010hn, Abazov:2012wh}, which
used subsamples of the data presented in this Article.
They also supersede a previous search
for Higgs boson production in the $\ell\nu jj$ final state by the D0 Collaboration~\cite{Abazov:2011bc}.

\section{The D0 Detector}

This analysis relies on all major components of the D0 detector: tracking detectors, calorimeters, and muon identification system. 
These systems are described in detail in Ref.~\cite{Abachi:1993em,Abazov:2005pn,Abolins:2007yz,Angstadt:2009ie}. 

Closest to 
the interaction point is the silicon microstrip tracker (SMT) followed by the central scintillating fiber tracker (CFT). These detector subsystems 
are located inside a 2~T magnetic field provided by a superconducting solenoid. They track charged particles and 
are used to reconstruct primary and secondary vertices for pseudorapidities~\cite{defseta}  of $|\eta|<3$. 
Outside the solenoid is the liquid argon/uranium calorimeter consisting of one central calorimeter (CC) covering $|\eta| \lesssim 1$ and two end calorimeters 
(EC) extending coverage to $|\eta| \approx 4$. 
Each calorimeter contains an innermost finely segmented electromagnetic layer followed by 
two hadronic layers, with fine and coarse segmentation, respectively.
The main functions of the calorimeters are to measure energies and help identify electrons, photons, and jets 
using coordinate information of significant energy clusters. They also give a measure of the $\MET$.
A preshower detector between the solenoidal magnet and central calorimeter consists of a cylindrical radiator and 
three layers of scintillator strips covering the region $|\eta|<1.3$.
The outermost system provides muon identification. It is divided into a central section that covers $|\eta| < 1$ and forward sections
that extend coverage out to $|\eta| \approx 2$. 
The muon system is composed of three layers of drift tubes and scintillation counters, one layer before and two
layers after a 1.8~T toroidal magnet.

\section{Event Trigger \label{sec:triggering}}

Events in the electron channel are triggered by a logical OR of triggers that require an 
electromagnetic object and jets, as described in Ref.~\cite{Abazov:2012wh}. 
Trigger efficiencies are modeled in the Monte Carlo (MC) simulation by applying the trigger efficiency, measured in data, as an event weight. This efficiency is parametrized as a function of electron $\eta$, azimuthal angle $\phi$~\cite{defphi}, 
and transverse momentum $p_T$. For the events selected in our analysis, these triggers have an efficiency of $(90-100)\%$ depending on the trigger and 
the region of the detector.

The muon channel uses an inclusive trigger approach, based on the logical OR of all available triggers,
except those containing lifetime-based requirements that can bias the performance of $b$-jet identification.
To determine the trigger efficiency, we compare data events selected with a well-modeled 
logical OR of the single muon and muon+jets triggers ($T_{\mu\text{OR}}$), which are 
about 70\% efficient, to events selected using all triggers. 
The increase in event yield in the inclusive trigger sample is used to determine an inclusive trigger correction for the MC trigger efficiency, $P_{\mathrm{corr}}$, relative to the $T_{\mu\text{OR}}$ trigger ensemble:
\begin{equation}
P_{\mathrm{corr}} = \frac{(N_{\mathrm{Data}} - N_{\mathrm{MJ}})_{incl} - (N_{\mathrm{Data}} - N_{\mathrm{MJ}})_{T_{\mu\text{OR}}}}{N_{MC}},
\end{equation}
where the numerator is the difference between the number of data events in the inclusive trigger sample and the $T_{\mu\text{OR}}$ trigger sample, after subtracting off instrumental multijet (MJ) backgrounds,
 and the denominator is the number of MC events (after the event selection and normalization to data described in Sec.~\ref{sec:eventsel} and the MC corrections are applied as described in Sec.~\ref{sec:mcrw}) with the trigger efficiency set to 1. 
The total trigger efficiency estimate for events in the muon channel is $T_{\mu\text{OR}} + P_{\mathrm{corr}}$, limited to be $\leq 1$.

Triggers based on jets and $\MET$\ make the most significant contributions to the inclusive set of triggers
beyond those included in the well-modeled $T_{\mu\text{OR}}$ trigger set.
To account for these contributions, 
the correction from $T_{\mu\text{OR}}$ triggers to the inclusive set of triggers is parametrized as a function of the scalar sum of the transverse momenta of all jets, $H_T$, and the $\MET$, and is derived for separate regions in muon $\eta$.

\begin{figure}[htbp]
\centering{
\includegraphics[width=0.4\textwidth]{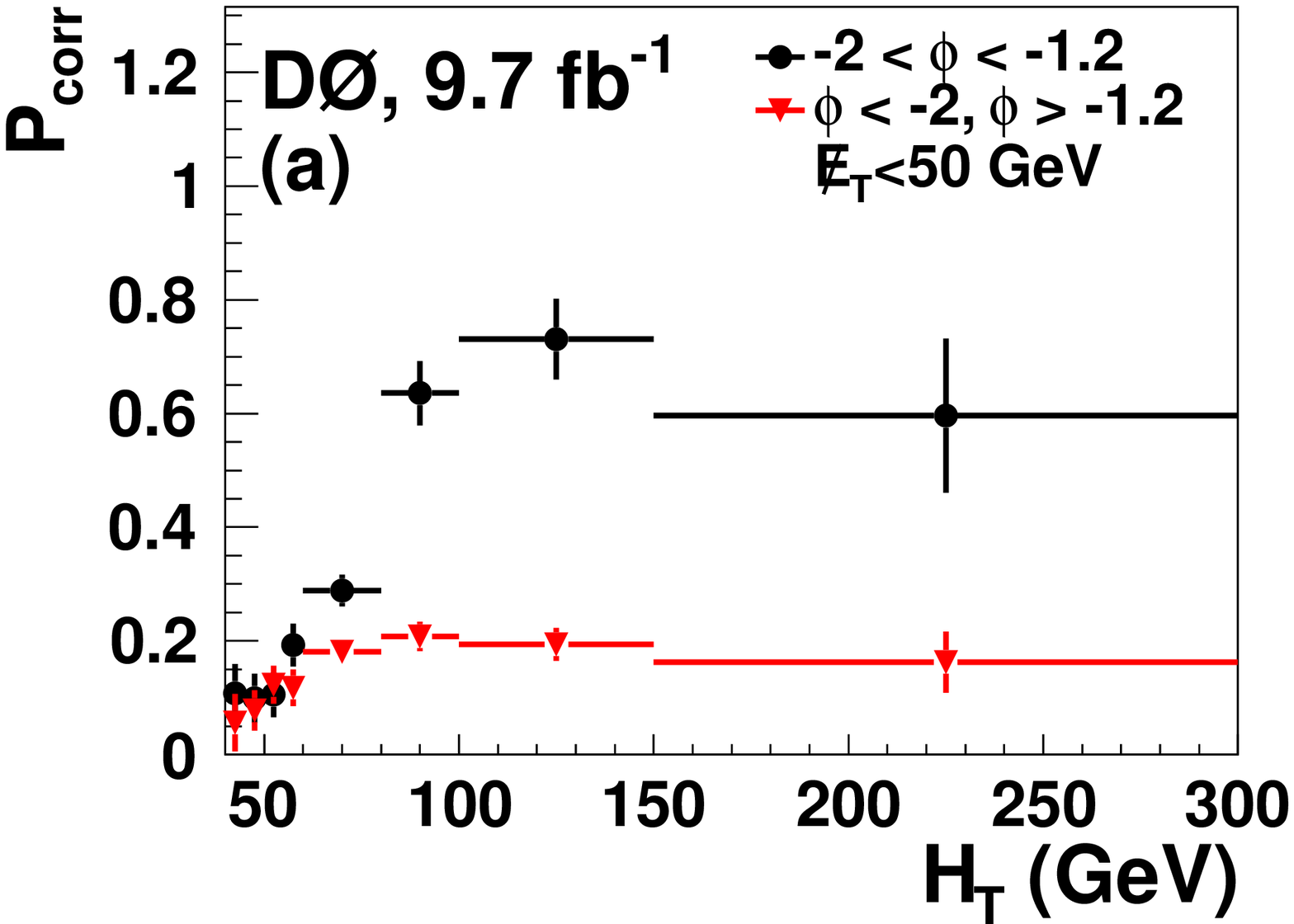}
\includegraphics[width=0.4\textwidth]{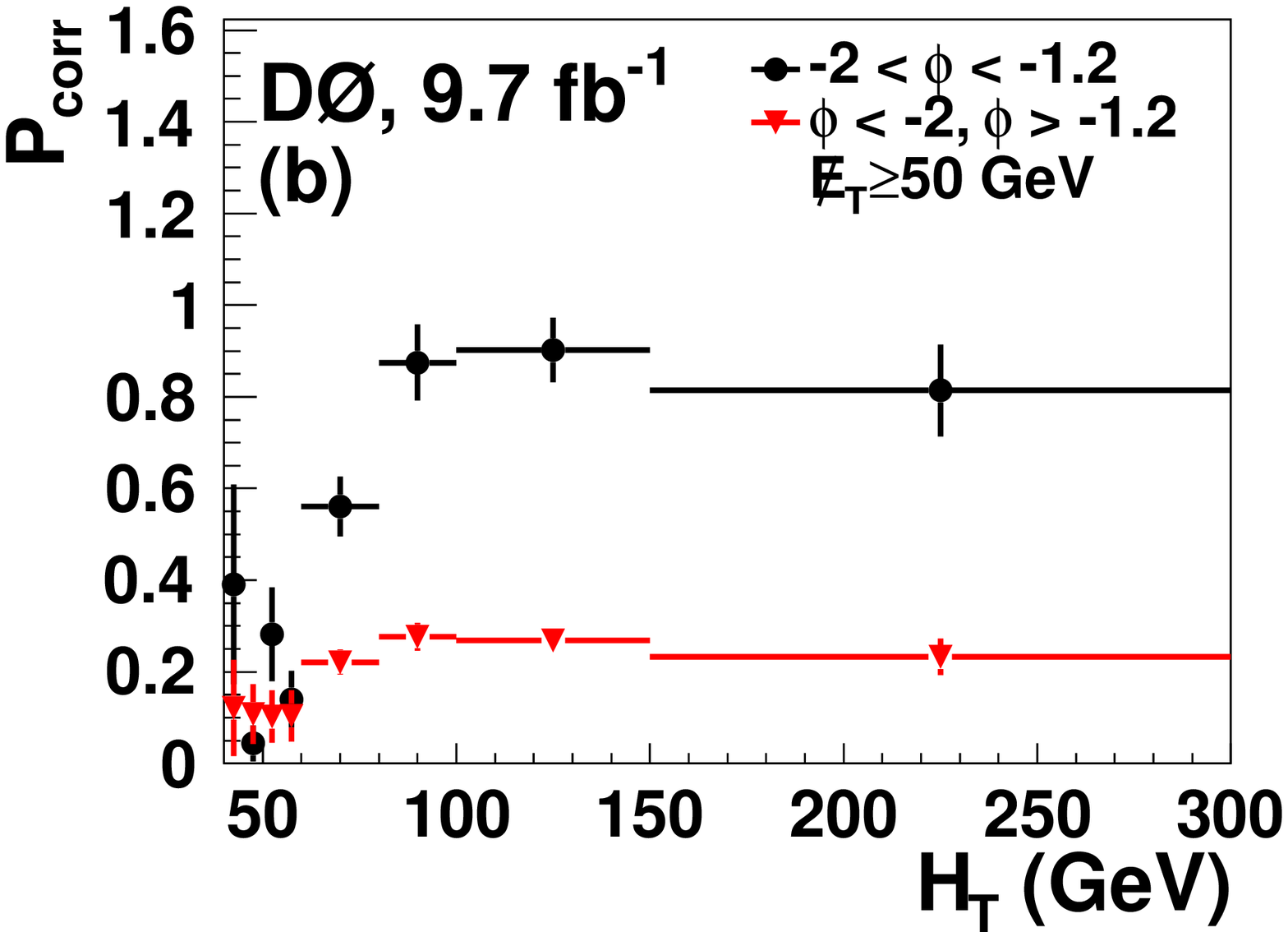}
}\caption{(color online) 
Data-derived muon trigger correction to account for the resulting efficiency gain in moving from single muon and muon+jets triggers to inclusive triggers as a function of
$H_{T}$ for $|\eta|<1.0$, shown (a) for events with $\MET < 50$~GeV and (b) for events with $\MET \geq 50$~GeV.
The black circles show the correction when the muon is in the region of $\phi$ ($-2<\phi<-1.2$) where there is a gap in the muon coverage for detector
supports, and the red triangles show the correction elsewhere in $\phi$.}
\label{fig:trigcorr}
\end{figure}

 For $|\eta|<1.0$, events are dominantly triggered by single 
muon triggers, while
 for $|\eta|>1.6$, triggers based on the logical OR of muon+jets prevail. 
The third region, $1.0<|\eta|<1.6$, is a mixture of single muon and muon+jets triggers. 
 In the $|\eta|<1.0$ and $1.0<|\eta|<1.6$ regions
the detector support structure allows only partial coverage by the muon
system.  This impacts the muon trigger efficiency in the region
$-2<\phi<-1.2$.  In these regions,  we therefore
derive separate corrections.
The inclusive trigger approach results in a gain of about 30\% in 
efficiency over using only muon and the muon+jets triggers.
Examples of these corrections, $P_{\mathrm{corr}}$ are shown in Fig.~\ref{fig:trigcorr}.

\section{Identification of Leptons, Jets, \newline and ${\bm{\MET}}$ \label{sec:leptonid}}
To reconstruct the candidate $W(\to\ell\nu)$ boson, our selected events are required to contain a single 
identified electron or muon together with significant \MET. To ensure statistical independence with channels that contain more than one 
lepton, we do not consider events with more than one electron or muon.
Two or more jets are also required in order to study
$WH\to\ell\nu b\bar{b}$, $H\to WW\to\ell\nu jj$, and $VH\to VWW \to \ell\nu jjjj$ production.
Two sets of lepton identification criteria are applied
for each lepton channel in order to form a ``loose'' sample, used to estimate the multijet background 
from data as described in Sec.~\ref{sec:MJ}, and a ``tight'' sample used to perform the search.
The event selection procedure, prior to $b$-jet categorization, is similar to that described in Ref.~\cite{Abazov:2012wh}
and described in more detail below.

Electrons with $p_T > 15$~GeV are selected in the pseudorapidity regions $|\eta| < 1.1$ and $1.5 < |\eta| < 2.5$,
corresponding to the CC and EC, respectively. 
A multivariate discriminant is used to identify electrons.
The discriminant is based
on a boosted decision tree~\cite{narsky-0507157,Breiman1984,schapire01boostapproach,schapireFreund,friedman}
(BDT) as implemented in the {\sc tmva} package~\cite{Hocker:2007ht} 
with input variables that are listed below.
The BDTs are discussed in more detail in Sec.~\ref{sec:MVA}.
The loose and tight electron samples are defined by different requirements on the response of this multivariate discriminant 
that are chosen to retain high electron selection efficiencies while suppressing backgrounds at differing rates.

Leptons coming from the leptonic decays of $W$ bosons tend to be isolated from jets.
Isolated electromagnetic showers are identified within a cone in $\eta$-$\phi$ space of $\deltaR = \sqrt{\Delta \eta ^{2} + \Delta \phi ^{2}} < 0.4$~\cite{defsDR}.
In the CC (EC), an electromagnetic shower is required to deposit $97\%$ $(90\%)$ of its total energy
within a cone of radius $\deltaR=0.2$ in the electromagnetic calorimeter.
The showers must have transverse and longitudinal distributions that are consistent with those expected from electrons. 
In the CC region, a reconstructed track, isolated from other tracks, is required to have a trajectory that extrapolates to the electromagnetic (EM) shower. 
The isolation criteria restrict the sum of the scalar $p_T$ of tracks with $p_T > 0.5$~GeV within a hollow cone of radius $0.05 < \deltaR < 0.4$ 
surrounding the electron candidate to be less than 2.5~GeV. 
The BDT is constructed using additional information such as: the  number and scalar $p_T$ sum of tracks in the cone of radius $\deltaR < 0.4$ surrounding the candidate cluster, 
track-to-cluster-matching probability, the ratio of the transverse energy of the cluster to the transverse momentum of the track 
associated with the shower, the EM energy fraction, lateral and longitudinal shower shape characteristics, as well as the number of hits in the various layers of the tracking detector, and information from the central preshower detector. 
The discriminants are trained using $Z / \gamma^* \to ee$ data events.

We select muons with  $p_T > 15$~GeV and $|\eta| < 2.0$. They are required to have reconstructed track segments 
in layers of the muon system both before and after the toroidal magnet, except where detector support structure limits
muon system coverage, for which the presence of track segments in any layer is sufficient.
The local muon system track must be spatially matched to a track in the central tracker. 

Muons originating from semi-leptonic decays of heavy flavored hadrons are typically not isolated due to jet fragmentation and 
secondary particles from the partial hadronic decays.
We employ a loose muon definition, requiring minimal separation of $\deltaR(\mu ,j)>0.5$ between the muon and any jet, while the tight
identification has additional isolation requirements.
For tight muons, the scalar sum of the $p_T$ of tracks with $\deltaR < 0.5$ around the muon candidate is required to be less than $0.4\times p_T^\mu$. 
Furthermore, the transverse energy deposits in the calorimeter in a hollow cone of $0.1 < \deltaR < 0.4$ around the muon must be less than 
$0.12\times p_T^\mu$.
To suppress cosmic ray muons, scintillator timing information is used to require hits in the detector to coincide with a beam crossing.

To reduce backgrounds from $Z/\gamma^* \to \ell\ell+$jets and $t\bar{t}$ 
production, we reject events containing more than one tight-isolated charged lepton.

Jets are reconstructed in the calorimeters in the region $|\eta| < 2.5$ using an iterative midpoint cone algorithm, with a cone size of 
$\deltaR=0.5$~\cite{Blazey:2000qt}. To minimize the possibility that jets are caused by noise or spurious energy deposits, the fraction of the total jet energy 
contained in the electromagnetic layers of the calorimeter is required to be between 5\% and 95\%, and  the energy fraction in the coarse hadronic layers of the 
calorimeter is required to be less than 40\%. To suppress noise, different energy thresholds are also applied to clustered and to isolated cells~\cite{Abazov:2011vi}. The energy of 
the jets is scaled by applying a correction determined from $\gamma +$jet events using the same jet-finding algorithm. This scale correction accounts for 
additional energy (e.g., residual energy from previous bunch crossings and energy from multiple $p\pbar$ interactions) that is sampled within the finite cone size, 
the calorimeter energy response to particles produced within the jet cone, and energy flowing outside the cone or moving 
into the cone via detector effects~\cite{Abazov:2011vi}. We also apply an additional correction that accounts for the flavor 
composition of jets~\cite{Abazov:2011ck}.

Jet energy calibration and resolution are adjusted in simulated events to match those measured in data. This correction is derived 
from $Z(\to ee)+$jet events from the 
$p_T$ imbalance between the $Z$ boson and the recoiling jet in MC simulation when compared to that observed in data, and applied
 to jet samples in MC events.
 Differences in reconstruction
thresholds in simulation and data are also taken into account, and
the jet identification efficiency and jet resolution
are adjusted in the simulation to match those measured
in data. All selected jets are required to have $p_T>20$~GeV and $|\eta|<2.5$. 
We require that jets originate from the primary $p\bar{p}$ vertex (PV), such that each selected jet is matched to at least two tracks with $p_T>0.5$~GeV
that have at least one hit in the SMT detector and  a distance of closest approach with respect to the PV of less than 0.5~cm in the transverse plane and less than 1~cm along the beam axis ($z$). Interaction vertices are reconstructed from tracks that have $p_T>0.5$~GeV with at least two hits in the SMT. The primary vertex is the reconstructed vertex with the highest average $p_T$ of its tracks. Vertex reconstruction is described in more detail in Ref.~\cite{Abazov:2010ab}. We also require that the PV be reconstructed within $z_{PV}=\pm 60~\rm cm$ of the center 
of the detector.

The \MET\ is calculated from individual calorimeter cell energies in the electromagnetic and fine hadronic sections of the calorimeter 
and is required to satisfy $\MET > 15$~GeV for the electron channel and $\MET > 20$~GeV for the muon channel. Energy from the coarse hadronic layers that is contained within 
a jet is also included in the \MET\ calculation. A correction for the presence of any muons 
and all energy corrections applied to electrons and jets 
are propagated to the value of \MET.

\section{Tagging of ${\bm{b}}$-Quark Jets \label{sec:bid}}

The $b$-tagging algorithm for identifying jets originating from $b$ quarks is based on a multivariate discriminant using a combination of variables sensitive to the presence of tracks or secondary vertices displaced significantly in the
$x$-$y$ plane from the $p\pbar$ interaction vertex. 
This algorithm provides improved performance over the neural network algorithm described in Ref.~\cite{Abazov:2010ab}.

Jets considered by the $b$-tagging algorithm are required to be ``taggable,'' i.e.,\ contain at least two tracks with each having at least one hit in the SMT.
The efficiency of this requirement accounts for variations in detector acceptance and track reconstruction efficiencies at different locations of the 
PV prior to the application of the $b$-tagging algorithm, and depends on 
the $z$ position of the PV and the $p_{T}$ and $\eta$ of the jet. For jets that pass through the geometrical acceptance of the tracking system, this efficiency is typically about 97\%.
The efficiency for $b$-tagging is determined with respect to taggable jets.
The correction for taggability is measured in the selected data sample, while the
corrections for $b$-tagging are determined in an independent heavy-flavor jet enriched sample of events that include a jet containing a muon, as described in Ref.~\cite{Abazov:2010ab}. The efficiency for jets to be taggable and to satisfy $b$-tagging requirements in the simulation is corrected to reproduce the respective efficiencies in data.

\begin{figure}[htbp]
\centering{
\includegraphics[width=0.38\textwidth]{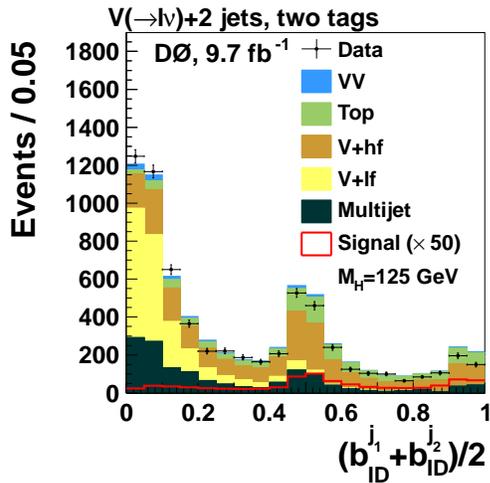}
}\caption{(color online) 
Average of the $b$-identification discriminant outputs of each jet in events with two jets.}
\label{fig:bid}
\end{figure}

We define six independent tagging samples with zero, one loose, one tight, two loose, two medium, or two tight 
$b$-tagged jets.
An inclusive ``pretag'' sample is also considered for parts of this analysis.
Events with no jets satisfying the $b$-tagging criteria are included in the zero $b$-tag sample. 
If exactly one jet is $b$-tagged, and the $b$-identification discriminant output for that jet, $\bidblI$, satisfies the tight selection threshold 
($\bidblI > 0.15$), that event is 
considered part of the one tight $b$-tag sample. Events with exactly one $b$-tagged jet that fails the tight selection threshold, but passes the loose selection threshold ($\bidblI > 0.02$) are included in
the one loose $b$-tag sample.
Events with two or more $b$-tagged jets are assigned to either the two loose $b$-tags, two medium $b$-tags, 
or two tight $b$-tags category, depending on the value of the average 
$b$-identification discriminant of the two jets with the highest discriminant values, i.e., the double tight category is required to satisfy $(\bidblA+\bidblB)/2>0.55$; the medium category is
$0.35<(\bidblA+\bidblB)/2\leq 0.55$; and
the loose category is $0.02<(\bidblA+\bidblB)/2\leq 0.35$ 
(see Fig.~\ref{fig:bid}). 
The operating point for the loose (medium, tight) threshold has an
identification efficiency of 79\%\ (57\%, 47\%) for individual $b$ jets, averaged over selected jet $p_T$ and $\eta$ distributions, 
with a $b$-tagging misidentification rate of 11\%\ (0.6\%, 0.15\%) for light quark jets ($lf$), calculated by the method described in Ref.~\cite{Abazov:2010ab}.

\section{Monte Carlo Simulation}

We account for all Higgs boson production and decay processes that can lead to a final state containing exactly one charged well isolated lepton, \MET, and two or more jets.
The signal processes considered are:
\begin{itemize} 

\item Associated production of a Higgs boson with a vector boson where the Higgs boson decays 
to $b\bar{b}$, $c\bar{c}$, $\tau\tau$, or $VV$. The associated weak vector boson decays leptonically in 
the case of $H\to b\bar{b}$
and either leptonically or hadronically
 in the case of $H\to WW$. 
Contributions from $Z(\to\ell\ell)H(\to b\bar{b})$ production arise from identifying only one charged lepton
in the detector, with the other contributing to the \MET.
\item Higgs boson production via gluon fusion  with the subsequent decay $H\to VV$, where one weak vector boson decays leptonically (with exactly one identified lepton).
\item Higgs boson production via vector boson fusion  with the subsequent decay $H\to VV$, where one weak vector boson decays leptonically (with exactly one identified lepton).
\end{itemize}

Various SM processes can mimic expected signal signatures, including 
$V+$jets, diboson ($VV$), MJ, $t\bar{t}$,  and single top quark 
production.

All signal processes and most of the background processes are estimated from 
MC simulation, while the MJ background is evaluated from data, as described in Sec.~\ref{sec:MJ}.
We use {\sc{pythia}}~\cite{Sjostrand:2006za} to simulate all signal processes 
and diboson processes. The $V+$jets and $t\bar{t}$ samples are simulated with the 
{\sc alpgen}~\cite{Mangano:2002ea} MC
generator interfaced to {\sc pythia} for parton showering
and hadronization, while the {\sc singletop} event generator~\cite{Boos:2004kh,Boos:2006af} interfaced to {\sc pythia} is used for single top quark 
events. To avoid overestimating the probability of further partonic emissions in {\sc pythia}, the MLM 
factorization (``matching'') scheme~\cite{Alwall:2007fs} is used. All of these simulations use CTEQ6L1~\cite{Lai:1996mg,Pumplin:2002vw}
parton distribution functions (PDF).

A full {\sc geant}-based~\cite{geant} detector simulation is used to process signal 
and background events. To account for residual activity from previous beam crossings and 
contributions from the presence of additional $\ppbar$ interactions, 
events from randomly selected beam crossings with the same instantaneous
luminosity profile as the data are overlaid on the simulated events. 
All events are then reconstructed using the same software as used for data.


The signal cross sections and branching fractions are normalized to the SM 
predictions~\cite{CDFandD0:2012aa}. The $WH$ and 
$ZH$ cross sections are calculated at next-to-next-to-leading order (NNLO)~\cite{Baglio:2010um},  
with MSTW2008 NNLO PDFs~\cite{Martin:2009iq}. The gluon fusion process uses the NNLO+next-to-next-to-leading log (NNLL) 
calculation~\cite{deFlorian:2009hc}, and the vector boson fusion process is calculated at 
NNLO in QCD~\cite{Bolzoni:2011cu}. The Higgs boson decay branching fractions are obtained with 
{\sc hdecay}~\cite{Djouadi:1997yw,Butterworth:2010ym}.
We use NLO cross sections to normalize single top quark~\cite{Kidonakis:2006bu} and diboson~\cite{Campbell:1999ah,mcfm_code} production, 
while we use an approximate NNLO calculation for $t\bar{t}$ production~\cite{Langenfeld:2009wd}.
The $p_T$ of the $Z$ boson in $Z+$jets events is corrected to match 
that observed in data~\cite{Abazov:2007jy}. The $p_T$ of the $W$ boson in $W$+jets 
events is corrected using the same dependence but taking into account 
the differences between the $p_T$ spectra of the $Z$ and $W$ bosons in NNLO QCD~\cite{Melnikov:2006kv}.
Additional scale factors to account for higher order terms in the {\sc alpgen} MC for the $V$+heavy flavor jets, $V+hf$, are obtained from {\sc mcfm}~\cite{Campbell:2001ik,mcfm_code}.
The $V$+jets processes are then normalized to data for each lepton flavor and jet multiplicity separately as described in Sec.~\ref{sec:eventsel}.

\subsection{MC Reweighting \label{sec:mcrw}}

Motivated by previous comparisons of {{\sc alpgen}} with data~\cite{Abazov:2008ez} and with other event generators~\cite{Alwall:2007fs}, 
we develop corrections to $W+\text{jets}$ and $Z+\text{jets}$ MC samples to correct for the shape discrepancies in kinematic distributions 
between data and simulation. The corrections are derived based on the direct comparison between data and MC samples prior to the application of $b$-tagging, where any contamination from signal is very small.

To improve the description of jet directions, we correct the $\eta$ distributions of the leading and second leading jets in $W/Z$+jets events. 
The correction function is a fourth-order polynomial determined from the ratio of the $V$+jets events in MC and data minus non-$V$+jets backgrounds.
The modeling of the lepton $\eta$ in $W$+jets events is adjusted by applying a second-order polynomial correction. 
Correlated discrepancies observed in the leptonically decaying $W$ boson transverse momentum, $\lnupt$, and the jet angular separation, $\jABdr$, are corrected through two
reweighting functions in the two-dimensional $\deltaR$-$\lnupt$ plane~\cite{Abazov:2012wh}. The $\lnupt$ reweighting is applied only to 
$W+\text{jets}$ events, while the $\deltaR$ reweighting is applied to both  $W+\text{jets}$ and $Z+\text{jets}$ events. 
Each of these corrections is designed to change differential distributions, but to preserve normalization.
Corrections are on the order of a few percent in the highly populated region of each distribution and may exceed 10\% for extreme values of each distribution.

All corrections are derived in events selected with muon+jets triggers to minimize
uncertainties due to  contamination from MJ events, and are applied to both the electron 
and muon channels. Additional $\lnupt$, $\deltaR$, and 
lepton $\eta$ corrections and corresponding systematic uncertainties are
determined from events selected with inclusive muon triggers
and  are applied to events containing muons, accounting for variations in modeling  distributions of the inclusively triggered events.

\section{Multijet Background \label{sec:MJ}}

The MJ background, events where a jet is misidentified as a lepton, is determined from the data prior to the application of $b$-tagging, using a method similar 
to the one used in Ref.~\cite{Abazov:2012wh}.
This method involves applying event weights that depend on the relative efficiency
$\varepsilon_{\text{LT}}^\ell$, of a lepton passing loose requirements to subsequently pass the tight requirements
and a similar relative probability, $P_{\text{LT}}^{\text{MJ}}$, for a MJ event to pass these sequential selections.  
A MJ template is constructed by selecting events from data in which the lepton passes the loose isolation requirement, but fails the
tight requirement, as described in Sec.~\ref{sec:leptonid}.  Each event in the MJ template is weighted by
\begin{equation}
w_{\text{MJ}} = \frac{P_{\text{LT}}^{\text{MJ}}}{1- P_{\text{LT}}^{\text{MJ}}},
\end{equation}
where $P^{\text{MJ}}_{\text{LT}}$ is a function of the event kinematics.
Since the MJ template contains a contribution from events with real leptons
originating from leptonic decays of $W/Z$ bosons, we correct the normalization of the $V$+jets MC using the event weight
\begin{equation}
w_{\text{VJ}} = 1- \frac{P_{\text{LT}}^{\text{MJ}}\left(1- \varepsilon_{\text{LT}}^{\ell}\right)}{\varepsilon_{\text{LT}}^{\ell}\left(1- P_{\text{LT}}^{\text{MJ}}\right)},
\end{equation}
where ${\epsilon^{\ell}_{\text{LT}}}$ and $P^{\text{MJ}}_{\text{LT}}$ are functions of event kinematics.
The efficiencies ${\epsilon^{\ell}_{\text{LT}}}$ are functions of lepton 
$p_{T}$, and they are determined from 
$Z/\gamma^{*} \rightarrow \ell\ell$ events.
The probabilities $P^{\text{MJ}}_{\text{LT}}$ are determined in the region
$5 < \MET < 15$~GeV
from the measured ratio of the number of events 
with tight leptons and those with loose leptons after correcting each 
sample for the expected MC contribution from real leptons in the specific kinematic interval. 
Electron channel probabilities are parametrized in $p_T$, calorimeter detector $\eta$, and  $\min\Delta\phi(\MET,j)$
while probabilities in the muon channel are parametrized in $p_T$ for different regions in muon detector $\eta$ and $\Delta\phi(\MET,\mu)$.

\section{Event Selection \label{sec:eventsel}}

\begin{figure}[htbp]
\centering{
\includegraphics[width=0.38\textwidth]{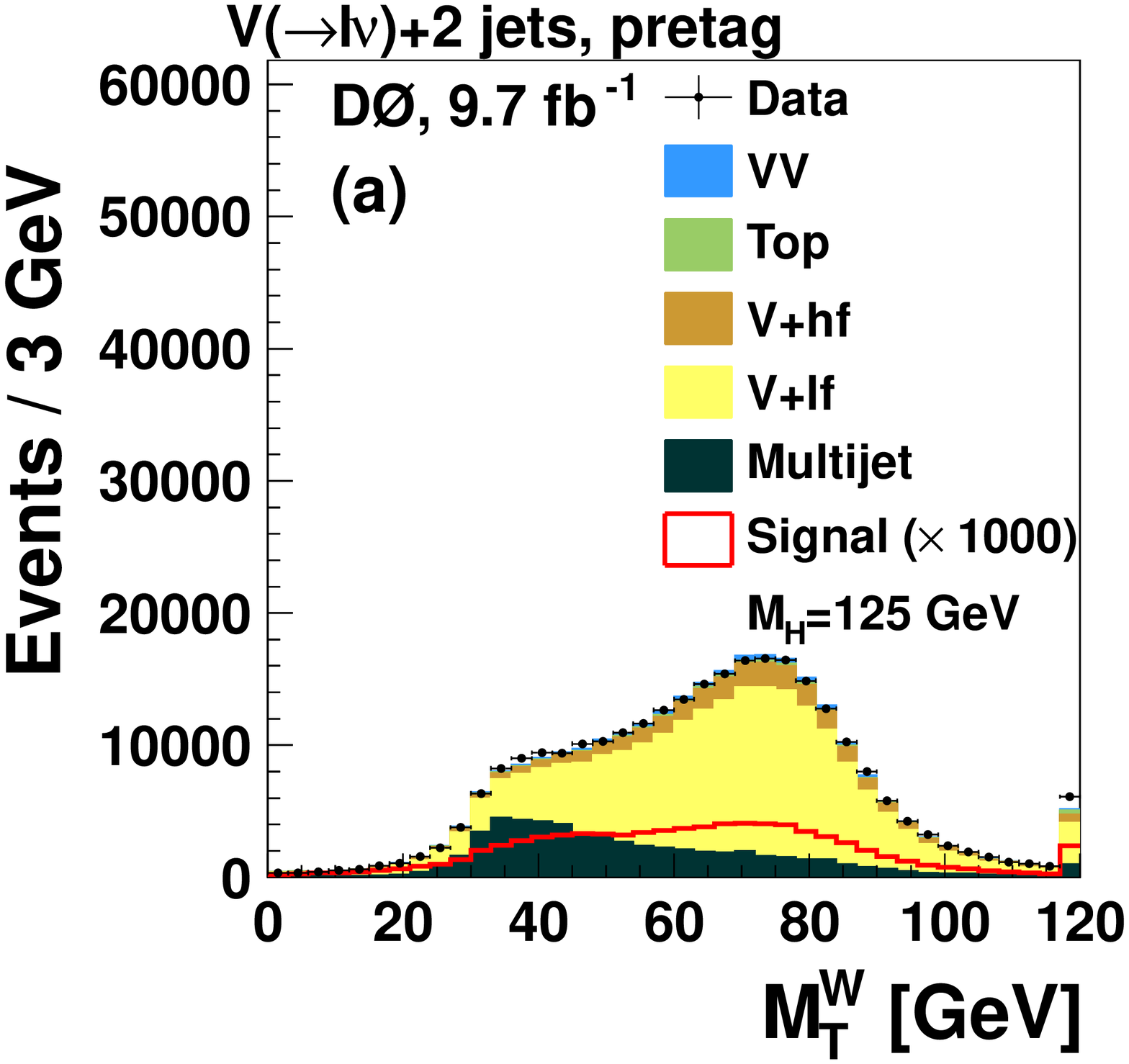}
\includegraphics[width=0.38\textwidth]{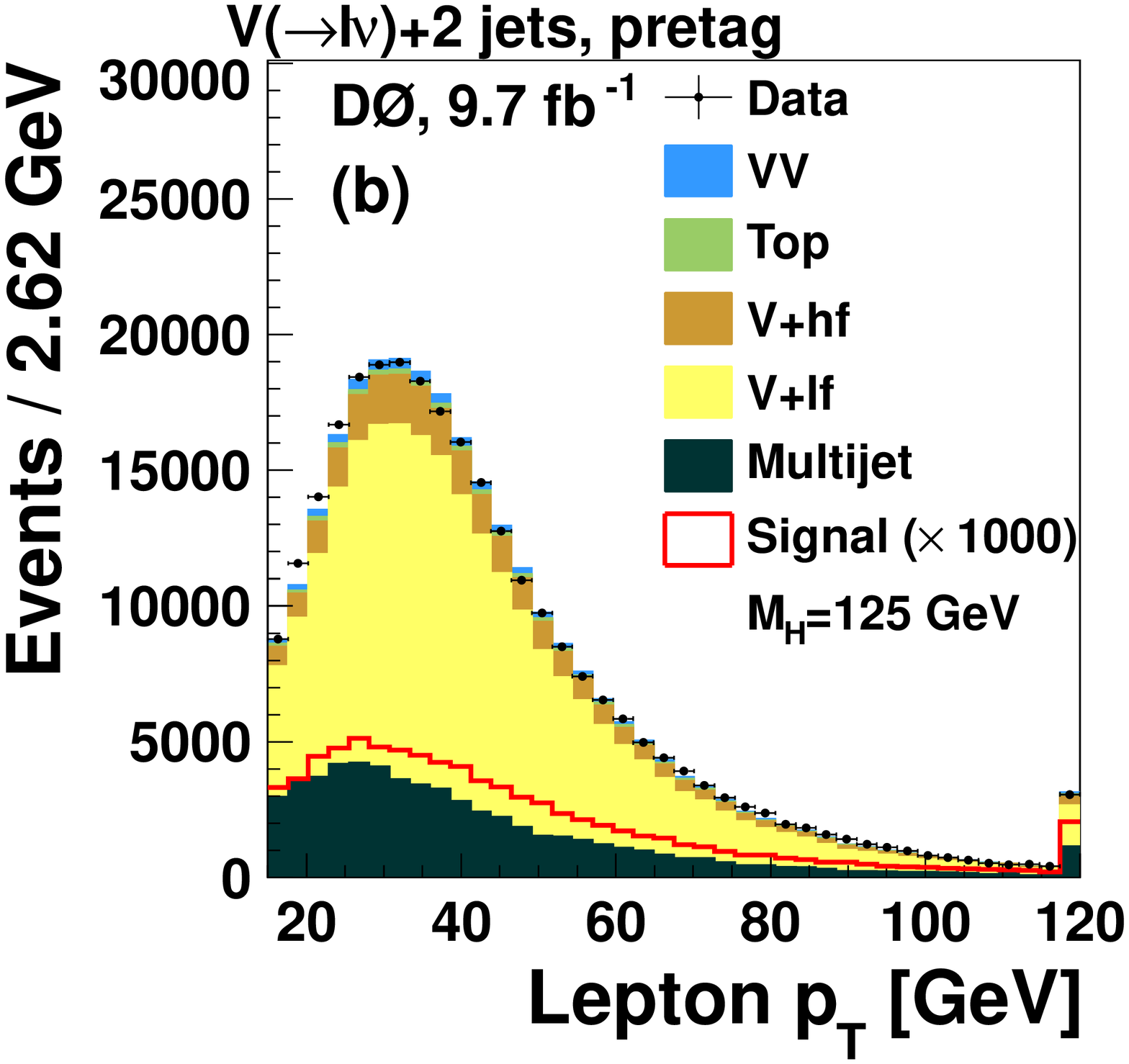}
\includegraphics[width=0.38\textwidth]{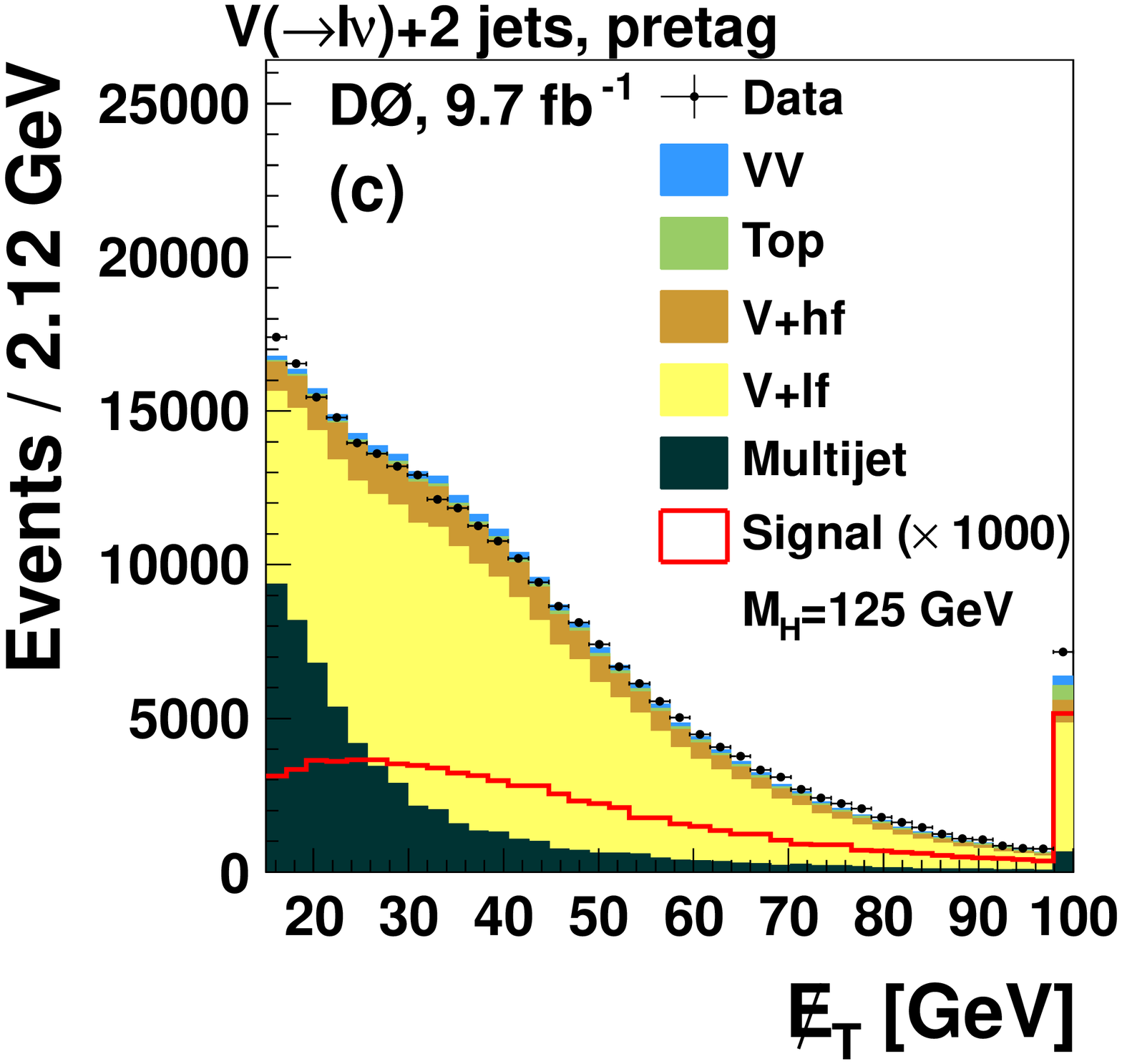}
}
\caption{(color online)
Distributions for all selected events with two jets of
(a) transverse mass of the lepton-\MET\ system,
(b) charged lepton $p_T$, and
(c) \MET.
The signal is multiplied by 1000. Overflow events are added to the last bin.
}
\label{fig:pretag}
\end{figure}

Events are required to have one isolated charged lepton, large \MET, and two or more jets, as described in Sec.~\ref{sec:leptonid}.
To suppress MJ backgrounds, events must satisfy the additional requirement that $\lnumt > 40\ \mathrm{GeV} - 0.5~\times\MET$, 
where $\lnumt$ is the transverse mass~\cite{defwtm} of the $W$ boson.  We then perform the final normalization of the $V$+jets and MJ backgrounds 
via a simultaneous fit to data in the $\lnumt$ distribution after subtracting the other SM background predictions from the 
data as described in Refs.~\cite{Abazov:2012wh97,Abazov:2010hn,Abazov:2012wh}.
The distribution of $\lnumt$ after this normalization procedure is shown in Fig.~\ref{fig:pretag}(a).
We perform separate fits for each lepton flavor and jet multiplicity category before dividing events into categories
based on the number and quality of identified $b$ jets, as described in Sec.~\ref{sec:bid}.  
All events passing these selection criteria constitute the pretag sample,
and each pretag event also belongs to exactly one of the six independent $b$-tag categories.
Only the zero and one-loose $b$-tag categories are considered when searching for the signal in events with four or more jets because
$t\bar{t}$ production dominates the small amount of signal present in higher $b$-tag categories.

\begin{figure}[htbp]
\centering{
\includegraphics[width=0.38\textwidth]{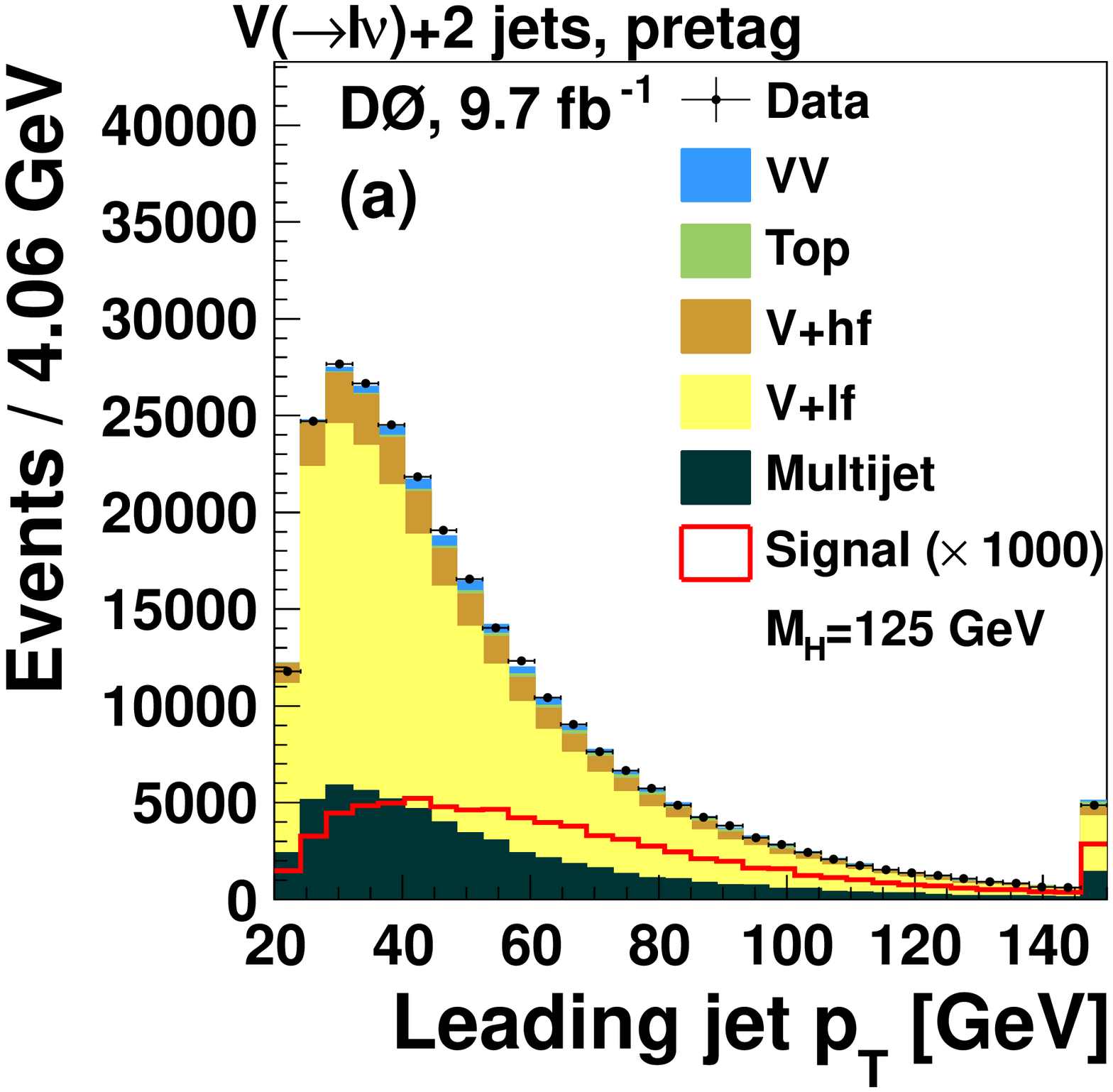}
\includegraphics[width=0.38\textwidth]{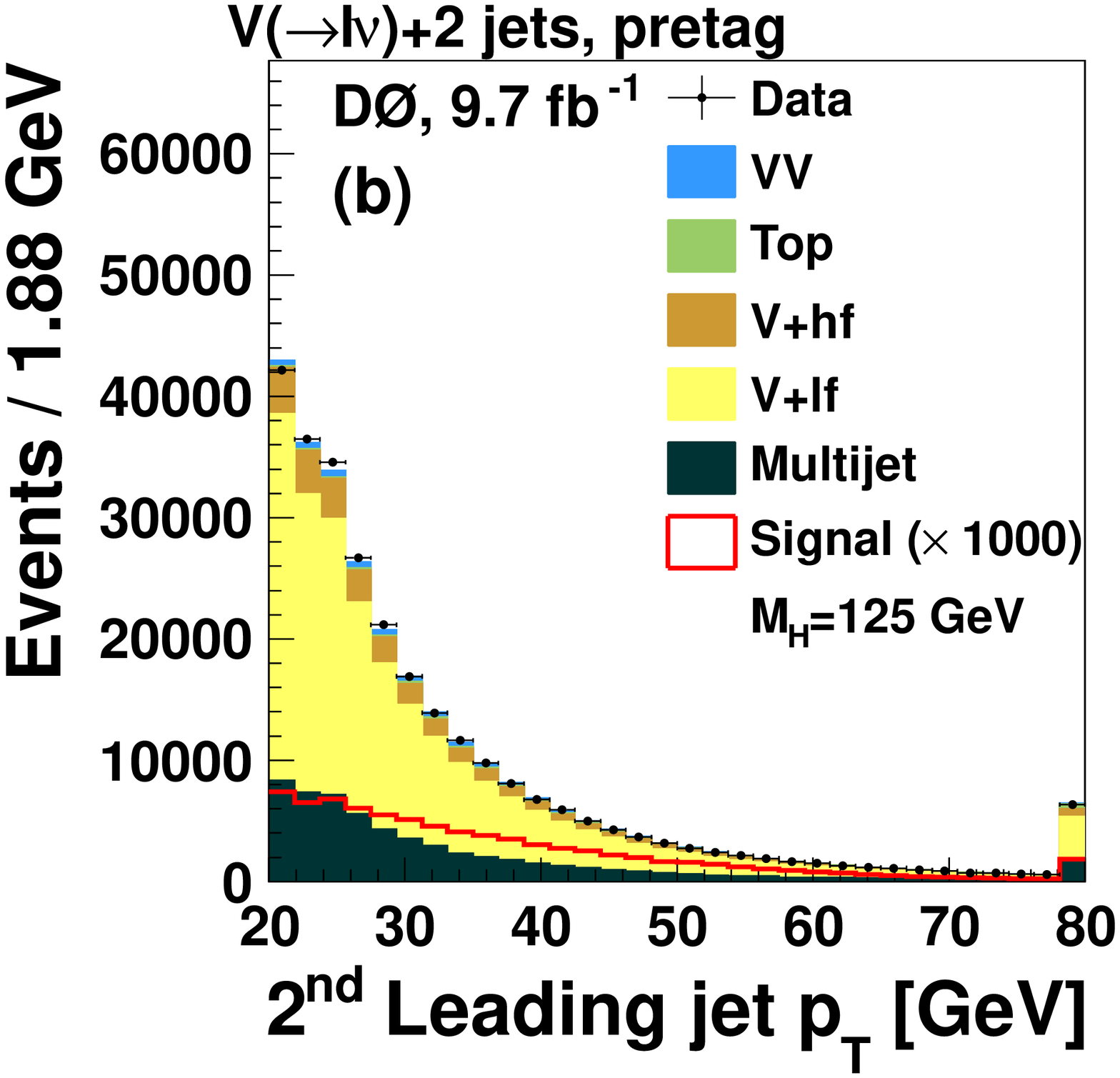}
\includegraphics[width=0.38\textwidth]{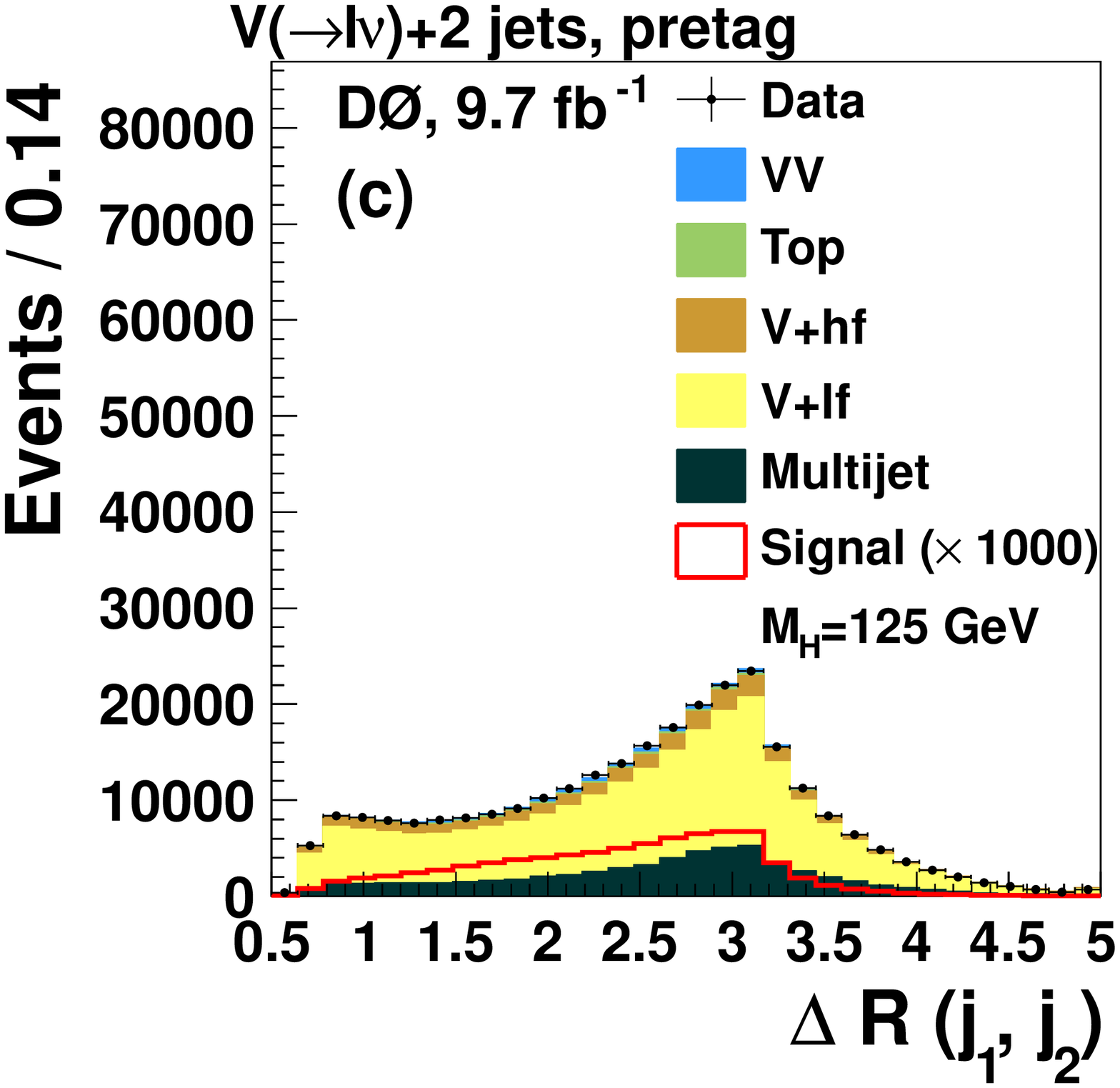}
}\caption{(color online)
Distributions for all selected events with two jets of
(a) leading jet $p_T$,
(b) second-leading jet $p_T$, and
(c) $\deltaR$ between the leading and second-leading jets.
The signal is multiplied by 1000. Overflow events are added to the last bin.
}
\label{fig:pretag-2}
\end{figure}

\begin{figure}[htbp]
\centering{
\includegraphics[width=0.38\textwidth]{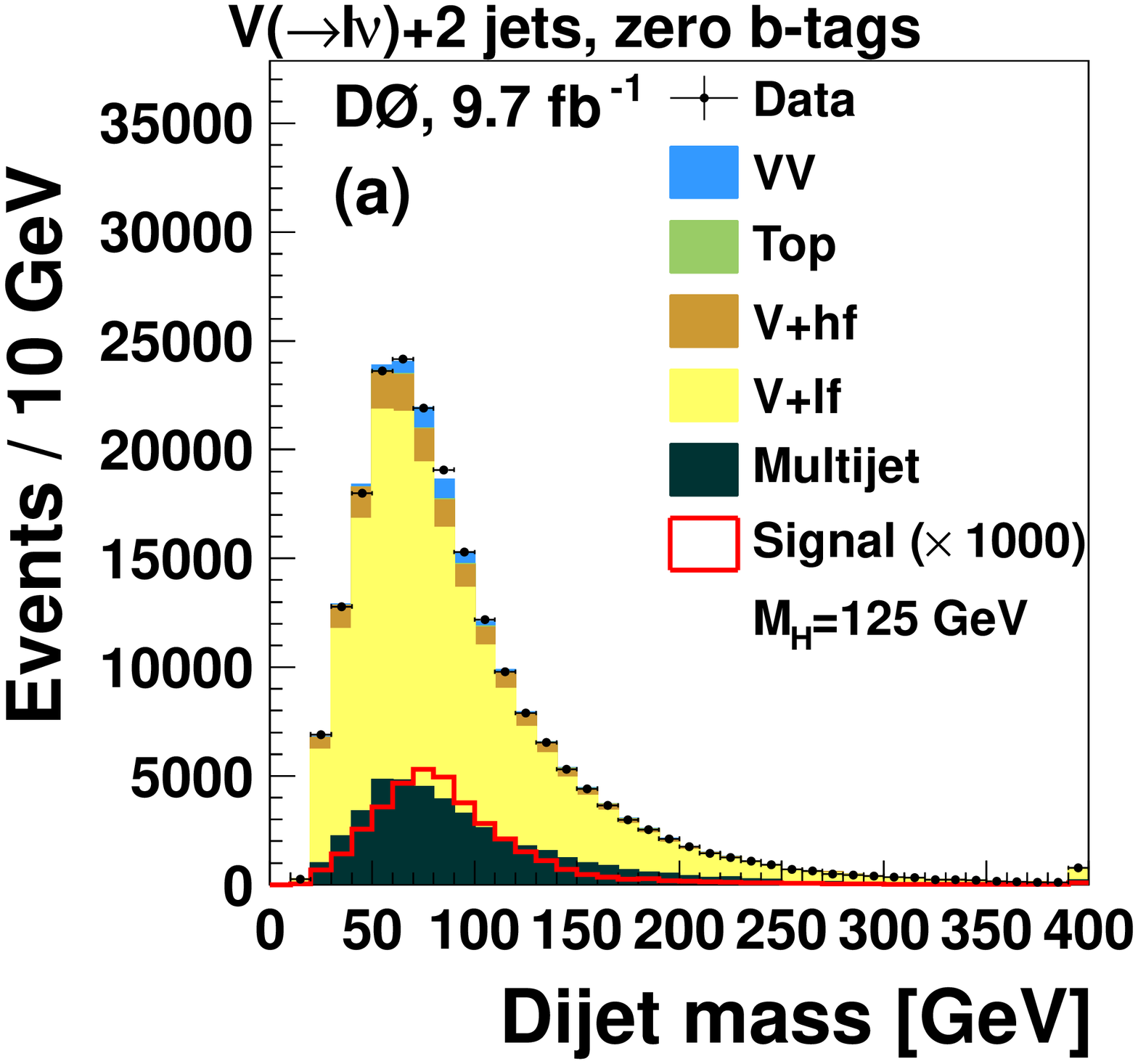}
\includegraphics[width=0.38\textwidth]{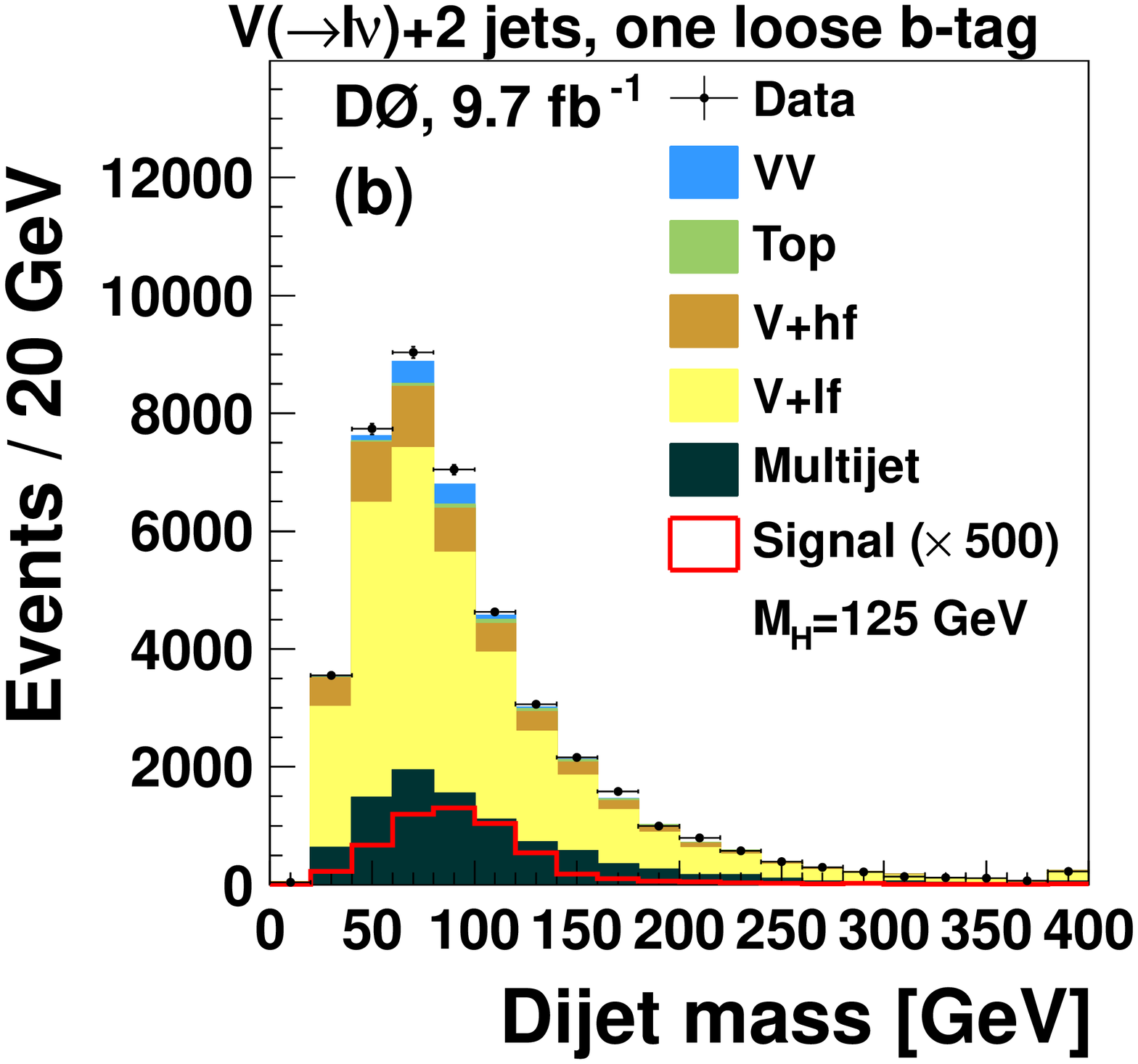}
\includegraphics[width=0.38\textwidth]{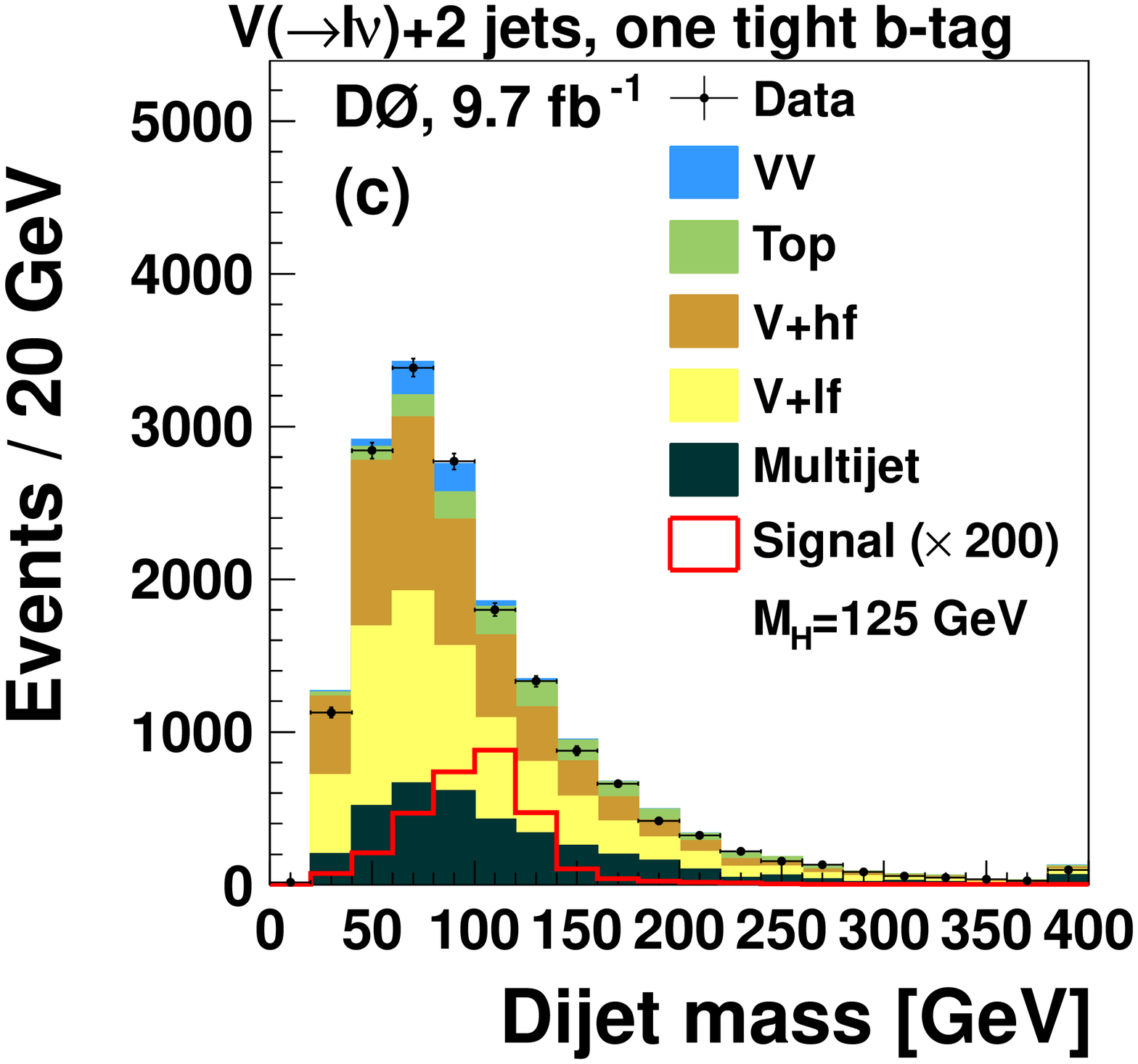}
}
\caption{(color online)
Invariant mass of the leading and second-leading jets in events with two jets and 
(a) zero $b$-tags,
(b) one loose $b$-tag, and
(c) one tight $b$-tag.
The signal is multiplied by 1000, 500, and 200, respectively. 
Overflow events are added to the last bin.
}
\label{fig:dijetm}
\end{figure}

\begin{figure}[htbp]
\centering{
\includegraphics[width=0.38\textwidth]{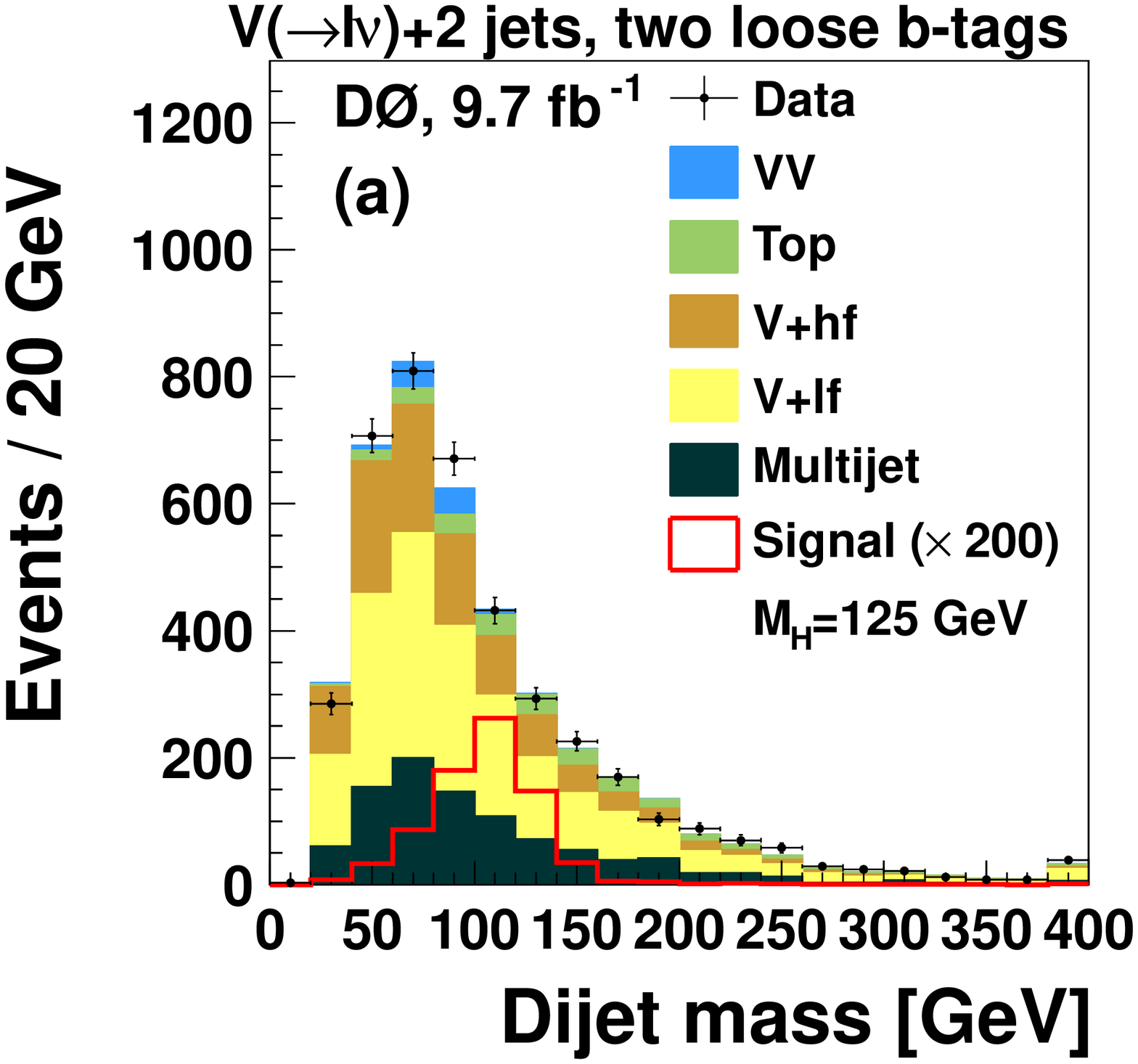}
\includegraphics[width=0.38\textwidth]{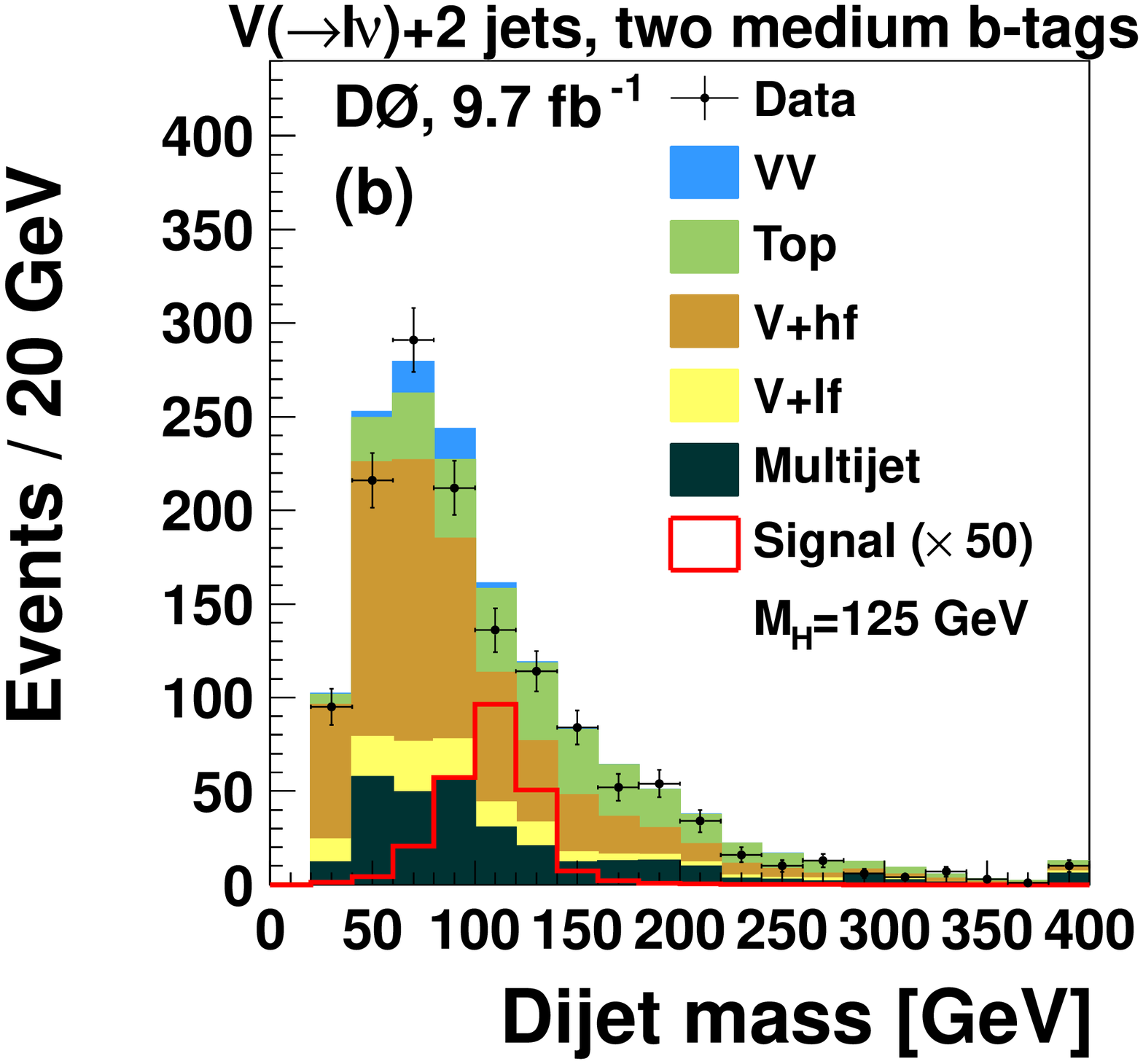}
\includegraphics[width=0.38\textwidth]{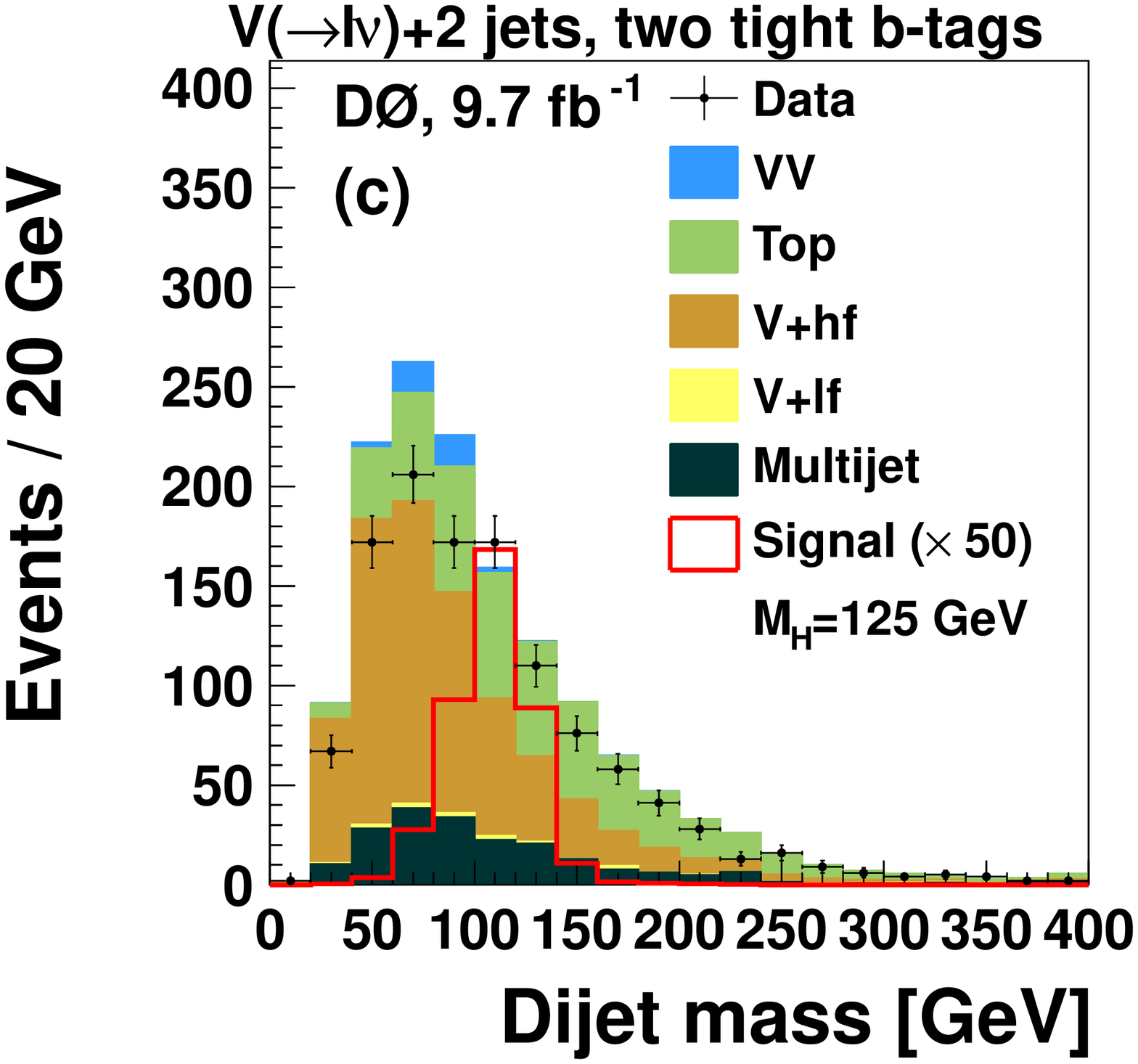}
}\caption{(color online)
Invariant mass of the leading and second-leading jets in events with two jets and 
(a) two loose $b$-tags,
(b) two medium $b$-tags, and
(c) two tight $b$-tags.
The signal is multiplied by 200, 50, and 50, respectively.
Overflow events are added to the last bin.
}
\label{fig:dijetm-2}
\end{figure}

The expected number of events from each signal and background category is compared to the observed data for each $b$-jet identification category 
for events with two jets, three jets, and four or more jets in Tables~\ref{tab:yield2j}, \ref{tab:yield3j}, and \ref{tab:yield4j}, respectively.  
Selected kinematic distributions are shown for all selected events in Figs.~\ref{fig:pretag} and \ref{fig:pretag-2}, and the dijet invariant mass for events with two jets 
is shown for all $b$-tag categories in Figs.~\ref{fig:dijetm} and \ref{fig:dijetm-2}. 
In all plots, data points are shown with error bars that reflect the statistical uncertainty only.
Discrepancies in data-MC agreement are within our systematic uncertainties described in Sec.~\ref{sec:syst}.

\begin{table*}[htbp]
\caption{\label{tab:yield2j}
Observed number of events in data and expected number of events from each signal and background source (where $V=W,Z$) for events with exactly two jets.
The expected signal is quoted at $M_H=125$~GeV.  The total background uncertainty includes all sources of systematic uncertainty added in quadrature.
}\centering{
\begin{tabular}{lrrrrrrr}
\hline
\hline
 & Pretag & 0 $b$-tags & 1 loose $b$-tag & 1 tight $b$-tag & 2 loose $b$-tags & 2 med.\ $b$-tags & 2 tight $b$-tags \\
\hline
$VH\to\ell \nu b\bar{b}$		& 37.3	  & 6.4		& 4.0		& 11.6	  & 3.2	& 4.6	& 7.7	  \\
$H\to VV\to\ell\nu jj$	                & 24.7	  & 18.8	& 3.9		& 1.8	  & 0.3	& 0.07	& 0	  \\
$VH\to VVV \to\ell\nu jjjj$	        & 13.0	  & 9.3		& 2.3		& 1.2	  & 0.3	& 0.04	& 0.01	  \\
\hline
Diboson					& 5686  & 4035	& 968		& 535	  & 109	& 42	& 38	 \\
$V+(g,u,d,s)$-jets                    	& 182\thinspace271 & 148\thinspace686 	& 26\thinspace421	& 6174  & 1762	& 132	& 13  \\
$V+(b\bar{b}/c\bar{c})$			& 27\thinspace443  & 15\thinspace089	& 4872		& 5236  & 978	& 691	& 691		\\
top ($t\bar{t}$ + single~top)		& 3528  & 758	& 455		& 1289	  & 247	& 333	& 462	  \\ 
Multijet				& 58\thinspace002  & 43\thinspace546	& 9316& 3700  & 946	& 298	& 195  \\ 
\hline
Total~expectation			& 276\thinspace930      & 212\thinspace114 	& 42\thinspace032	& 16\thinspace935  & 4043       & 1496        & 1400	 \\
Total~uncertainty                       & $\pm$ 14\thinspace998 & $\pm$ 11\thinspace352 & $\pm$ 2438            & $\pm$ 1696       &  $\pm$ 362 &  $\pm$ 117  & $\pm$ 175 \\
Observed~events				& 276\thinspace929 	& 211\thinspace169      & 42\thinspace774       & 16\thinspace406  & 4057       & 1358        & 1165  \\
\hline
\hline
\end{tabular}}
\end{table*}

\begin{table*}[htbp]
\caption{\label{tab:yield3j}
Observed number of events in data and expected number of events from each signal and background source (where $V=W,Z$) for events with exactly three jets.
The expected signal is quoted at $M_H=125$~GeV.  The total background uncertainty includes all sources of systematic uncertainty added in quadrature.
}\centering{
\begin{tabular}{lrrrrrrr}
\hline
\hline
 & Pretag & 0 $b$-tags & 1 loose $b$-tag & 1 tight $b$-tag & 2 loose $b$-tags & 2 med.\ $b$-tags & 2 tight $b$-tags \\
\hline
$VH\to\ell\nu b\bar{b}$		& 8.6		& 1.3		& 1.0	& 2.4	& 0.9	& 1.1	& 1.7	 \\
$H\to VV\to\ell\nu jj$	& 8.8		& 6.0		& 1.7	& 0.8	& 0.3	& 0.07	& 0.01	 \\
$VH\to VVV\to\ell\nu jjjj$	& 7.3		& 4.5		& 1.6	& 0.9	& 0.3	& 0.05	& 0.01	 \\
\hline
Diboson					& 1138		& 727		& 238	& 113	& 42	& 14 	& 10	\\
$V+(g,u,d,s)$-jets                      & 24\thinspace086	& 18\thinspace078	& 4577& 976	& 582	& 34	& 3	\\
$V+(b\bar{b}/c\bar{c})$			& 6625& 3213	& 1349& 1250& 411	& 228	& 164	 \\
top ($t\bar{t}$ + single~top)		& 3695	& 563		& 419	& 1123  & 365	& 460	& 570	  \\
Multijet				& 10\thinspace364	& 6629		& 2162& 933	& 367	& 130	& 82	\\
\hline
Total~expectation			& 45\thinspace908 	& 29\thinspace209 & 8746      & 4395       & 1768      & 867       & 830	      \\
Total~uncertainty                       & $\pm$ 2582            & $\pm$ 1619      & $\pm$ 587 & $\pm$ 528  & $\pm$ 209 & $\pm$ 118 & $\pm$ 113  \\
Observed~events				& 45\thinspace907       & 28\thinspace924 & 8814      & 4278       &  1815     &  879      & 797 \\
\hline
\hline
\end{tabular}}
\end{table*}

\begin{table}[tbp]
\caption{\label{tab:yield4j}
Observed number of events in data and expected number of events from each signal and background source (where $V=W,Z$) for events with four or more jets.
The expected signal is quoted at $M_H=125$~GeV.  The total background uncertainty includes all sources of systematic uncertainty added in quadrature.
}\centering{
\begin{tabular}{lrrr}
\hline
\hline
\multicolumn{2}{r}{Pretag} & 0 $b$-tags & 1 loose $b$-tag \\
\hline
$VH\to\ell\nu b\bar{b}$			& 1.4	&   0.2	&  0.2 \\ 
$H\to VV\to\ell\nu jj$			& 2.4	&   1.4	&  0.6 \\
$VH\to VVV\to\ell\nu jjjj$	        & 3.6	&   2.0	&  0.8 \\
\hline
Diboson			                &  199	&   112	&   46 \\
$V+(g,u,d,s)$-jets			& 3055	&  2143	&  679 \\
$V+(b\bar{b}/c\bar{c})$			& 1280	&   542	&  286 \\
top ($t\bar{t}$ + single~top) 	        & 2889	&   311	&  268 \\
Multijet			        & 2092	&  1110	&  450 \\
\hline
Total~expectation			& 9516	&  4217	& 1729 \\
Total~uncertainty                       & $\pm$  530  & $\pm$   231 & $\pm$  144 \\
Observed~events			        & 9685	&  3915	& 1786 \\
\hline
\hline
\end{tabular}}
\end{table}

\section{Multivariate Signal Discriminants}\label{sec:MVA}

We employ multivariate analysis (MVA) techniques to separate signal from background events. To separate signal from the MJ events, we
use a boosted decision tree implemented with the {\sc tmva} package~\cite{Hocker:2007ht} . 
This multivariate analysis is 
described in Sec.~\ref{sec:mjmva}.
A BDT is also used to separate signal from other specific background sources in events with four or more jets (see Sec.~\ref{sec:4jmva}).
For the final multivariate analysis, we use a BDT in the
 one tight $b$-tag channel and all three two $b$-tag channels, and we use a random forest decision tree (RF)~\cite{Breiman:2001:RF:570181.570182} implemented in the {\sc statpatternrecognition} (SPR) 
package~\cite{narsky-0507143,narsky-0507157}
for events in the zero and one loose $b$-tag channels.

The BDT and the RF are forms of machine learning techniques known as decision trees. Decision trees operate on a series of yes/no splits on events that are known to be classified as either signal or background. The splitting is done to maximally separate signal from background. The resulting nodes are continually split to optimally separate signal from background until either a minimum number of events in a node is reached or the events in a node are pure signal or pure background. The technique of boosting in the BDT builds up a series of trees where each tree is retrained, boosting the weights for events that are misclassified in the previous training. The RF technique creates a collection of decision trees where each tree is trained on a subset of the training data that is randomly sampled.

 We train separate BDTs and RFs for each lepton flavor, jet multiplicity, and tagging category, and for each hypothesized Higgs 
 boson mass in steps of 5~GeV. Since the branching fraction for the Higgs decay to $b$ quarks is only significant over the mass range 90--150~GeV, we restrict the search in the one tight and two $b$-tag channels to this range of $M_H$. In the zero and one loose $b$-tag channels, the primary signal contribution is from Higgs decays to vector bosons, the search is performed over the mass range of 100--200~GeV.

Each of the final BDTs and RFs are trained to distinguish the signal from all of the backgrounds. 
We choose 
variables to train the BDTs and RFs that have good agreement between data and 
background simulation (since the expected contribution from signal events is small), and so that there is a good separation between signal and at least one background.
Background and signal samples are each split into three independent samples for use in training, testing, and performing the final statistical analysis
with each multivariate discriminant. We ensure that the discriminant is not biased towards statistical fluctuations in the training
sample by comparing the training output to the testing sample. The independent sample used for the limit setting procedure ensures
that any optimizations performed based on the output of the training and testing samples do not bias the final limits.

\subsection{Multivariate multijet discriminators \label{sec:mjmva}}

\begin{figure*}
\centering{
\includegraphics[width=0.38\textwidth]{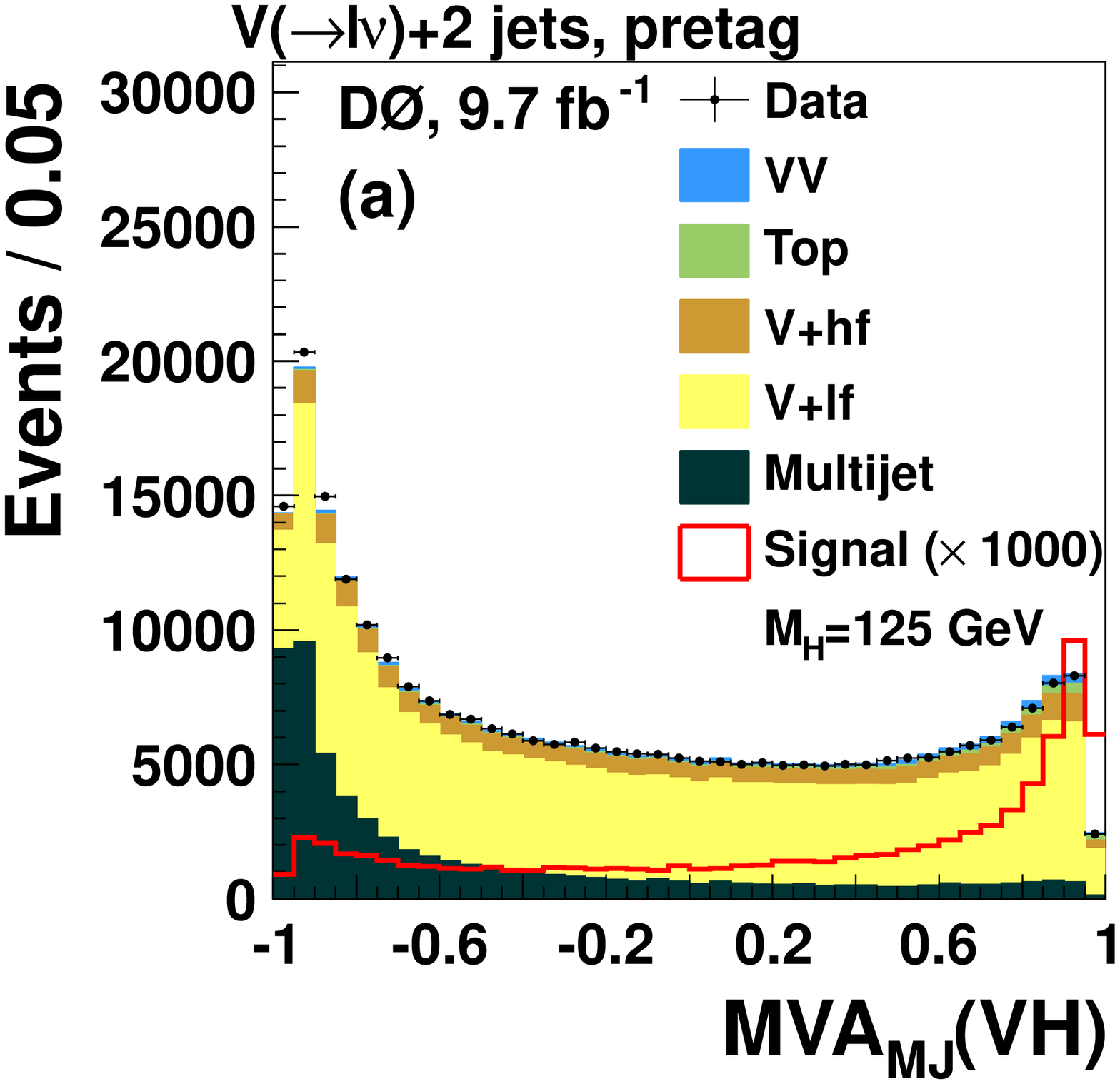}
\includegraphics[width=0.38\textwidth]{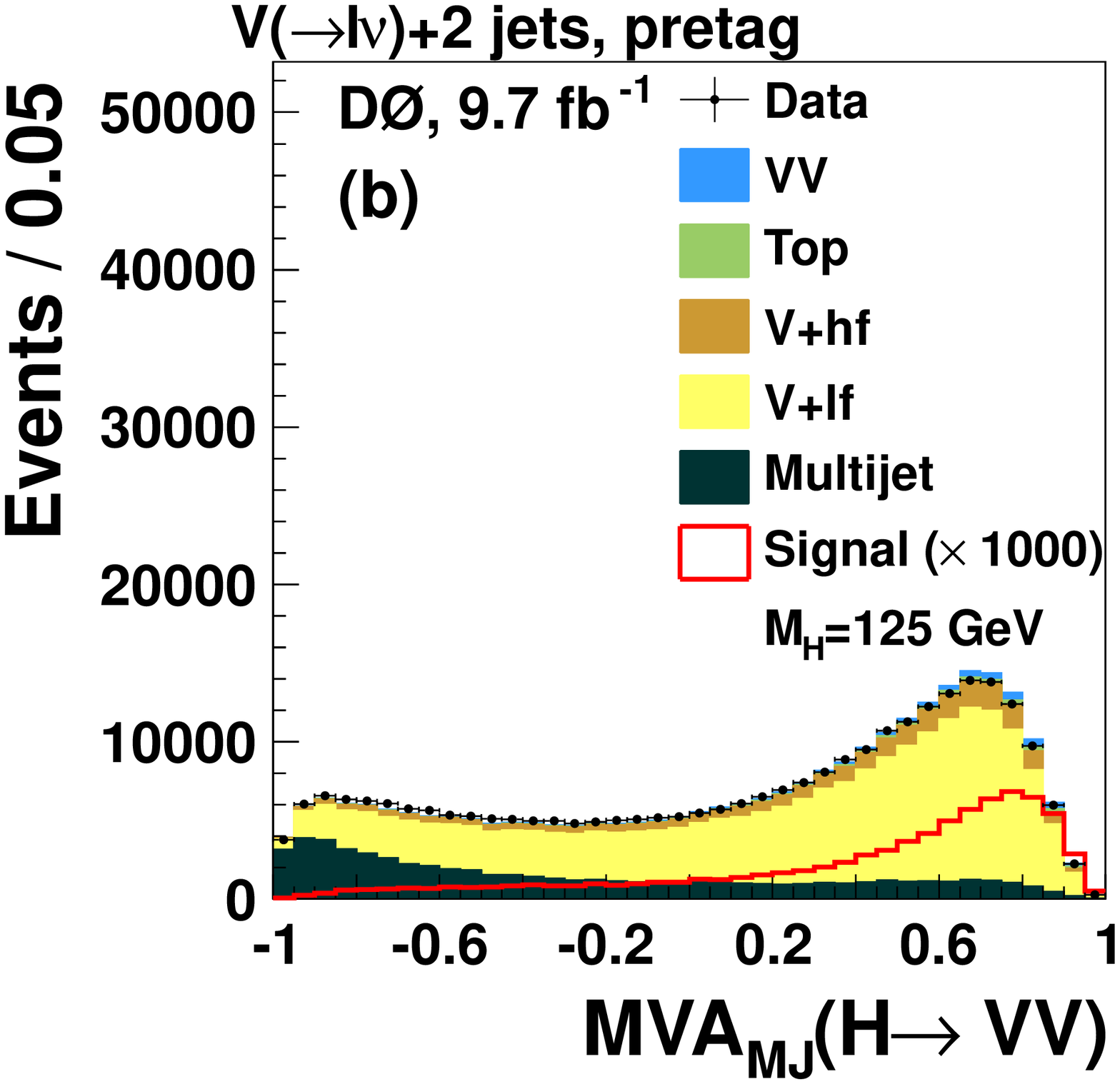}
}\caption{(color online) The multivariate discriminant output for 
(a) $\mvamjvsvh$ and 
(b) $\mvamjvshvv$, for all events.
The signal is multiplied by 1000.
}
\label{fig:mvaqcd}
\end{figure*}

We train two separate BDTs to separate the MJ background 
from signal events: one for 
$VH(\to b\bar{b},c\bar{c},\tau\tau)$ signals, $\mvamjvsvh$, and one for $H\to VV$ signals, $\mvamjvshvv$.
The variables used in training these BDTs are chosen to exploit kinematic differences between 
the MJ and signal events, and are documented in Appendix~\ref{app:input_var_def}. 
To improve the training statistics, we combine signal events for $M_{H}=120$, 125, and 130~GeV in training. 
We find that a BDT trained on this combination of Higgs boson masses has a similar performance when applied to other masses, 
eliminating the need for a mass dependent MJ discriminant.
The BDT outputs $\mvamjvsvh$ and $\mvamjvshvv$
 are shown in Fig.~\ref{fig:mvaqcd}.  
The $\mvamjvsvh$ and $\mvamjvshvv$ discriminant outputs are used as input variables to the final MVAs, as detailed in Appendix~\ref{app:input_var_def}.

\subsection{Final $\bm{WH\to\ell\nu b\bar{b}}$ MVA analysis}

\begin{figure*}[htbp]
\centering{
\includegraphics[width=0.38\textwidth]{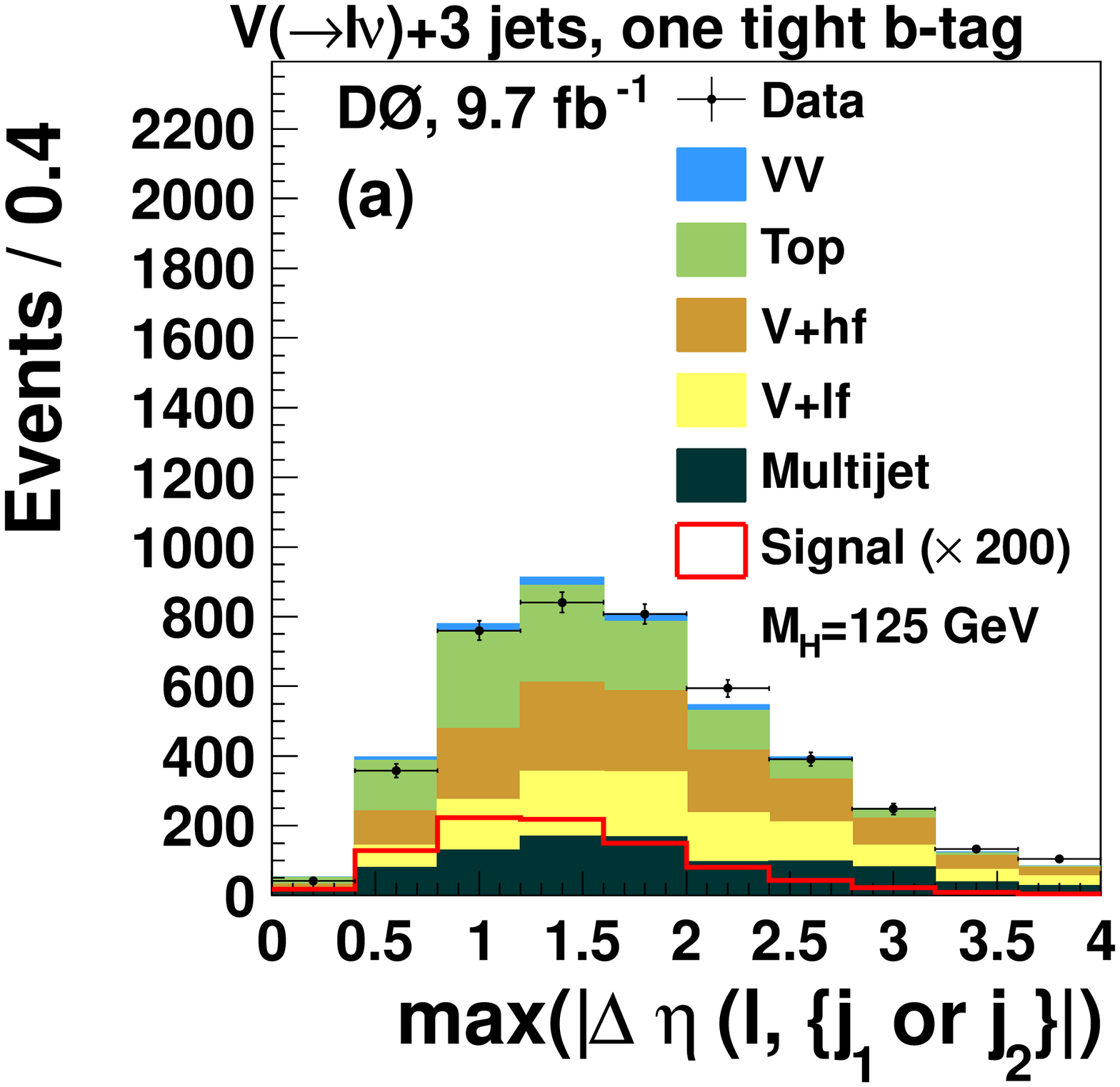}
\includegraphics[width=0.38\textwidth]{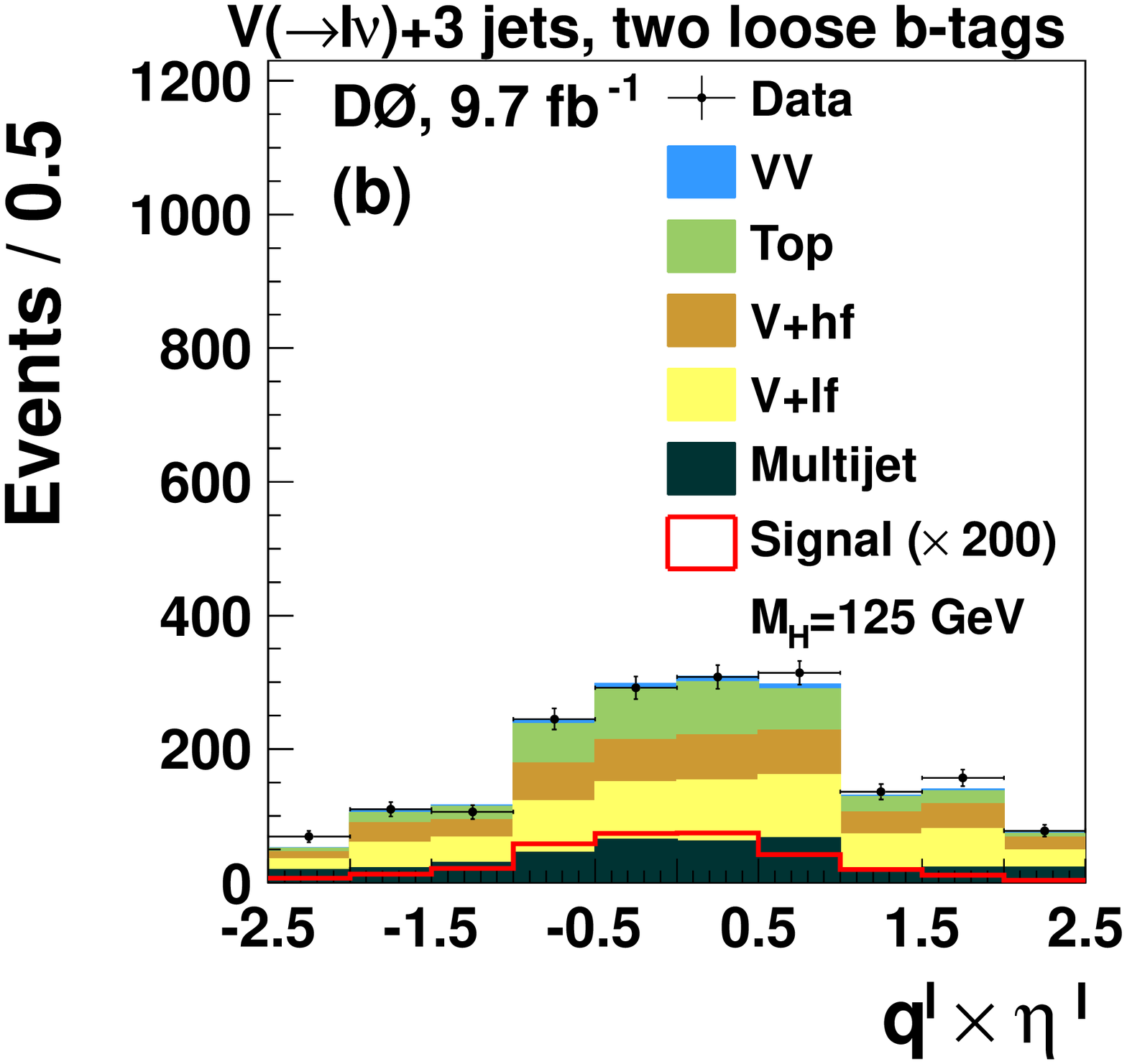}
\includegraphics[width=0.38\textwidth]{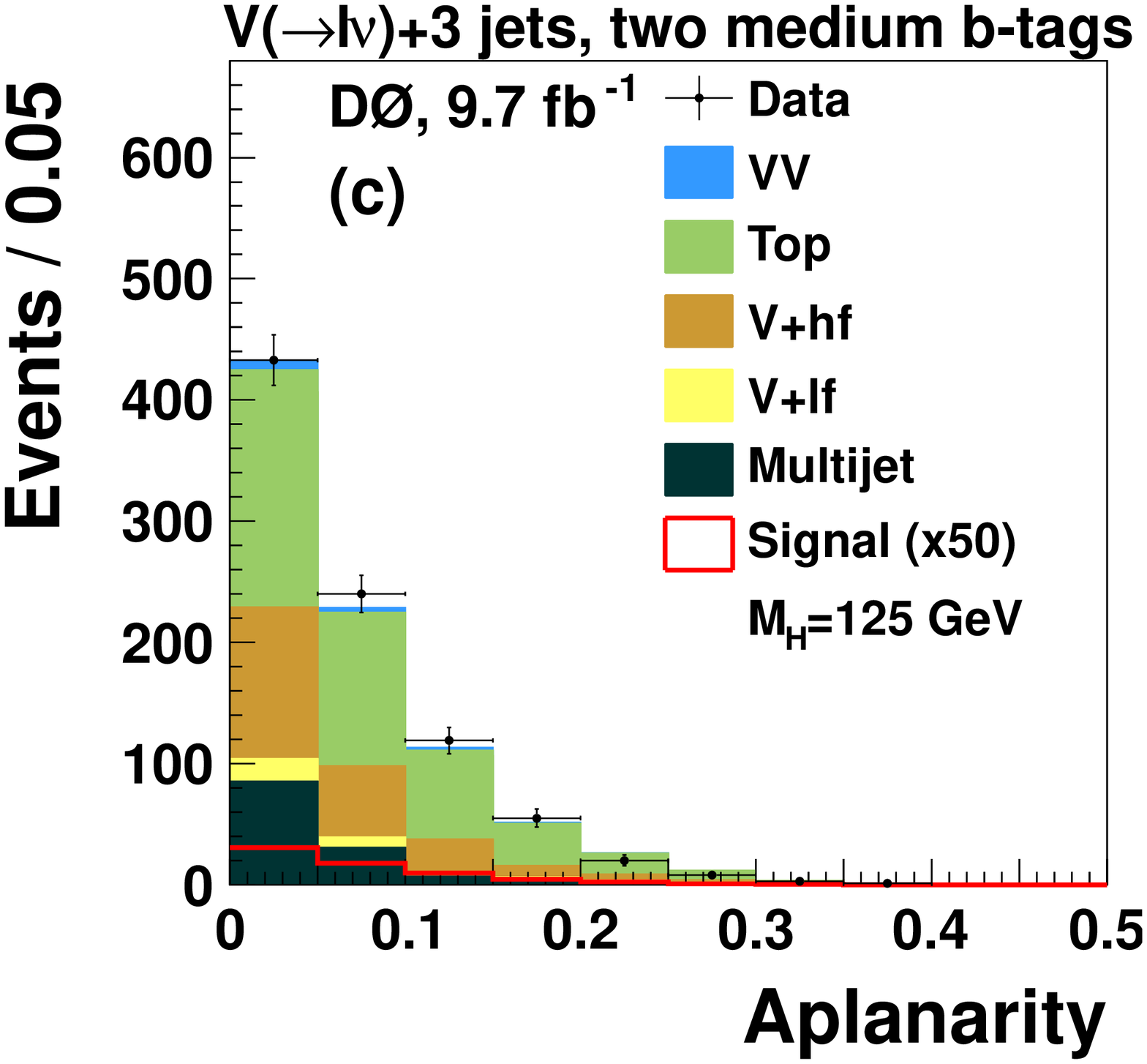}
\includegraphics[width=0.38\textwidth]{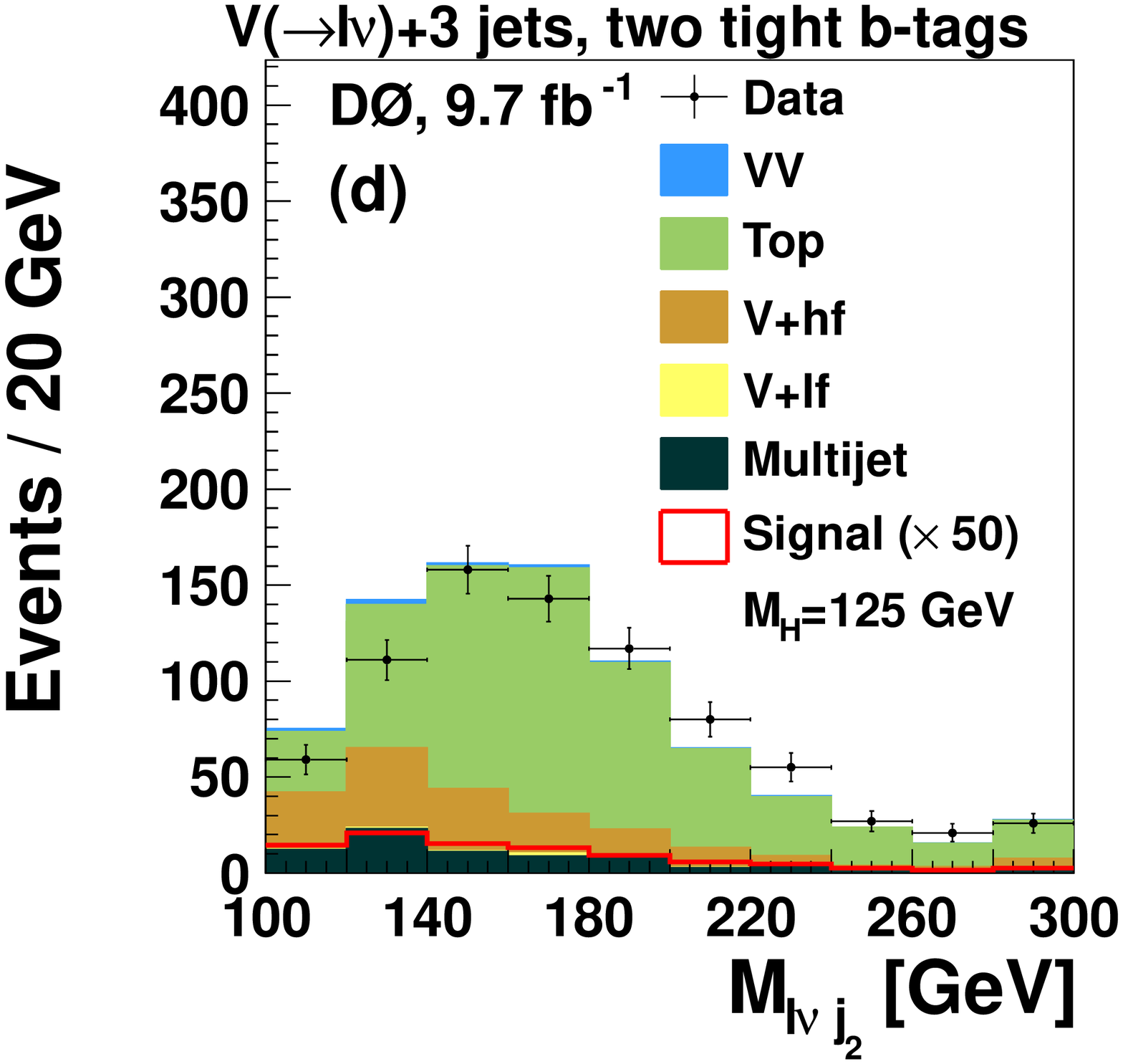}
}\caption{(color online)
Distributions of some of the most significant inputs to the final discriminant in events with exactly three jets
and either one tight $b$-tag, two loose $b$-tags, two medium $b$-tags, or two tight $b$-tags:
(a) \jetlepdetamax~\cite{defsjetlepdetamax}, shown for events with one tight $b$-tag;
(b) \lepqeta~\cite{defslepqeta}, shown for events with two loose $b$-tags;
(c) aplanarity~\cite{defstopoaplanarity}, shown for events with two medium $b$-tags;
(d) \lnujmB~\cite{defslnujmB,defsnupz}, shown for events with two tight $b$-tags.
The signal is multiplied by 200, 50, 50, and 50, respectively.
Overflow events are added to the last bin.
}
\label{fig:lvbb3jin}
\end{figure*}

\begin{figure*}[htbp]
\centering{
\includegraphics[width=0.38\textwidth]{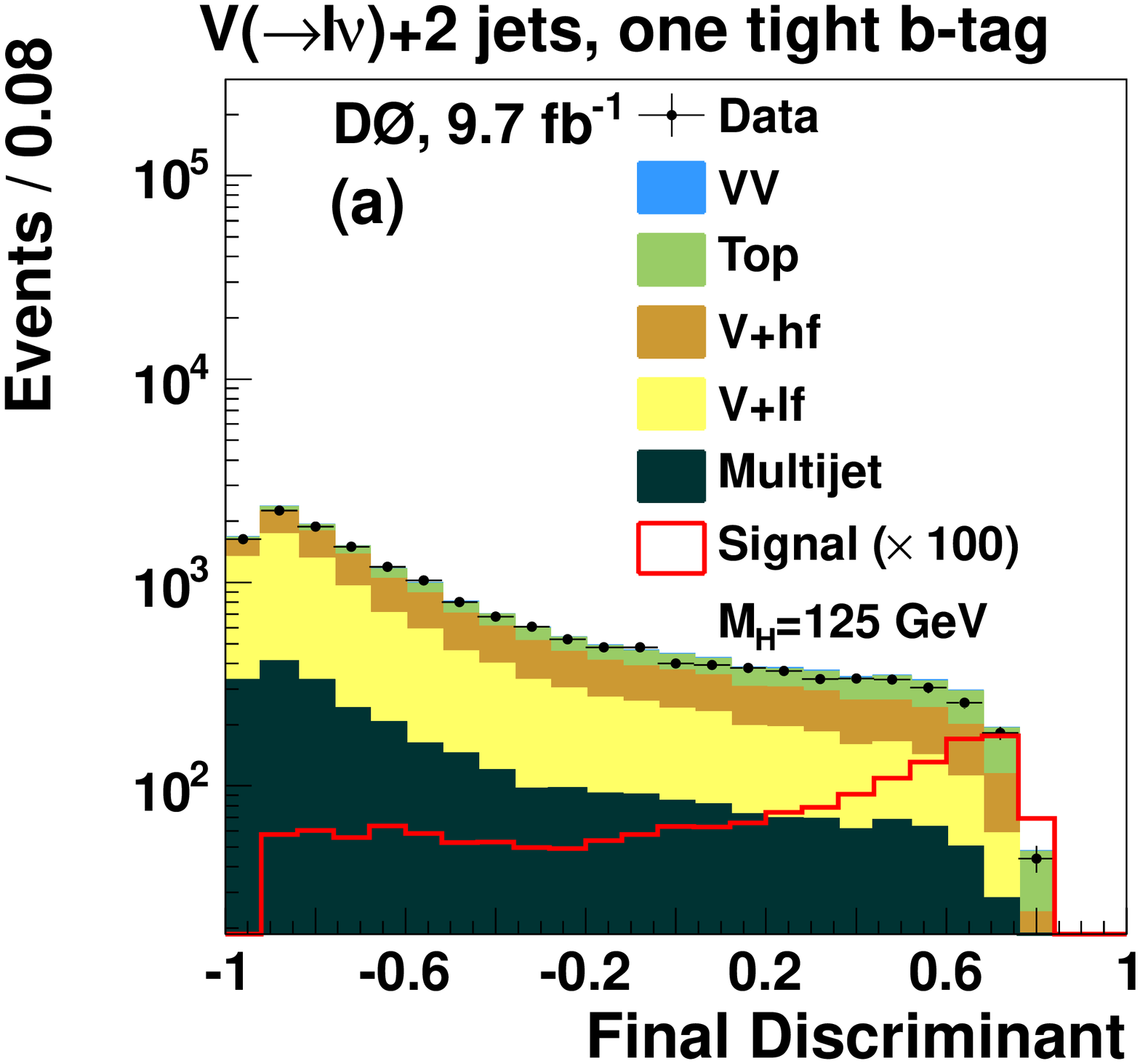}
\includegraphics[width=0.38\textwidth]{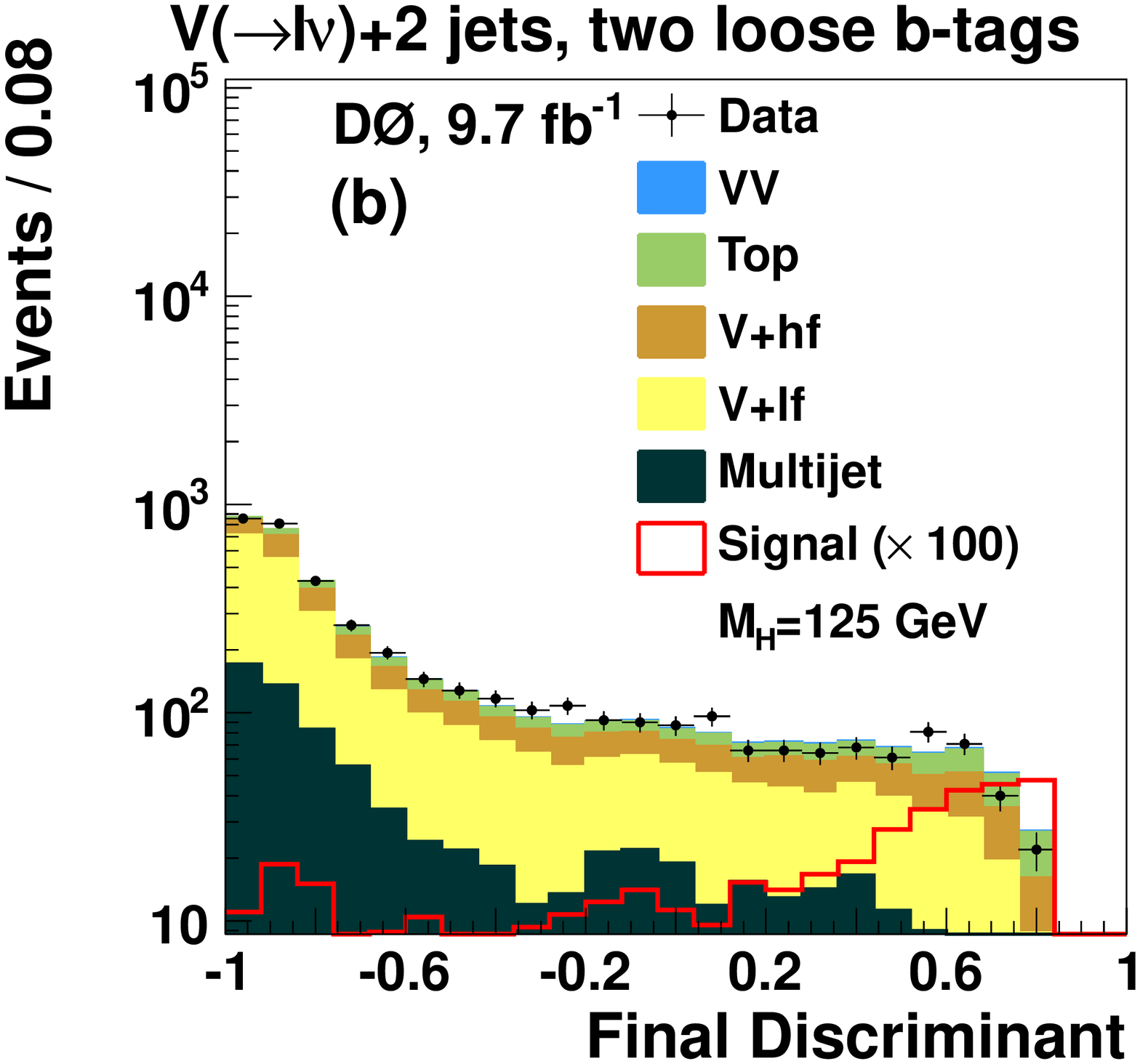}
\includegraphics[width=0.38\textwidth]{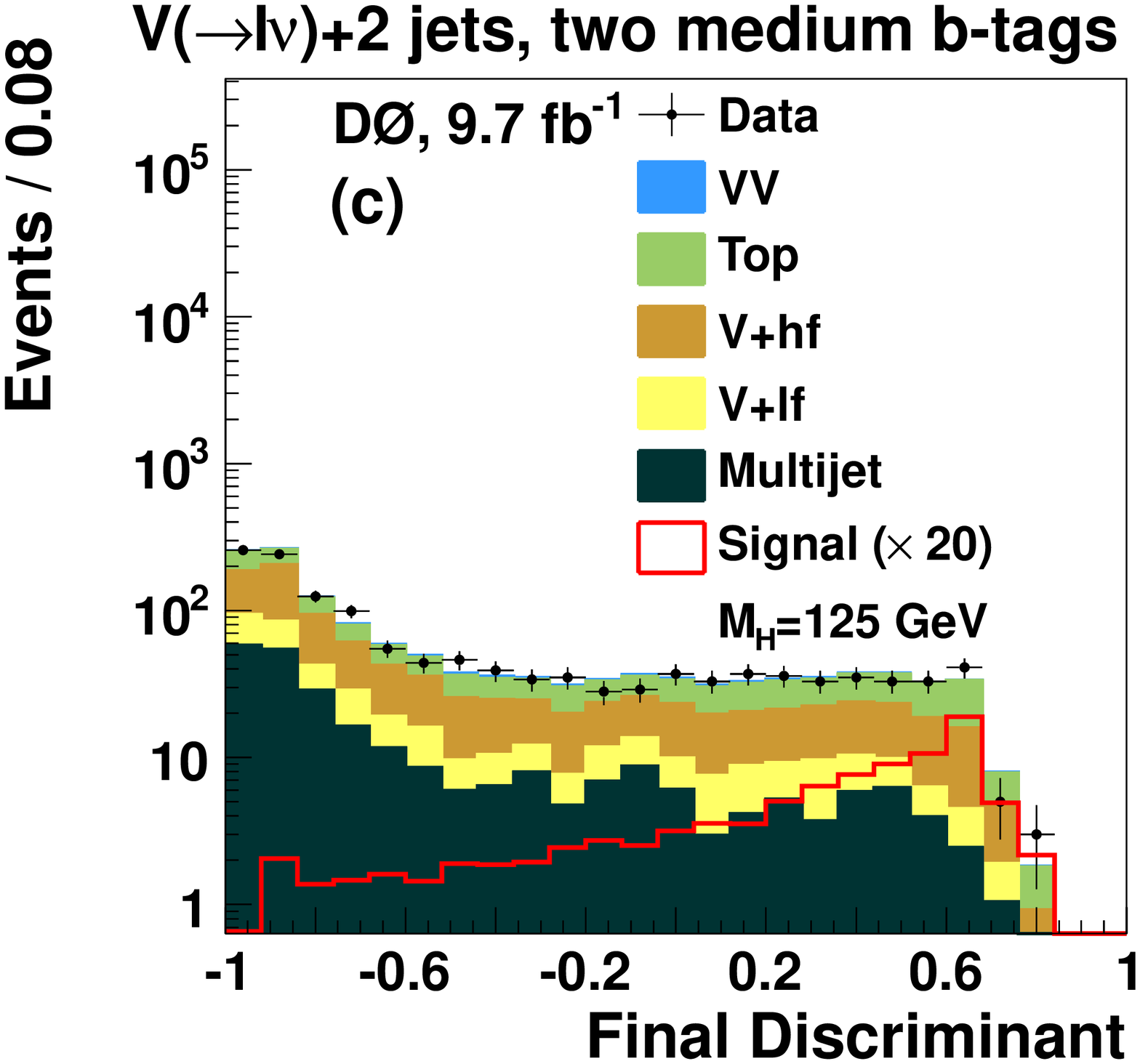}
\includegraphics[width=0.38\textwidth]{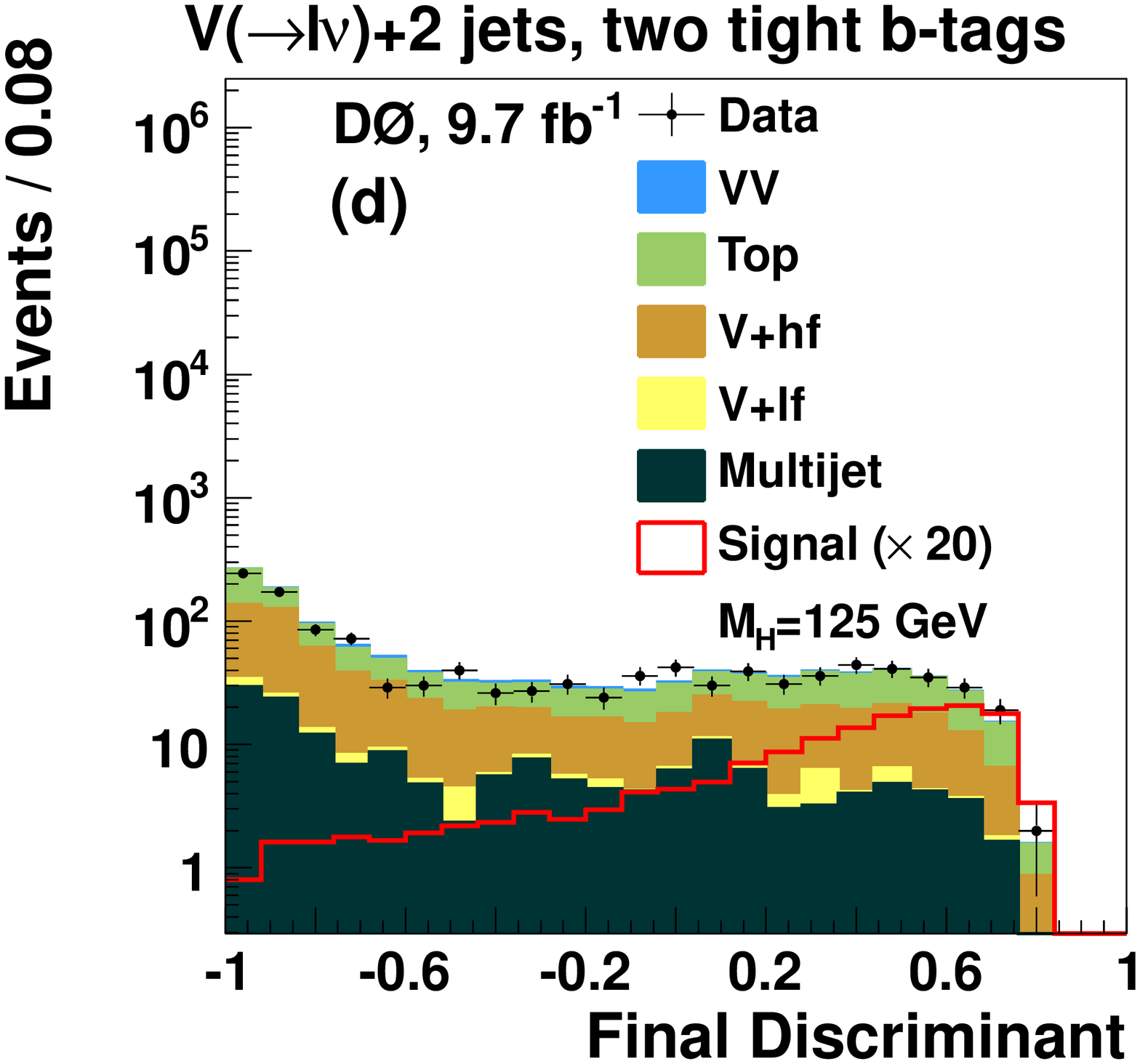}
}\caption{(color online)
Distributions of the final discriminant output, after the maximum likelihood fit (described in Sec.~\ref{sec:limits}), in events with exactly two jets and:
(a) one tight $b$-tag, (b) two loose $b$-tags, (c) two medium $b$-tags, and (d) two tight $b$-tags.
The signal is multiplied by 100, 100, 20, and 20, respectively.
}
\label{fig:lvbb2jout}
\end{figure*}

\begin{figure*}[htbp]
\centering{
\includegraphics[width=0.38\textwidth]{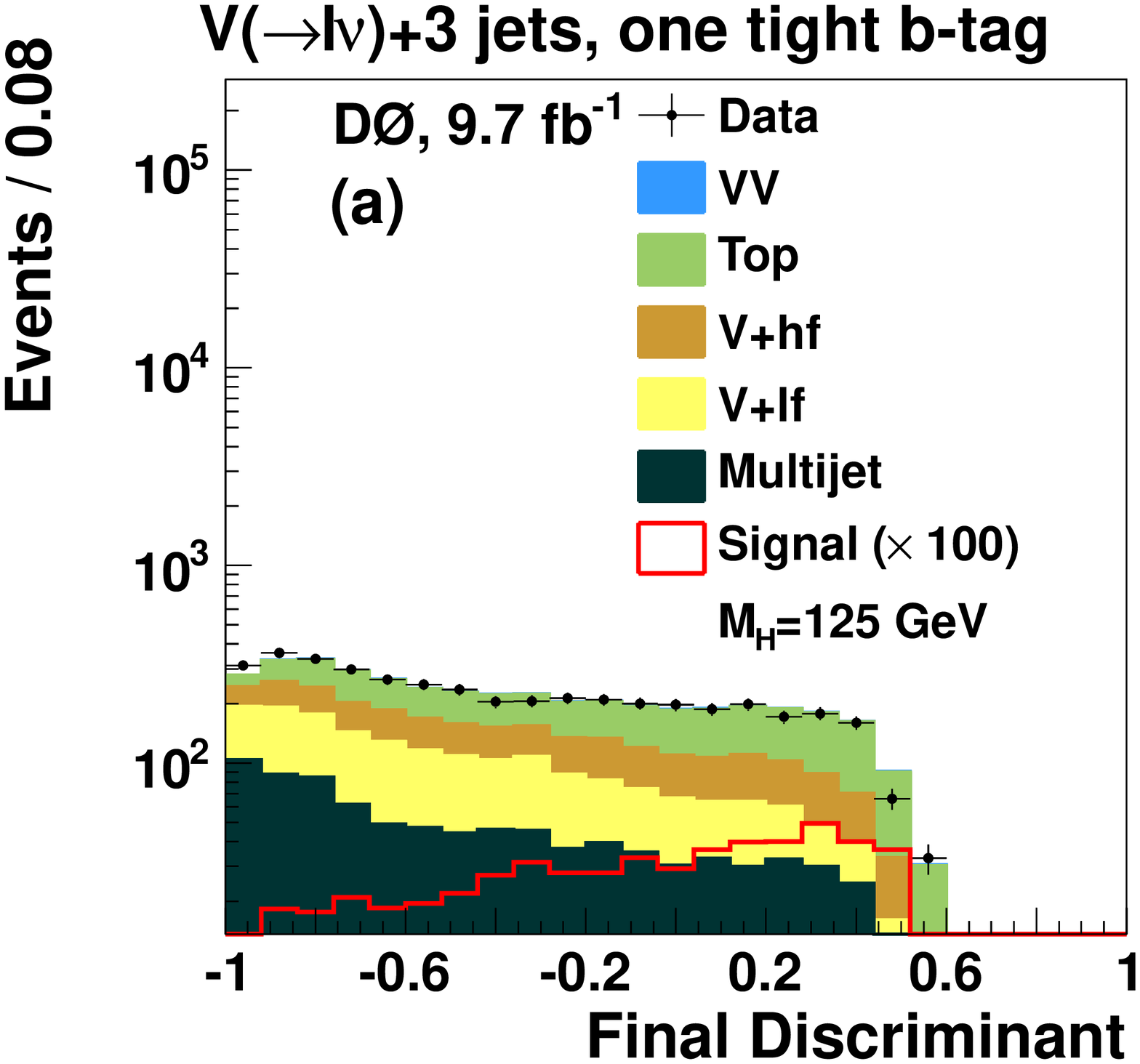}
\includegraphics[width=0.38\textwidth]{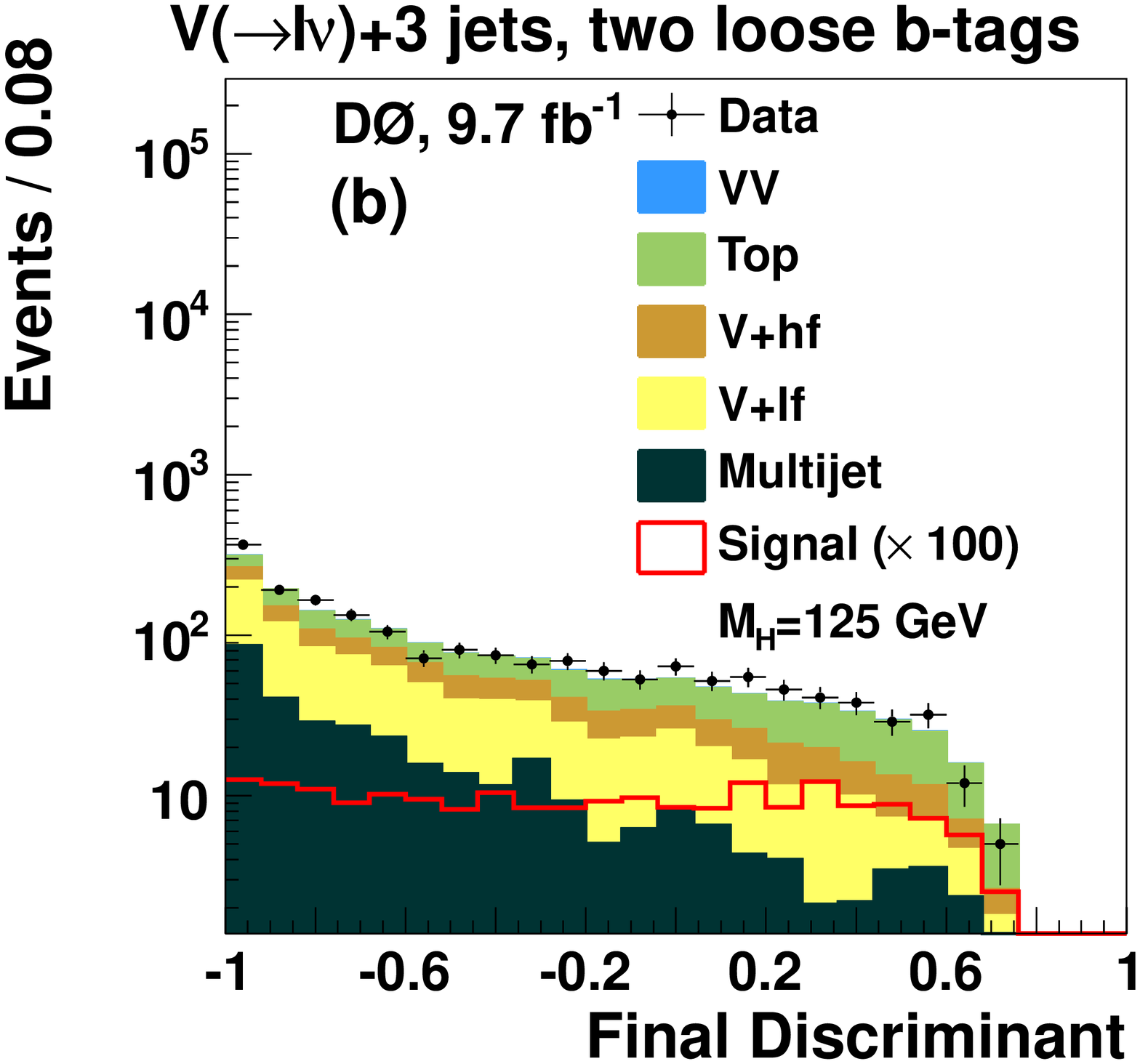}
\includegraphics[width=0.38\textwidth]{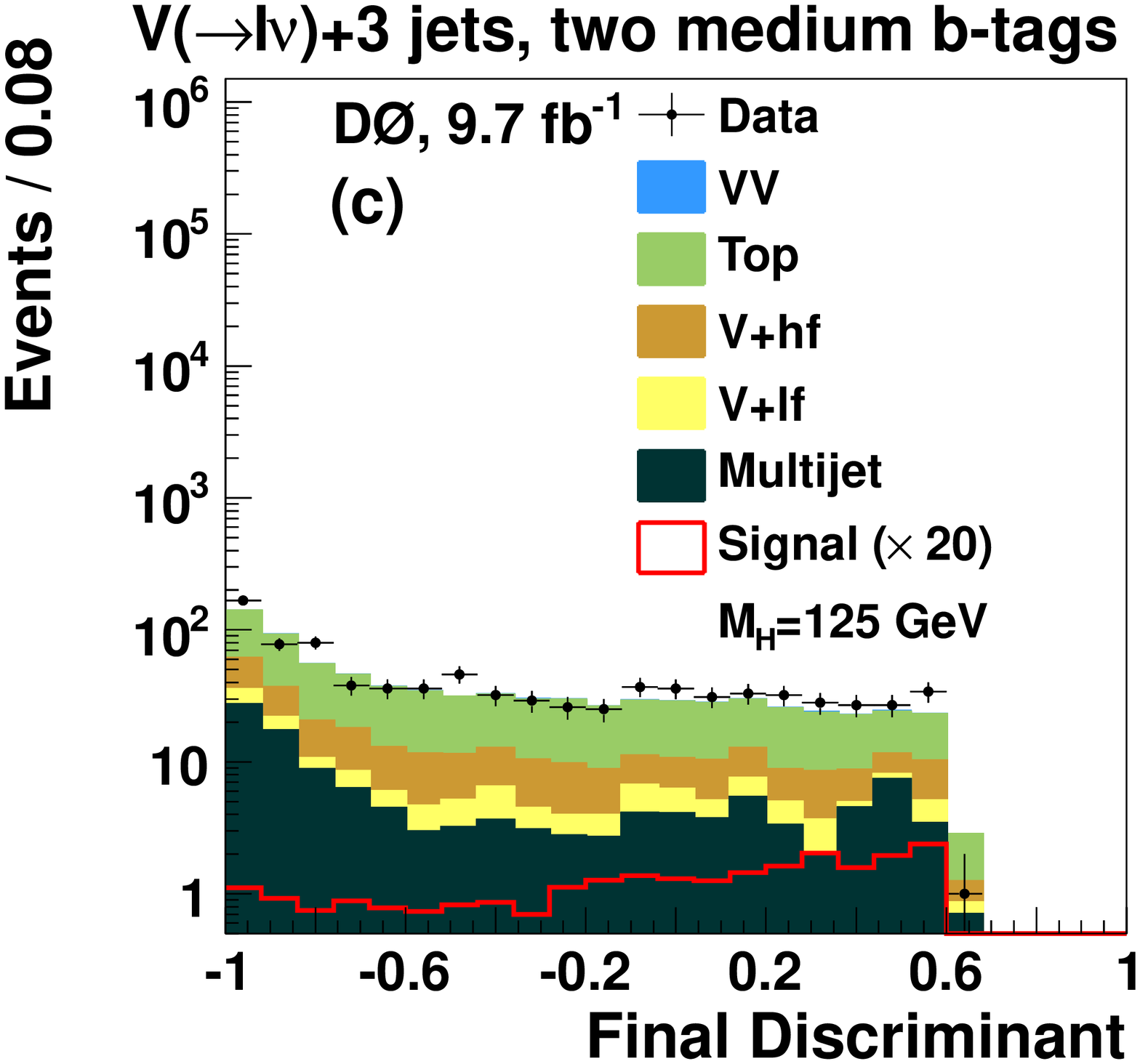}
\includegraphics[width=0.38\textwidth]{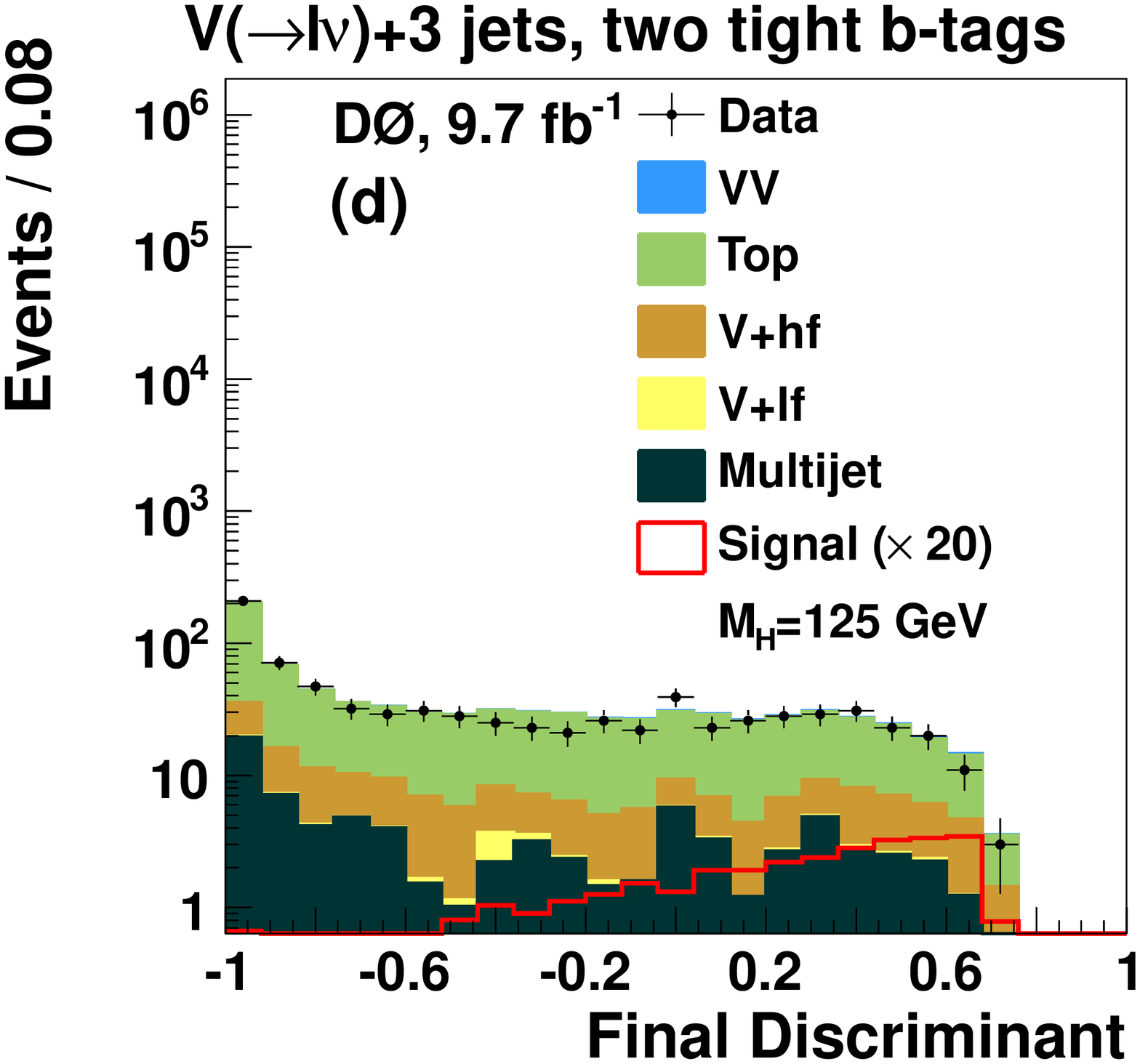}
}\caption{(color online)
Distributions of the final discriminant output, after the maximum likelihood fit (described in Sec.~\ref{sec:limits}), in events with exactly three jets and:
(a) one tight $b$-tag, (b) two loose $b$-tags, (c) two medium $b$-tags, and (d) two tight $b$-tags.
The signal is multiplied by 100, 100, 20, and 20, respectively.
}
\label{fig:lvbb3jout}
\end{figure*}

In events with two or three jets and one tight $b$-tag or two $b$-tags, the $WH\to\ell\nu b\bar{b}$ process provides the dominant signal contribution. 
To separate signal from background, we train a BDT on the $WH\to \ell \nu b\bar{b}$ 
signal and all backgrounds. The lists of input variables to the MVA and their descriptions are included in Appendix~\ref{app:input_var_def}.
Figures~\ref{fig:lvbb2jin} and \ref{fig:lvbb3jin} show examples of some of the most effective 
discriminating variables used in our BDTs for the two-jet and three-jet channels, respectively, in the one 
tight $b$-tag and all two $b$-tags channels. Figures~\ref{fig:lvbb2jout} and \ref{fig:lvbb3jout} show the 
BDT output for the two and three-jet channels, respectively, in the one tight $b$-tag and all the two $b$-tag channels.

\begin{figure*}[htbp]
\centering{
\includegraphics[width=0.38\textwidth]{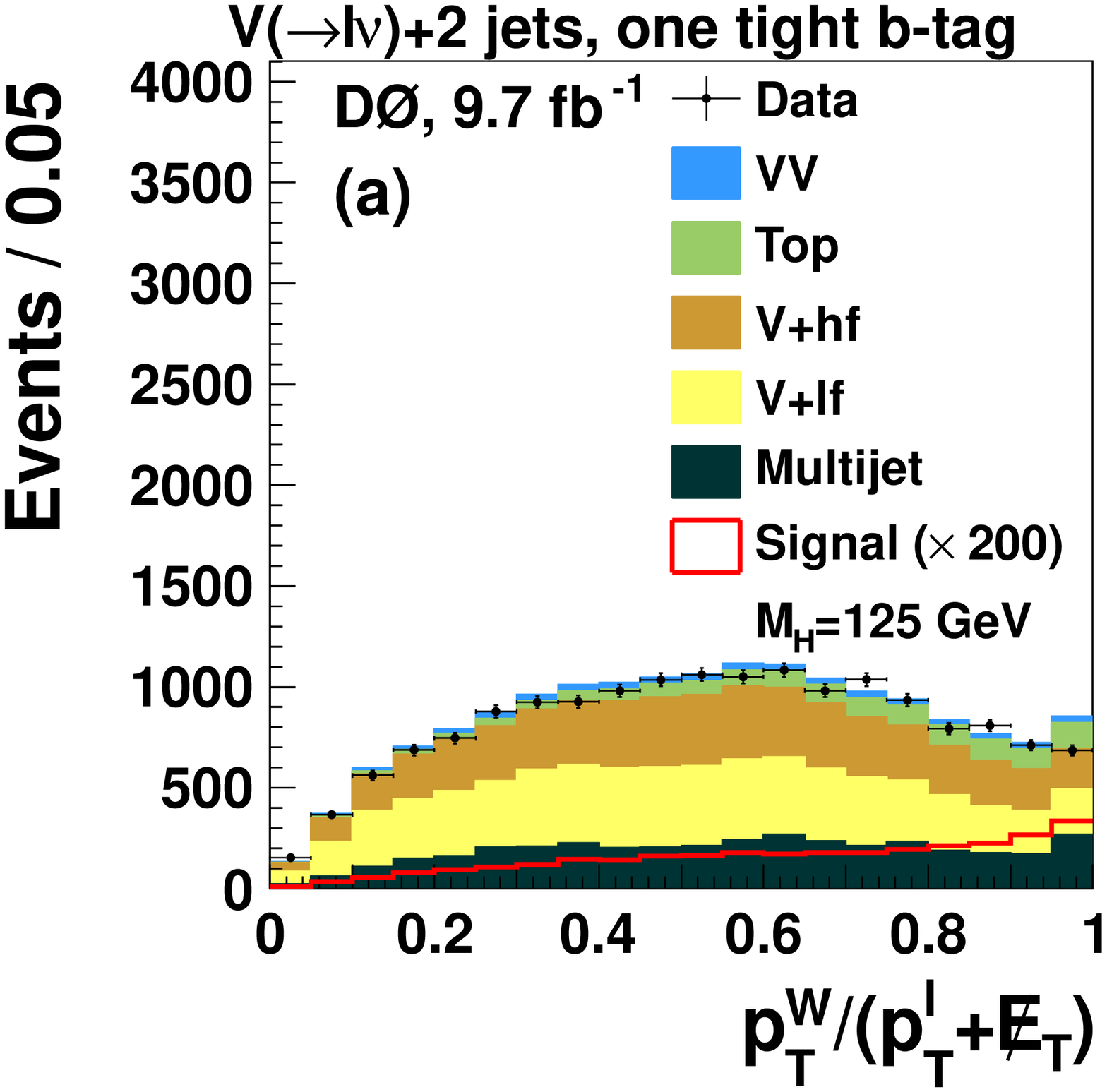}
\includegraphics[width=0.38\textwidth]{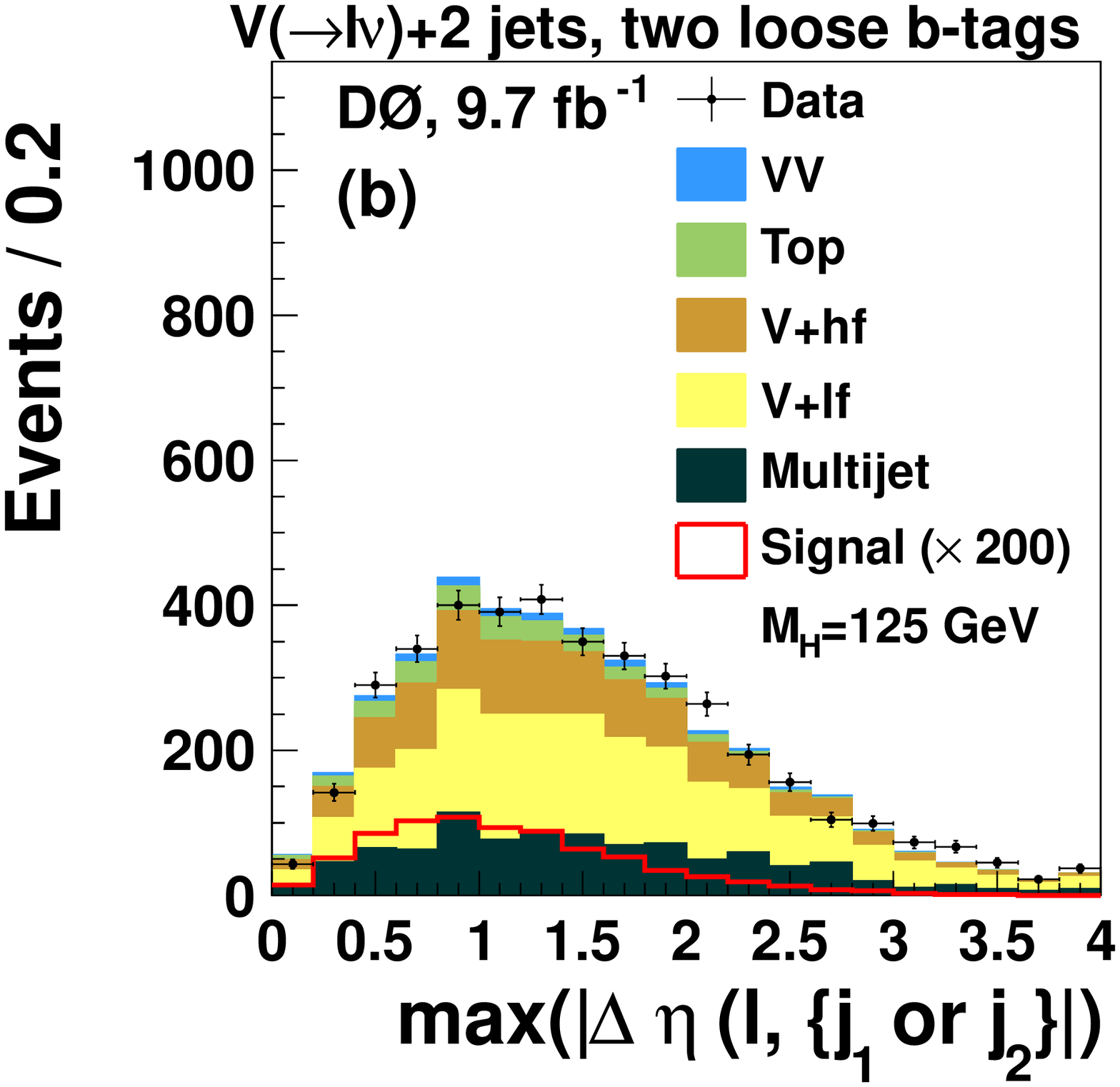}
\includegraphics[width=0.38\textwidth]{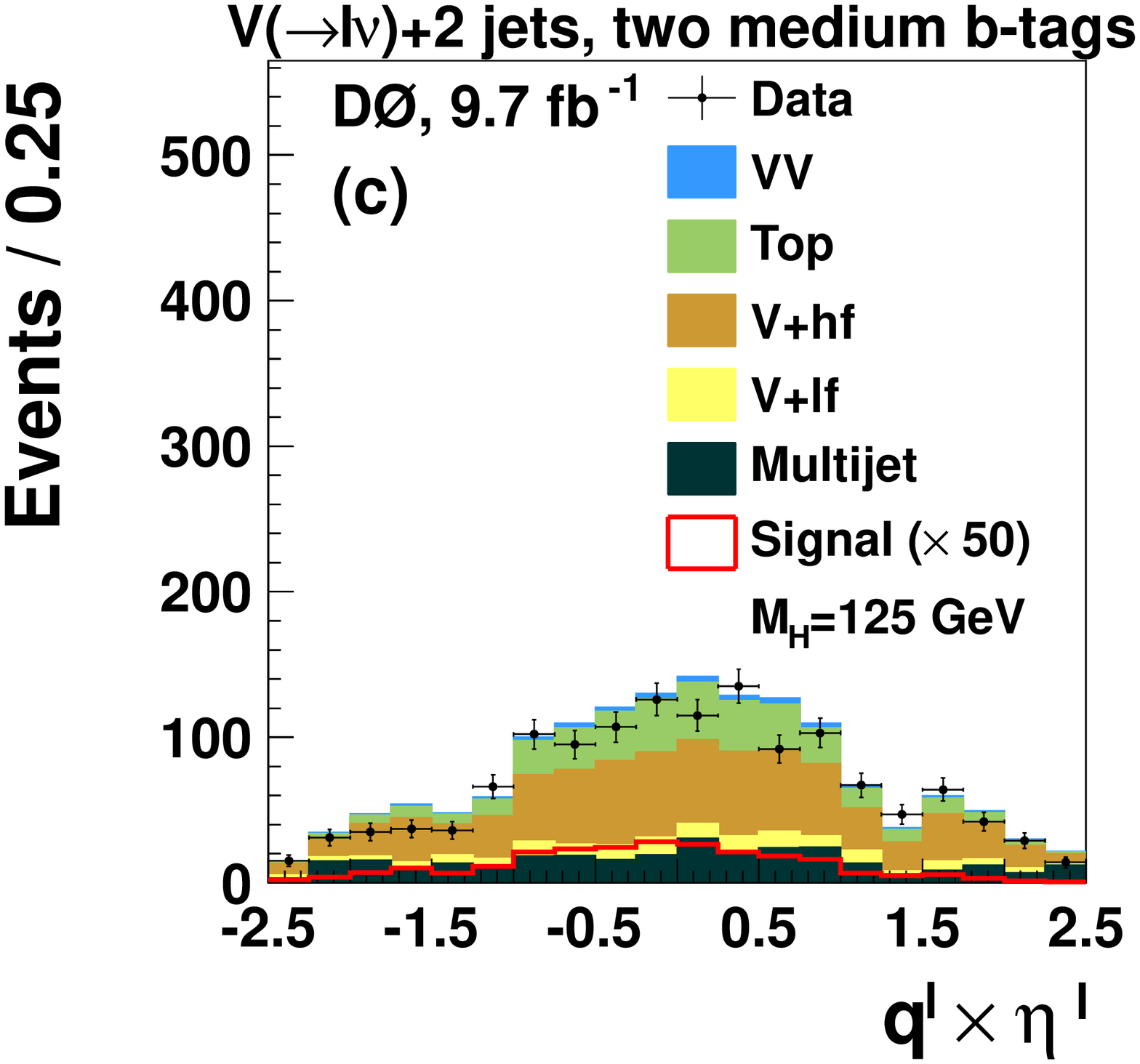}
\includegraphics[width=0.38\textwidth]{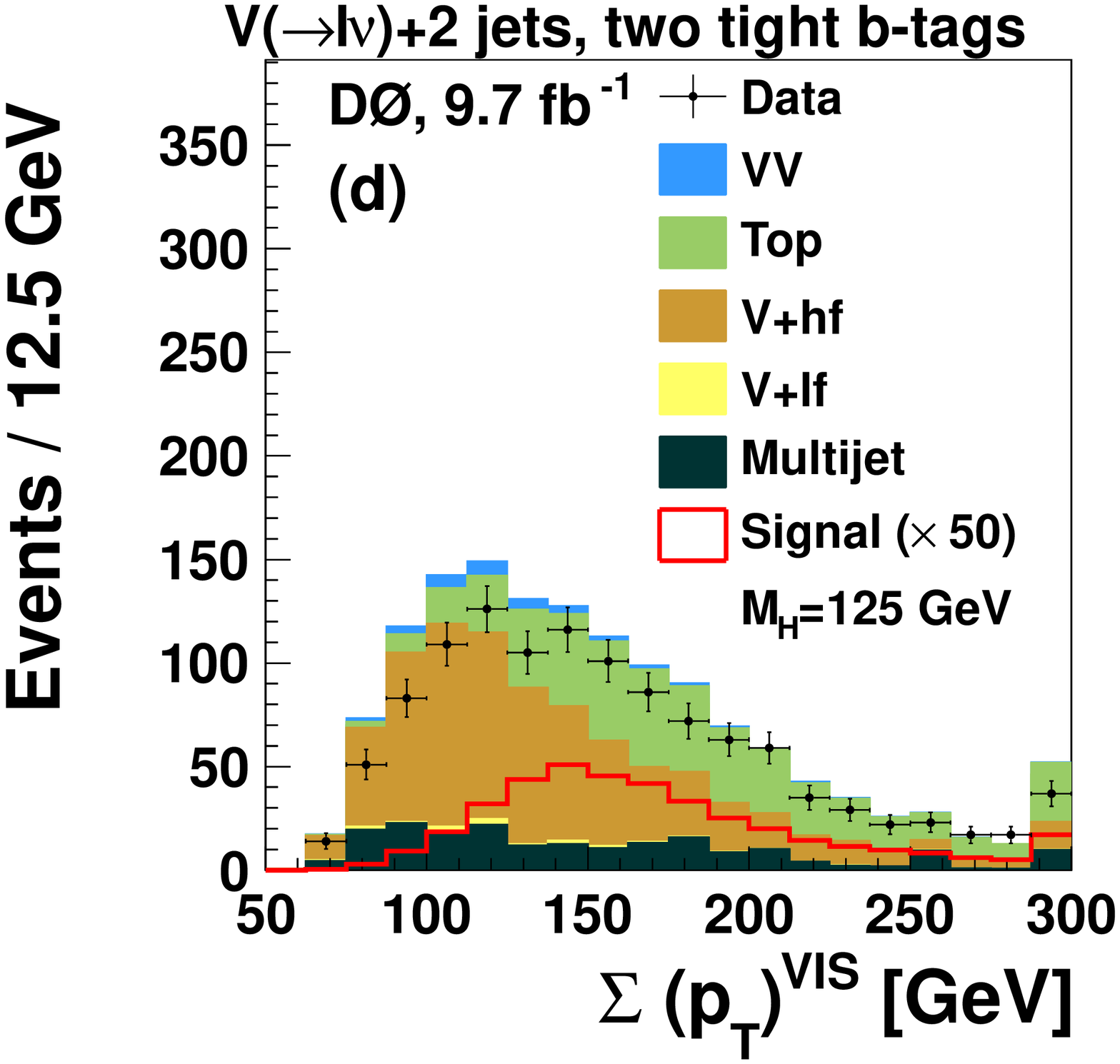}
}\caption{(color online)
Distributions of some of the most significant inputs to the final discriminant in events with exactly two jets and either 
one tight $b$-tag, two loose $b$-tags, two medium $b$-tags, or two tight $b$-tags:
(a) \lnuptoversumpt, shown for events with one tight $b$-tag;
(b) \jetlepdetamax~\cite{defsjetlepdetamax}, shown for events with two loose $b$-tags;
(c) \lepqeta~\cite{defslepqeta}, shown for events with two medium $b$-tags;
(d) \topovissumpt~\cite{defstopovissumpt}, shown for events with two tight $b$-tags.
The signal is multiplied by 200, 200, 50, and 50, respectively.
Overflow events are added to the last bin.
}
\label{fig:lvbb2jin}
\end{figure*}

\subsection{Final $\bm{H\to WW\to\ell\nu jj}$ MVA analysis}

The $H\to WW \to\ell\nu jj$ process provides the dominant signal in events with two or three 
jets and zero $b$-tags or one loose $b$-tag, since the $W$ boson decays producing a $b$ quark are rare. For signal searches 
in these channels,
we apply a multivariate technique based on the RF discriminant. Events  
in the above tagging categories are examined for $100 \leq M_H \leq 150$~GeV. Since we do not perform the search in the one tight and two $b$-tag channels for $M_H >150$~GeV, events having exactly two 
or three jets in all $b$-tagging categories (i.e.\ pretag events) are used in the search for $155 \leq M_H \leq 200$~GeV. 

To suppress MJ background in the electron channel in these subchannels, 
we select events with $\mvamjvshvv > -0.4$
for $M_H \leq 150$~GeV in events with zero or one loose tag, and $\mvamjvsvh>-0.5$ 
for $M_H \geq 155$~GeV in all events. These requirements were optimized to maximize the ratio of number of signal events to the square root of the number of background events. The MJ component in the zero or one loose $b$-tag muon channel is small, so there is no cut applied to the MJ MVA outputs. 

\begin{figure}[htbp]
\centering{
\includegraphics[width=0.38\textwidth]{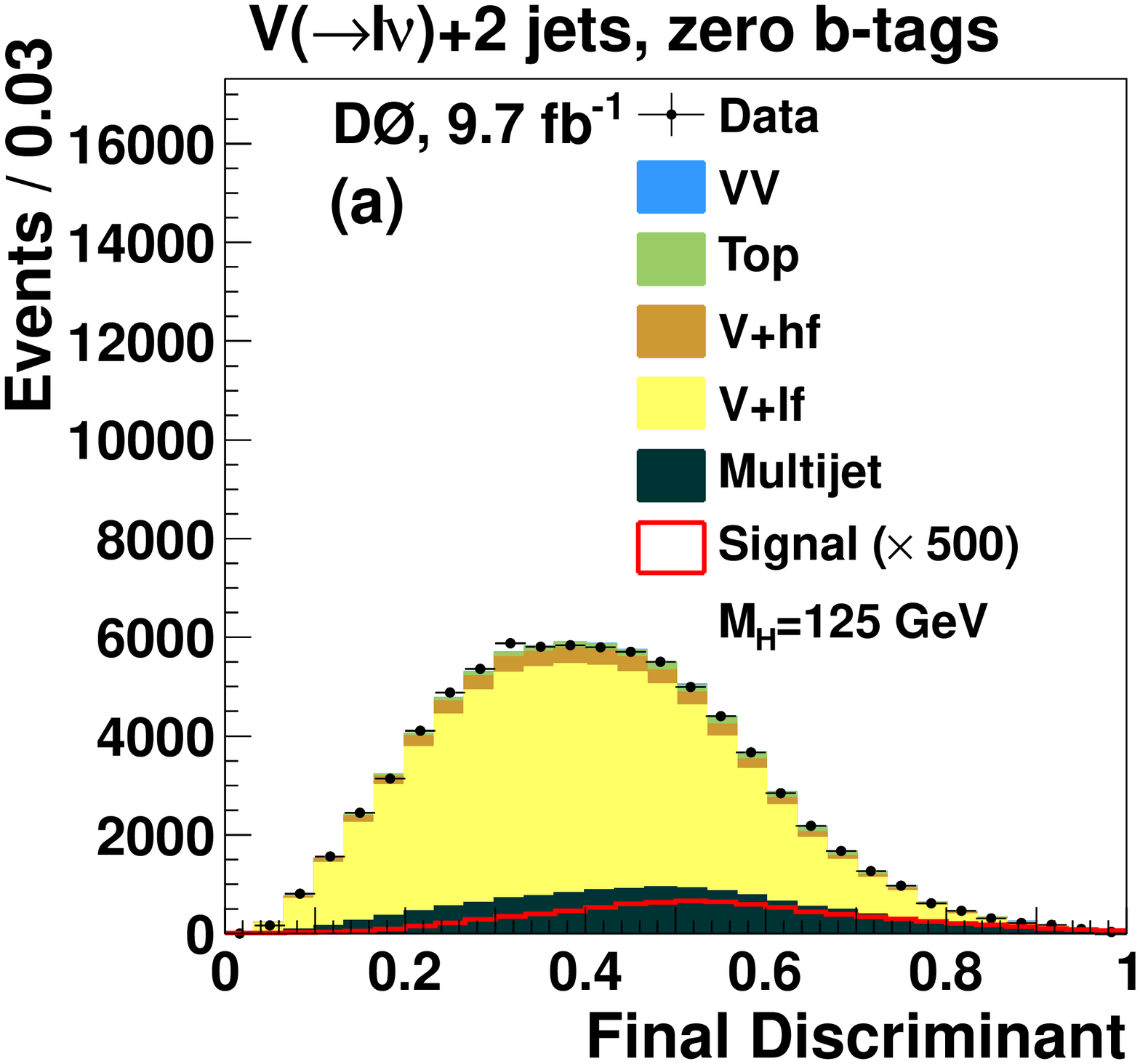}
\includegraphics[width=0.38\textwidth]{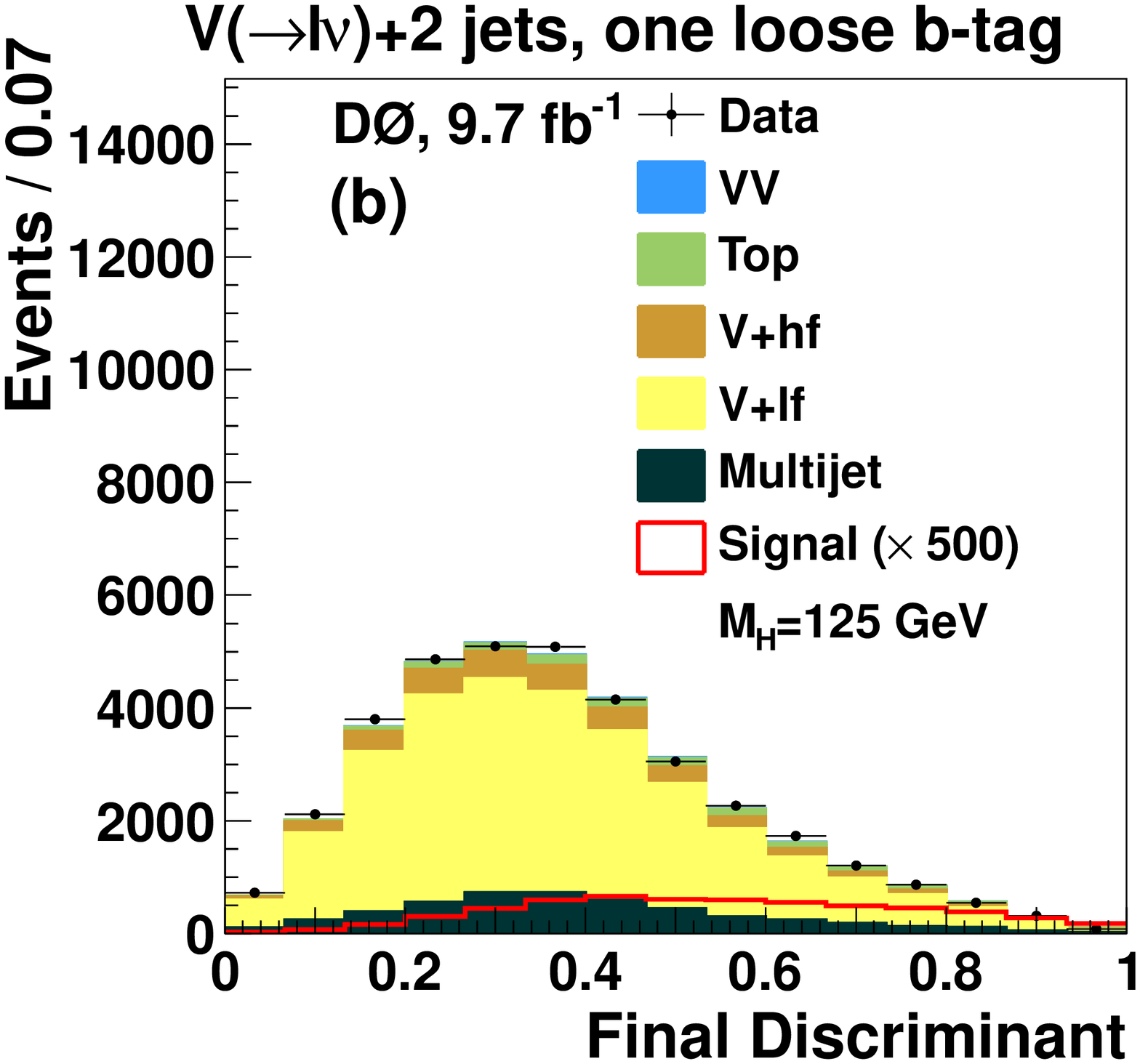}
\includegraphics[width=0.38\textwidth]{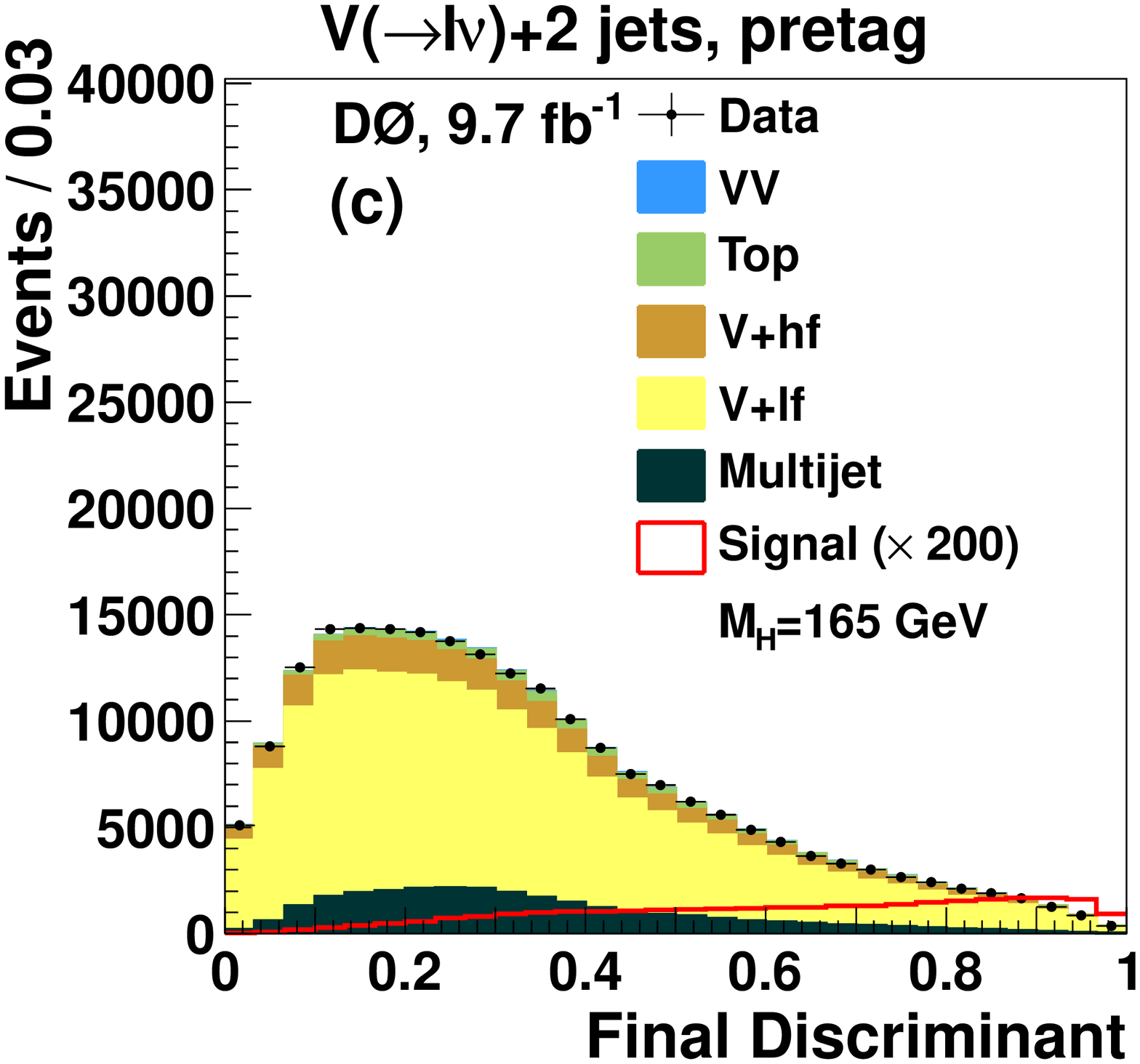}
}\caption{(color online)
Distributions of the final discriminant output, after the maximum likelihood fit (described in Sec.~\ref{sec:limits}), for events in the following channels:
(a) two jets, zero $b$-tags for $M_H=125$~GeV, (b) two jets, one loose $b$-tag for $M_H=125$~GeV, (c) two jets, all tags $M_H=165$~GeV.
The signal is multiplied by 500, 500, and 200, respectively.
}
\label{fig:lvjj_2jetout}
\end{figure}

\begin{figure}[htbp]
\centering{
\includegraphics[width=0.38\textwidth]{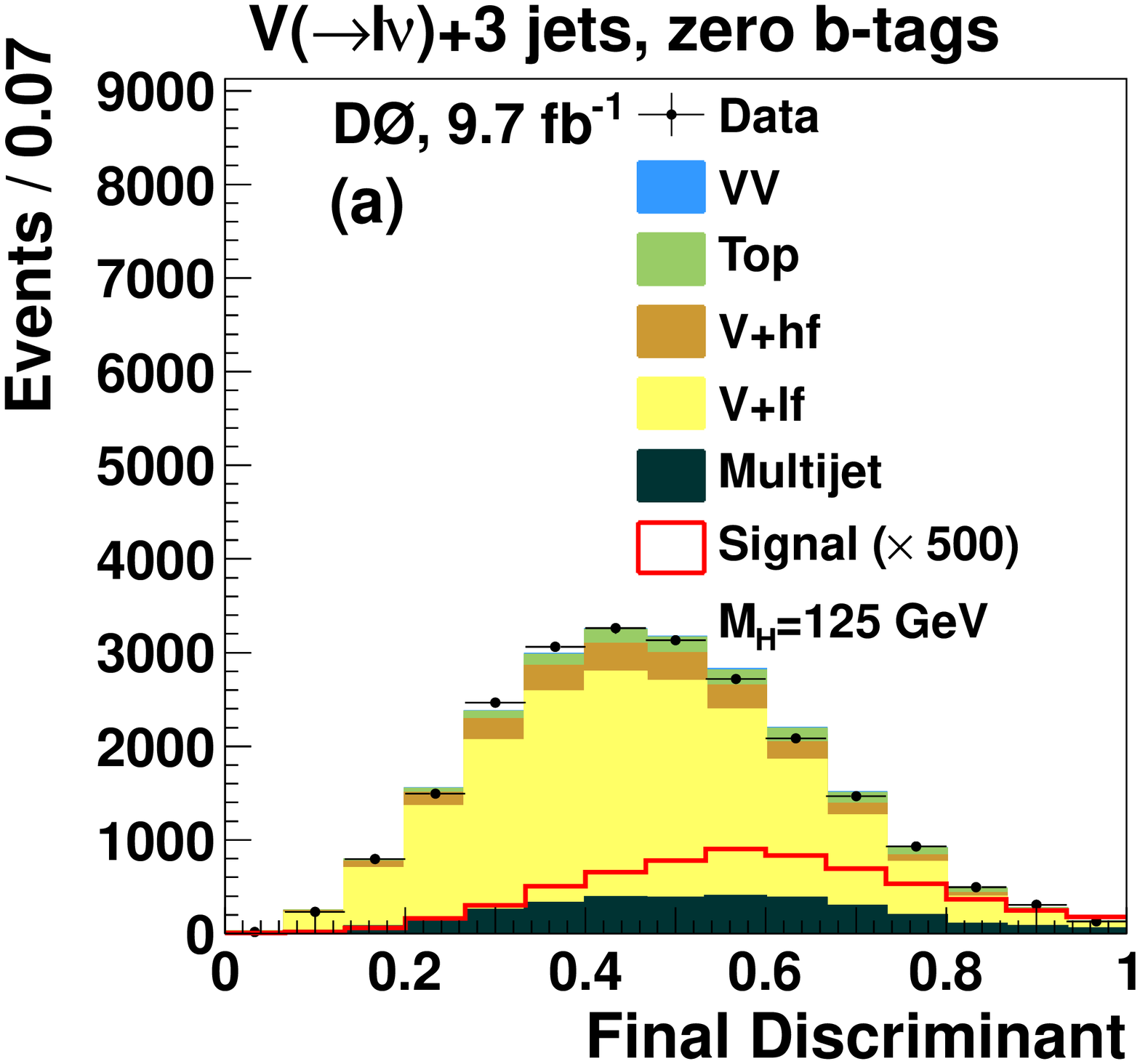}
\includegraphics[width=0.38\textwidth]{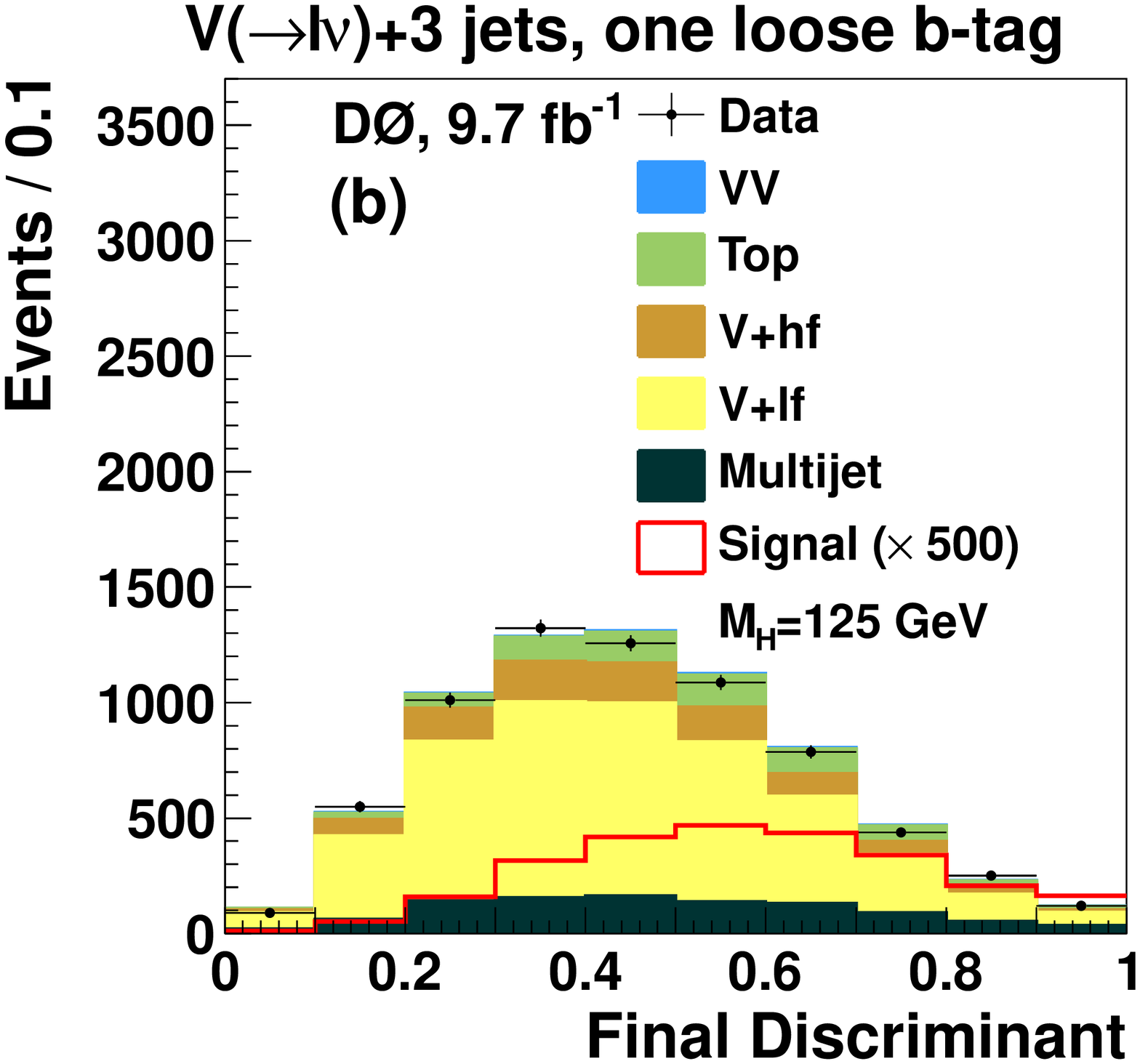}
\includegraphics[width=0.38\textwidth]{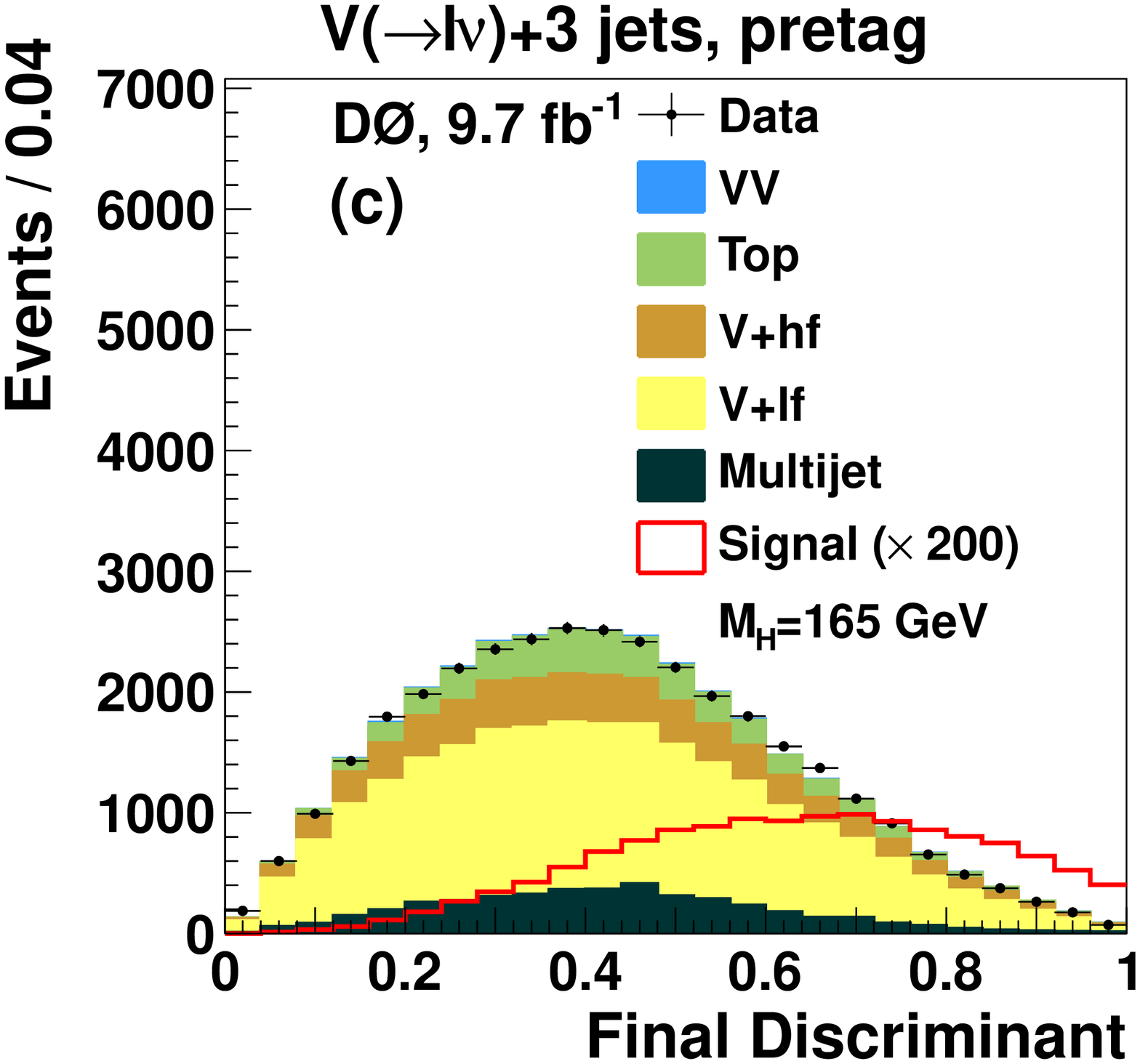}
}\caption{(color online)
Distributions of the final discriminant output, after the maximum likelihood fit (described in Sec.~\ref{sec:limits}), for events in the following channels:
(a) three jets, zero $b$-tags for $M_H=125$~GeV, (b) three jets, one loose $b$-tag for $M_H=125$~GeV, (c) three jets, all tags $M_H=165$~GeV.
The signal is multiplied by 500, 500, and 200, respectively.
}
\label{fig:lvjj_3jetout}
\end{figure}

We train a RF on the total signal and background from all considered physics 
processes. We optimize the RF independently in the electron and muon channels for each $b$-tag and jet multiplicity 
category. As the signal shape is strongly driven by the signal mass hypothesis, we optimize the MVA variable list 
at two different mass points: at $M_H=125$~GeV for masses below 150~GeV and at $M_H=165$~GeV for masses
above 150~GeV. Because the resolution of the reconstructed Higgs boson mass is about 20~GeV for channels presented in this Article, optimizing the input variable list at only these mass points is sufficient. 
Each RF is trained using between 14 and 30 well modeled discriminating variables formed from kinematic properties 
of either elementary objects like jets or leptons, or composite objects, such as reconstructed $W$ boson candidates 
(see Figs.~\ref{fig:evjj2jin} and \ref{fig:evjj3jin}).  The lists of input variables and their descriptions are included in Appendix~\ref{app:input_var_def}.
The final RF discriminants for the electron and muon 
channels are shown in Figs.~\ref{fig:lvjj_2jetout} and \ref{fig:lvjj_3jetout}.

\begin{figure}[htbp]
\centering{
\includegraphics[width=0.38\textwidth]{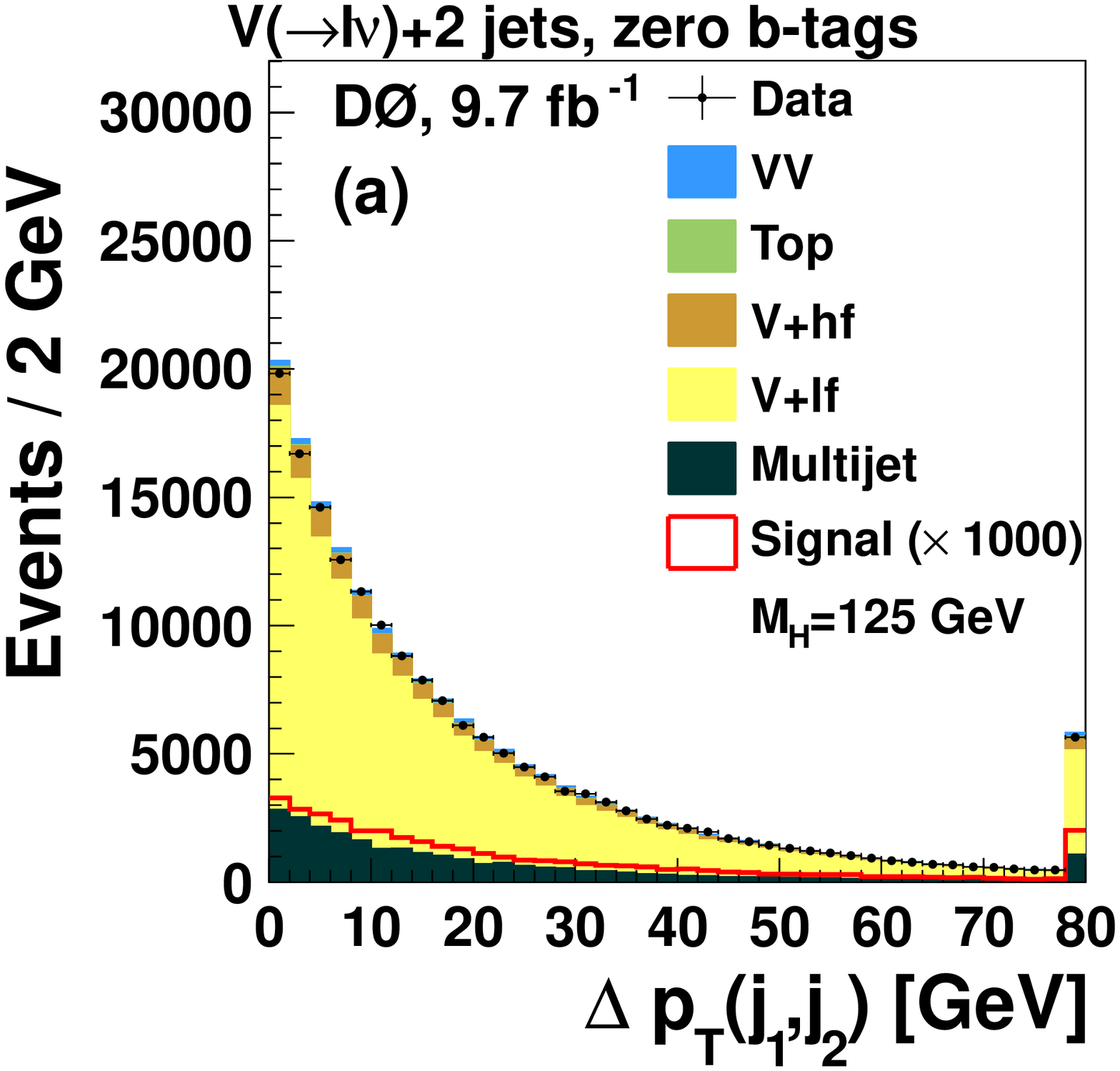}\vspace{-0.15cm}
\includegraphics[width=0.38\textwidth]{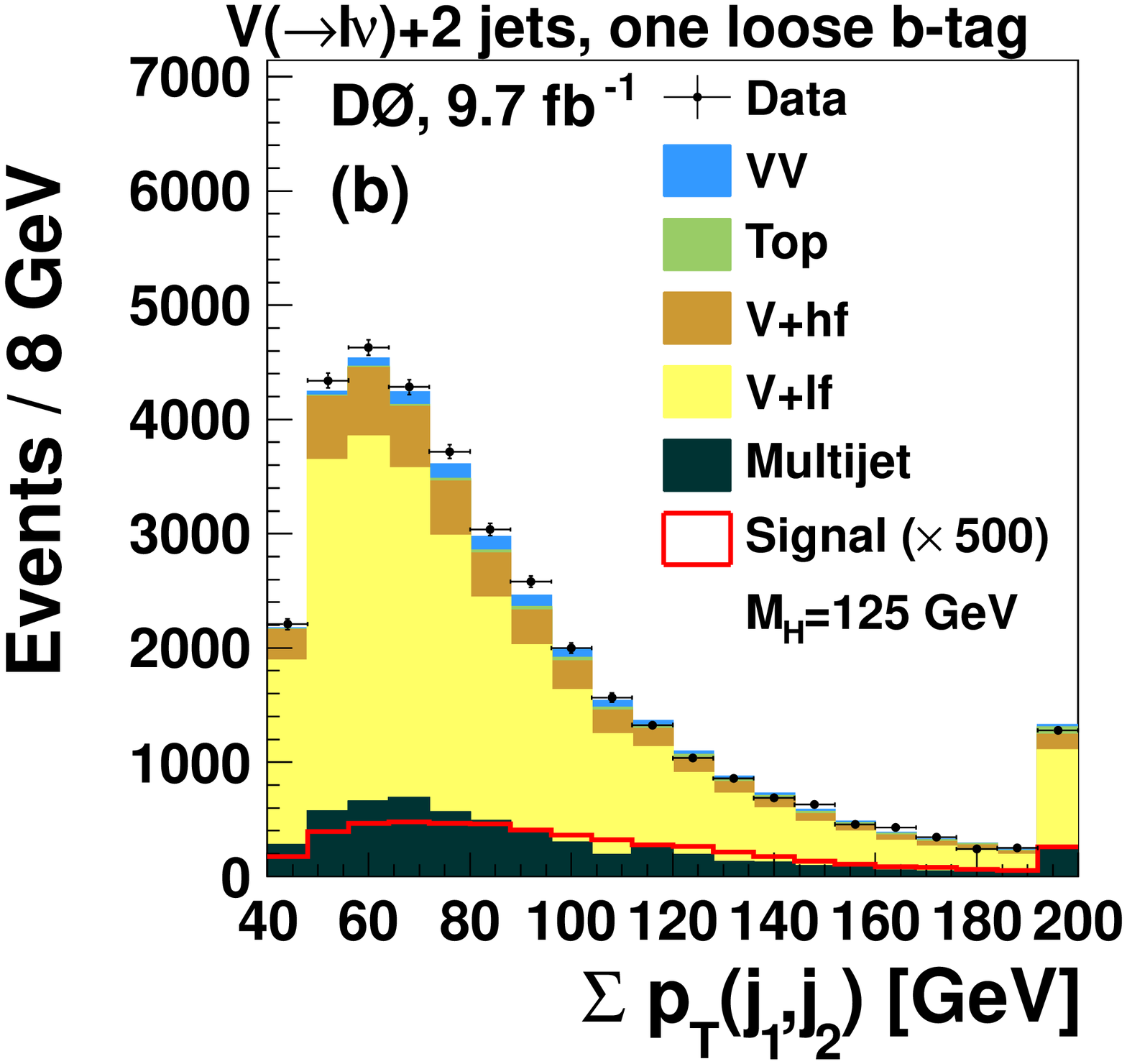}\vspace{-0.15cm}
\includegraphics[width=0.38\textwidth]{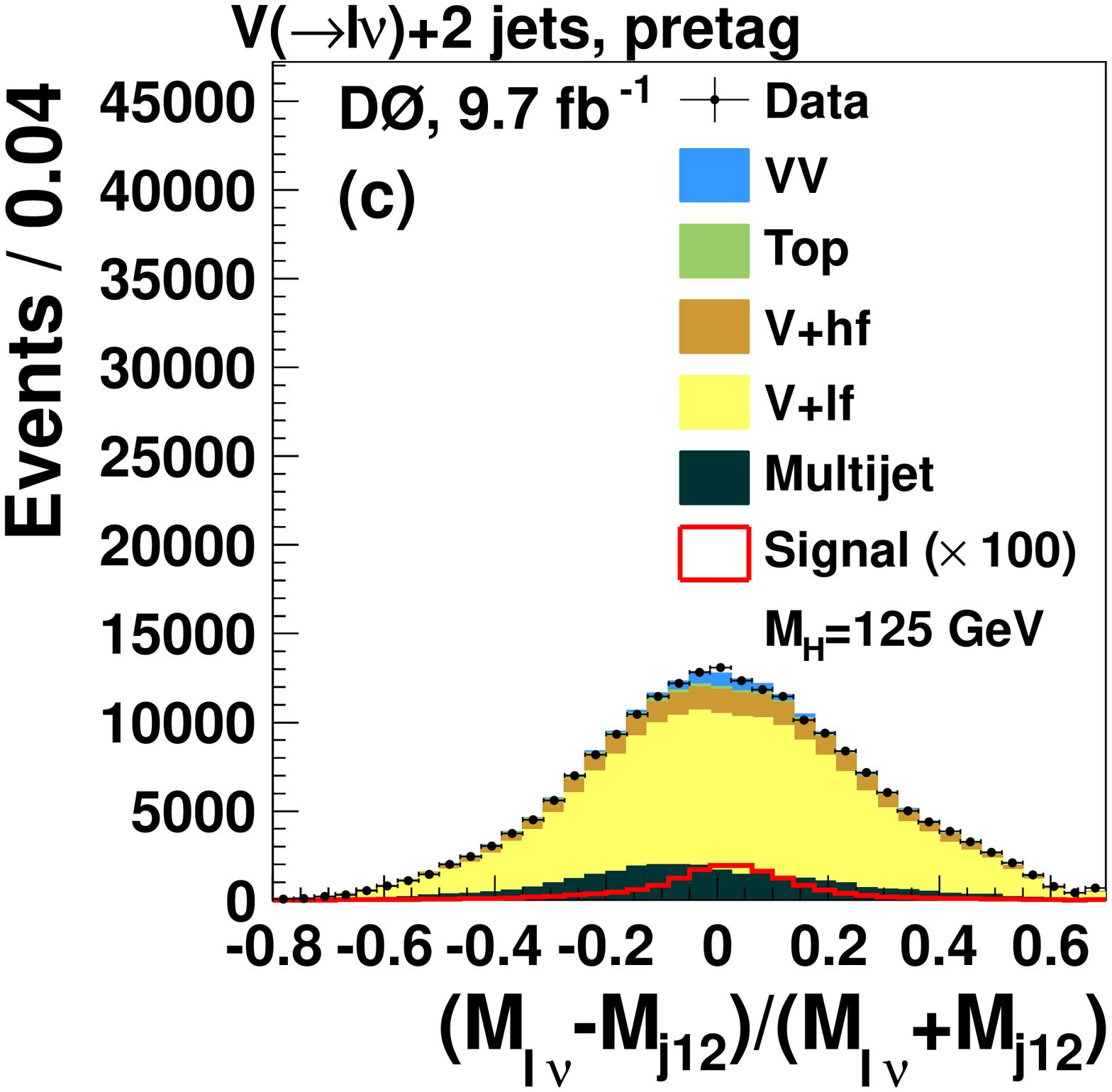}\vspace{-0.15cm}
}\caption{(color online)
Distributions of the most significant inputs to the final multivariate discriminants for the 
two-jets zero and one loose $b$-tag channels:
(a) $\jABdpt$, shown for events with zero $b$-tags for $M_H=125$~GeV;
(b) $\sum _{i=1}^{2} p_T^{j_i}$, shown for events with one loose $b$-tag for $M_H=125$~GeV;
(c) $(M_{\ell\nu} - M_{j_{12}})/(M_{\ell\nu} + M_{j_{12}})$, shown for all tags for $M_H=165$~GeV.
The signal is multiplied by 1000, 500, and 100, respectively. 
Overflow events are added to the last bin.
}
\label{fig:evjj2jin}
\end{figure}

\begin{figure}[htbp]
\centering{
\includegraphics[width=0.38\textwidth]{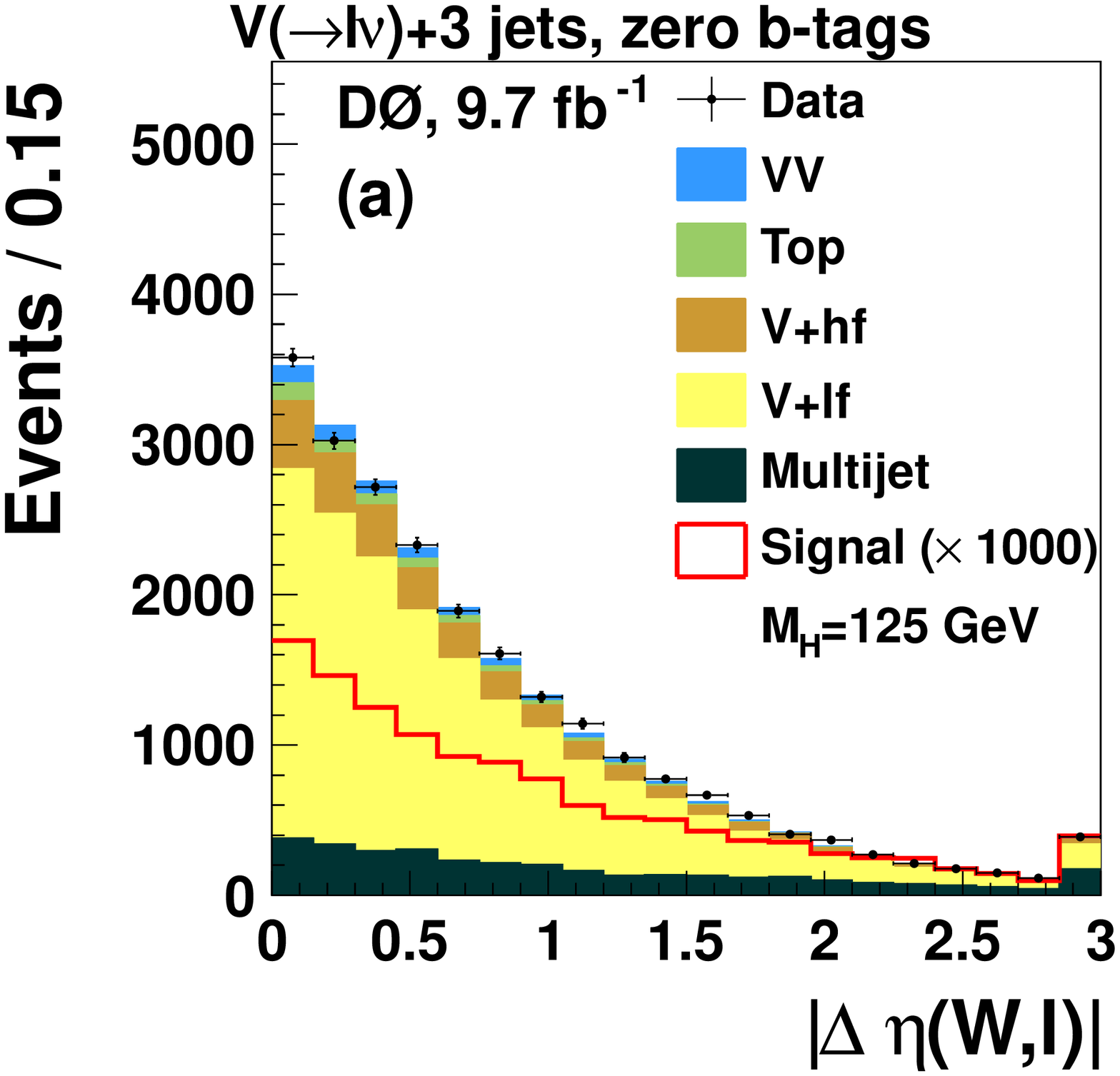}\vspace{-0.15cm}
\includegraphics[width=0.38\textwidth]{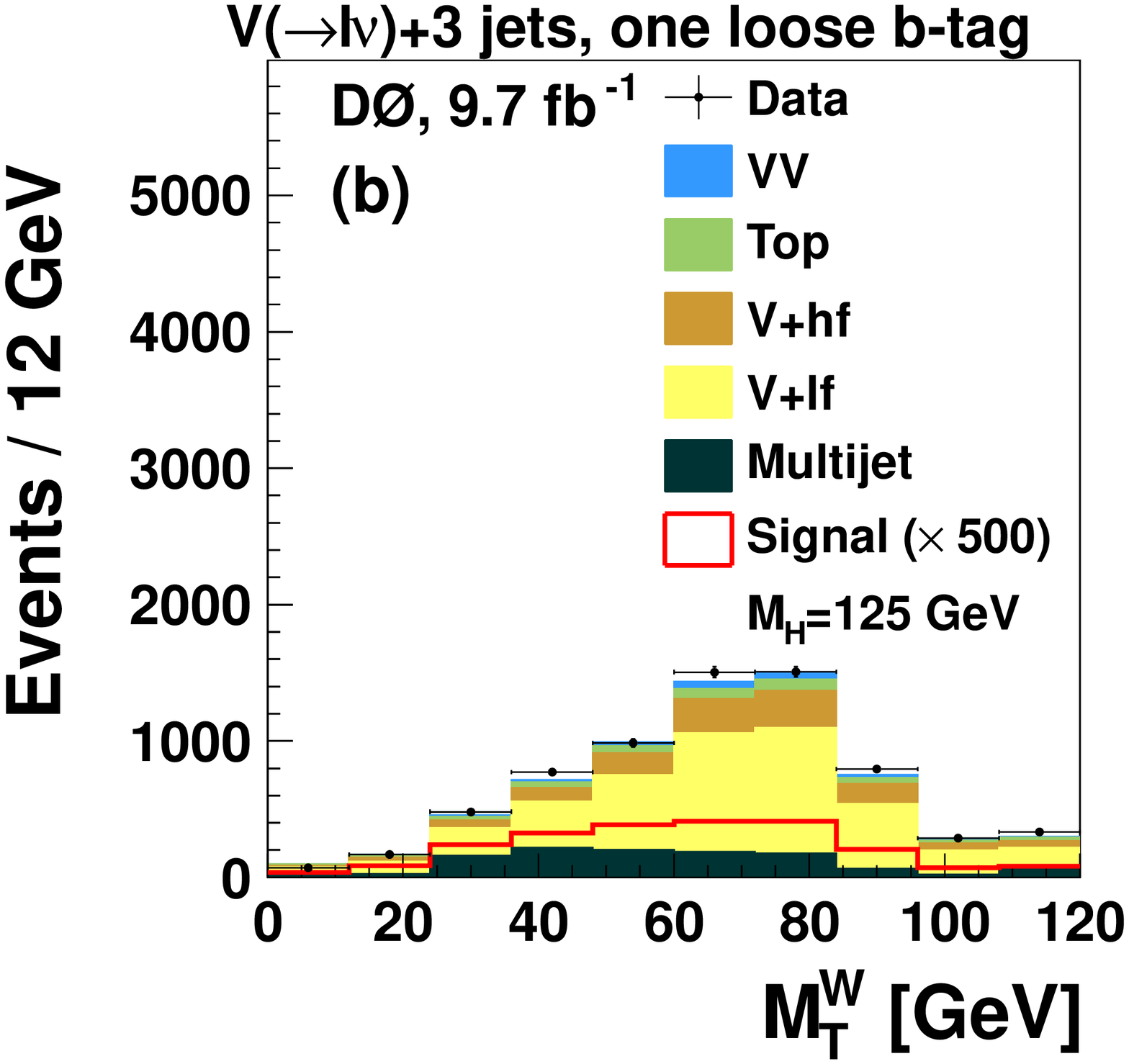}\vspace{-0.15cm}
\includegraphics[width=0.38\textwidth]{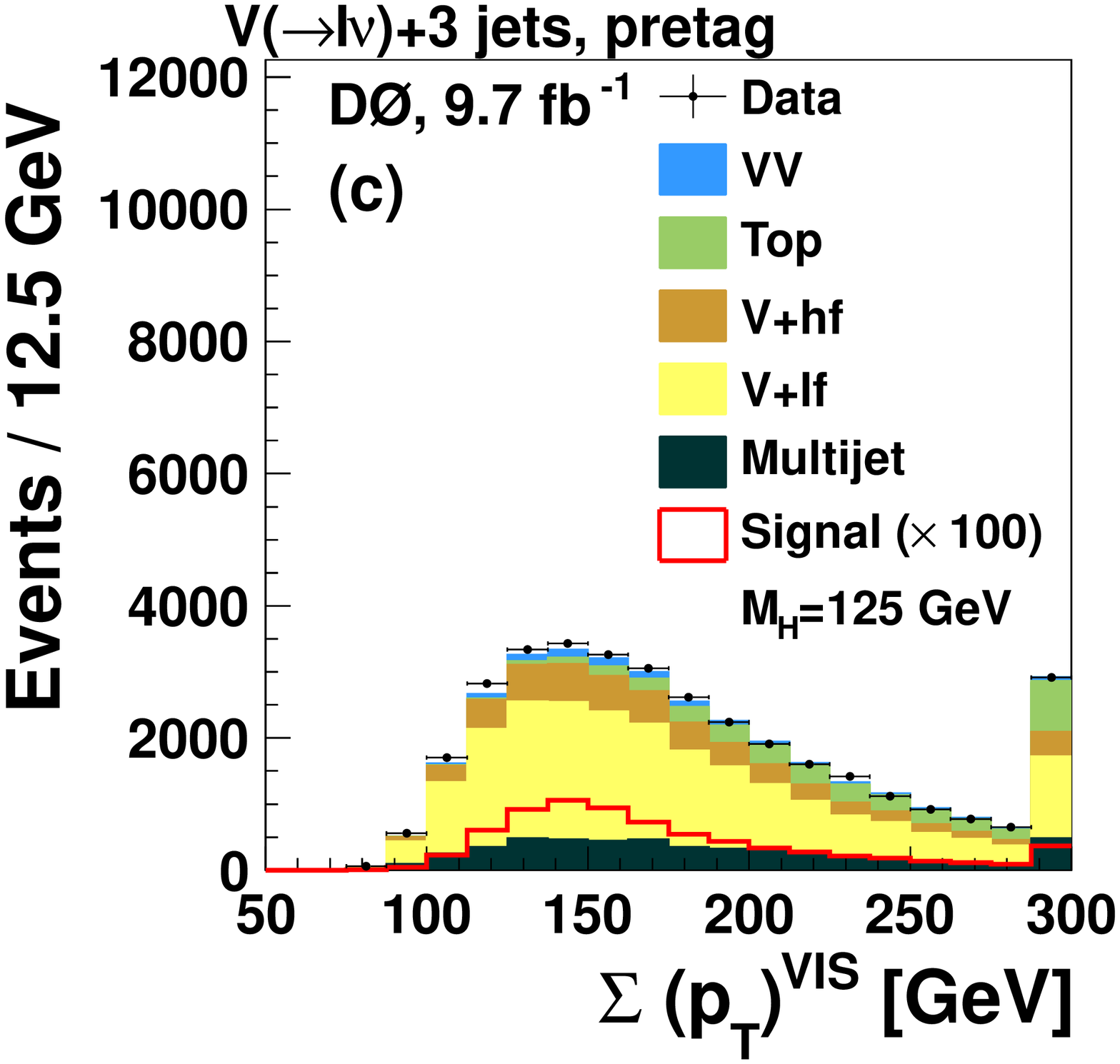}\vspace{-0.15cm}
}\caption{(color online)
Distributions of the most significant inputs to the final multivariate discriminants for 
the three-jets zero and one loose $b$-tag channels:
(a) $\lnulepdeta$~\cite{defslnulepdeta,defsnupz}, shown for events with zero $b$-tags for $M_H=125$~GeV;
(b) $\lnumt$, shown for events with one loose $b$-tag for $M_H=125$~GeV;
(c) $\topovissumpt$~\cite{defstopovissumpt}, shown for all tags for $M_H=165$~GeV.
The signal is multiplied by 1000, 500, and 100, respectively. 
Overflow events are added to the last bin.
}
\label{fig:evjj3jin}
\end{figure}

\subsection{Final $\bm{VH\to VWW\to\ell\nu jjjj}$ MVA analysis\label{sec:4jmva}}

The majority of signal events with four or more jets and zero $b$-tags or one loose $b$-tag are from the $VH\to VWW\to\ell\nu jjjj$ process,
but there are significant contributions from direct production via gluon fusion and vector boson fusion.
Identification of the Higgs boson decay products in $VH\to VWW$ events is
complicated by the combinatorics of pairing four jets into two hadronically decaying vector boson candidates and 
then two of the three total vector boson candidates into the Higgs boson candidate.
The discriminating variables are different for fully hadronic and semileptonic Higgs boson decays,
and determining the Higgs boson candidate for an event also determines which of these two decay scenarios is considered.
Variables unique to a particular decay scenario are set to a default value outside of the physical range of that variable in events
reconstructed under the alternate decay scenario.
To reconstruct the two hadronically decaying vector boson candidates, we examine the leading four jets in an event and choose the jet pairings that
minimize:
\begin{equation}
  E_{ab,cd} = |m_{ab}-M_W| + |m_{cd}-M_W|,
\end{equation}
where $m_{ab}$ ($m_{cd}$) is the invariant mass of the $a^{\rm th}$ and $b^{\rm th}$ ($c^{\rm th}$ and $d^{\rm th}$) jets,
and $M_W = 80.4$~GeV ~\cite{Beringer:2012}.  
The Higgs boson candidate is then determined by considering the semileptonically decaying $W$ boson and the
two hadronically decaying vector bosons and selecting the vector boson candidate pair with the minimum $\deltaR$ separation
in an event, out of the three possible pairings.

\begin{figure}[htbp]
\centering{
\includegraphics[width=0.38\textwidth]{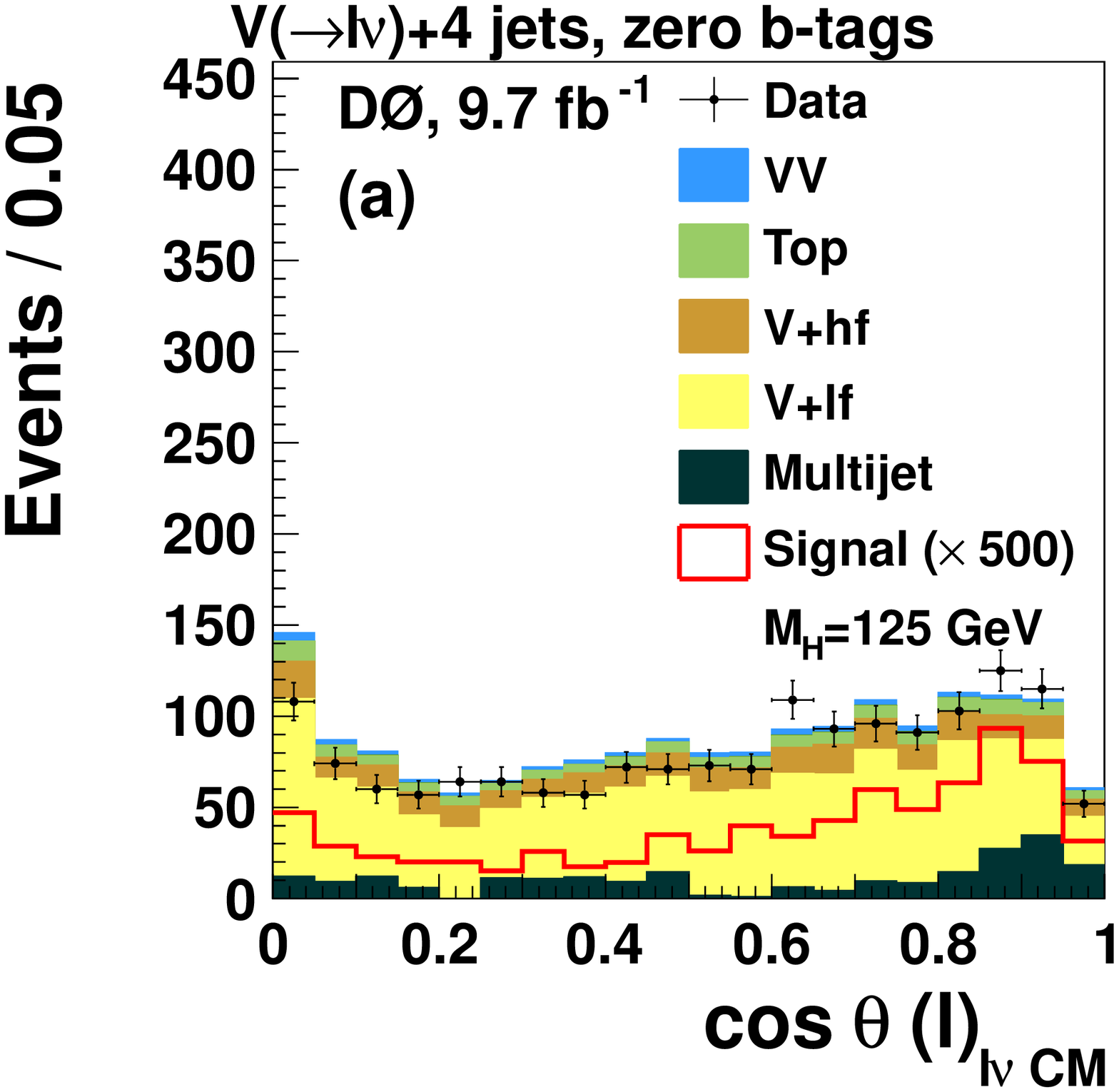}\vspace{-0.15cm}
\includegraphics[width=0.38\textwidth]{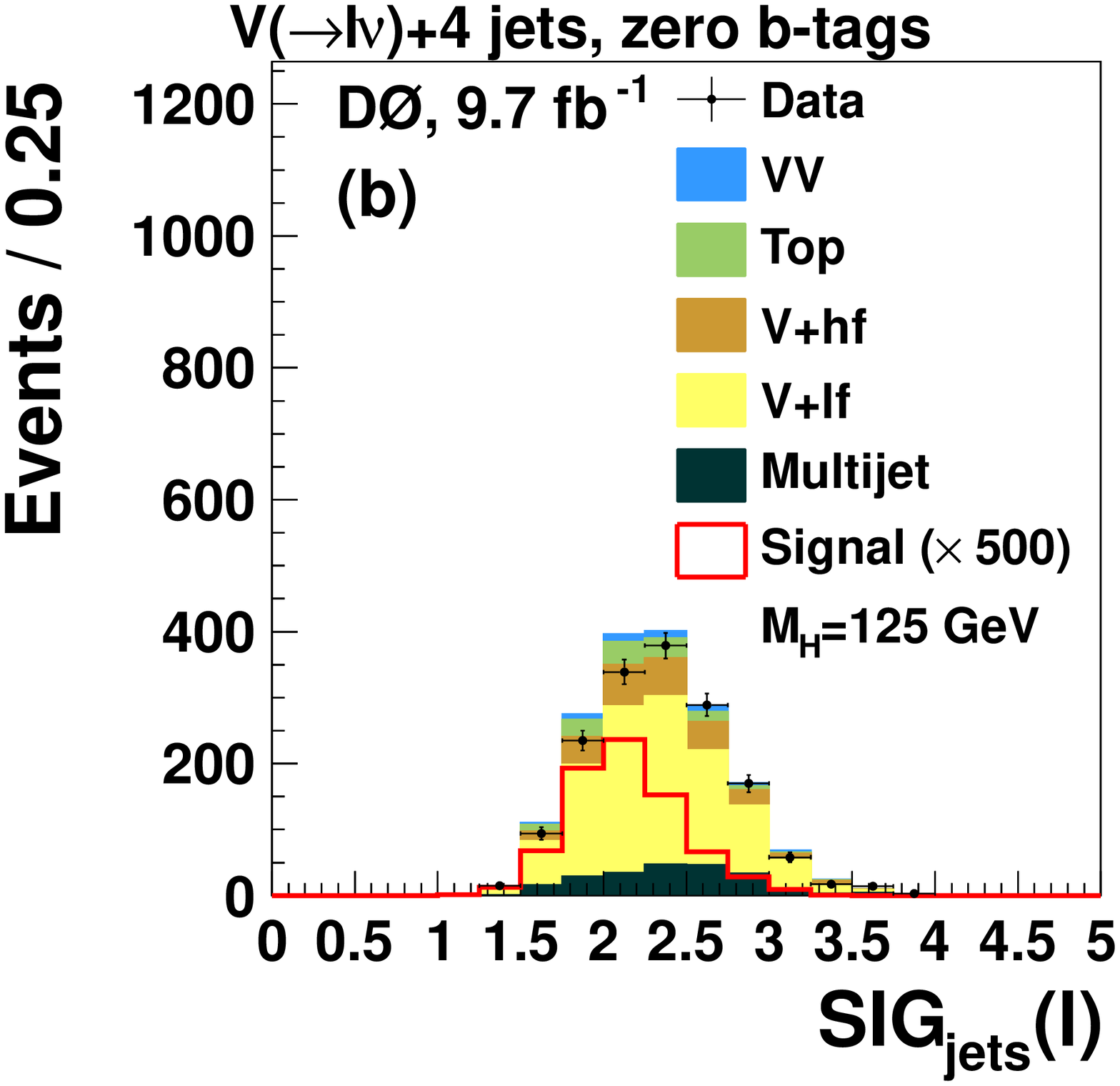}\vspace{-0.15cm}
\includegraphics[width=0.38\textwidth]{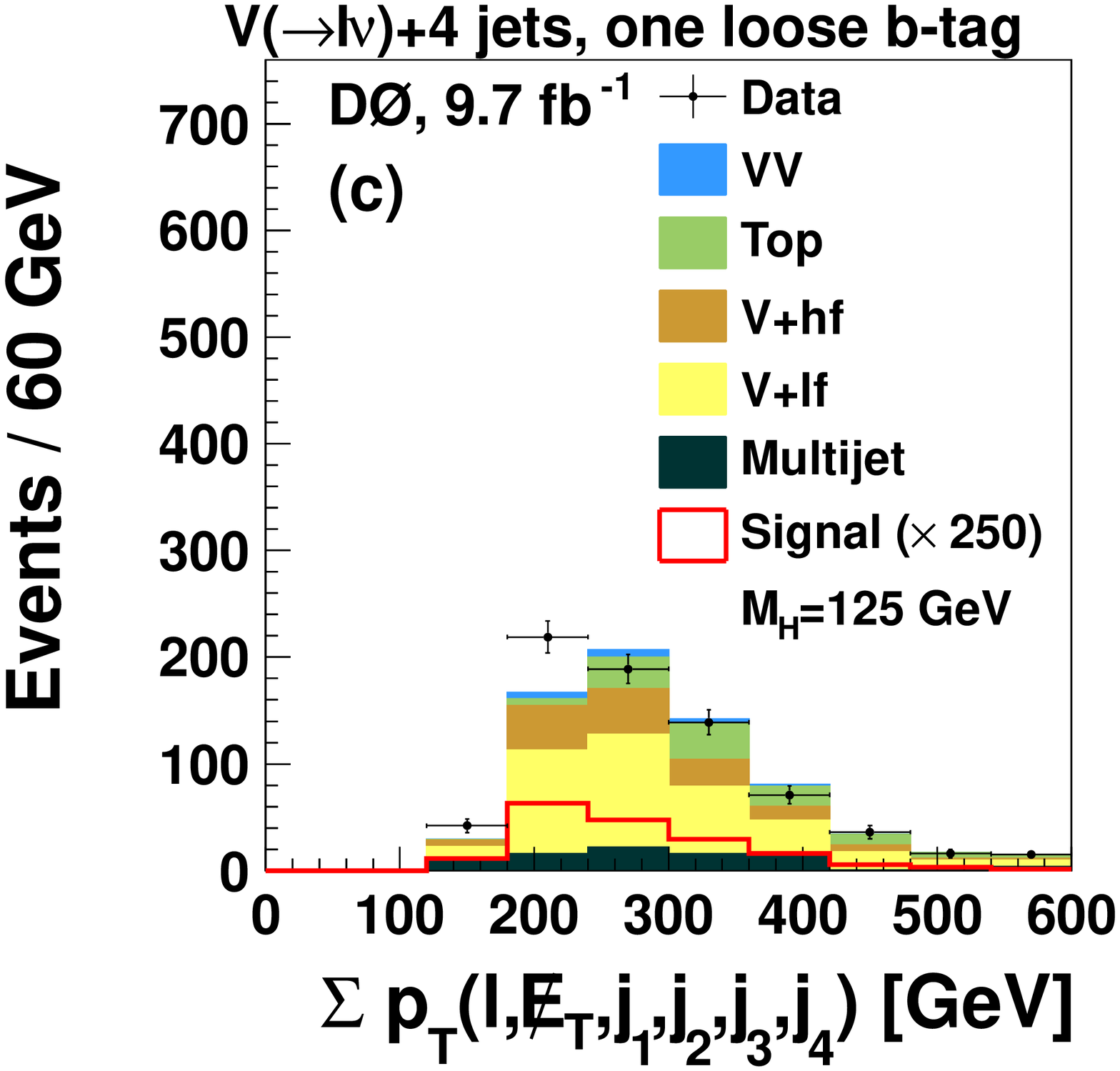}\vspace{-0.15cm}
}
\caption{(color online)
Distributions of the most significant inputs to background-specific multivariate discriminants for the
$\geq 4$-jet subchannels:
(a) \lnucmcostheta~\cite{defslnucmcostheta,defsnupz}, input to discriminant against $V$+jets backgrounds, shown for events with zero $b$-tags;
(b) \lepsigma~\cite{defslepsigma}, input to discriminant against diboson backgrounds, shown for events with zero $b$-tags;
(c) \lnujjjjsumpt, input to discriminant against top quark backgrounds, shown for events with one loose $b$-tag.
The $M_H=125$~GeV signal is multiplied by 250 in (c) and by 500 in (a) and (b). 
Overflow events are added to the last bin.
}
\label{fig:lv4jBSMVA}
\end{figure}

\begin{figure}[htbp]
\centering{
\includegraphics[width=0.38\textwidth]{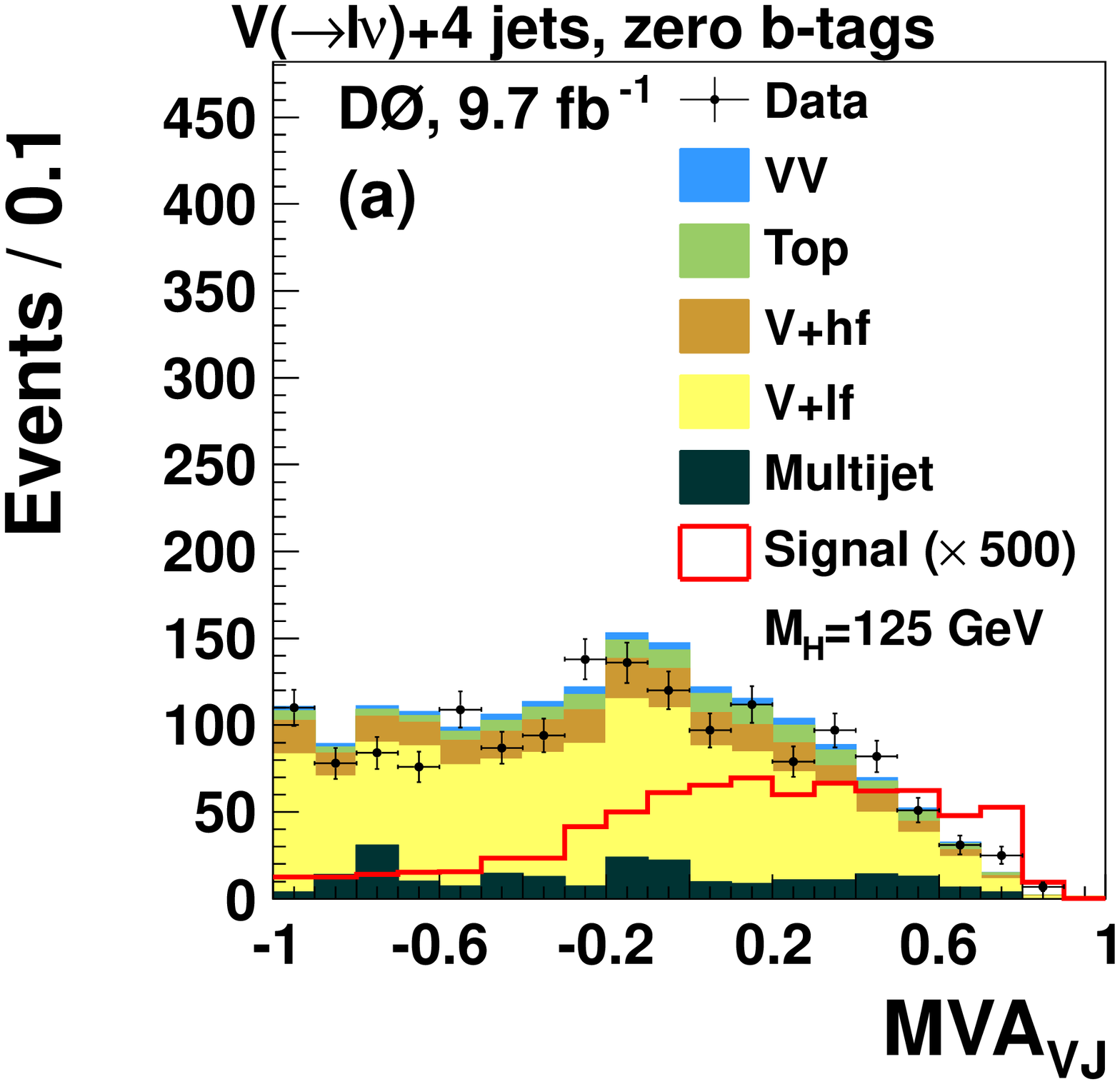}\vspace{-0.15cm}
\includegraphics[width=0.38\textwidth]{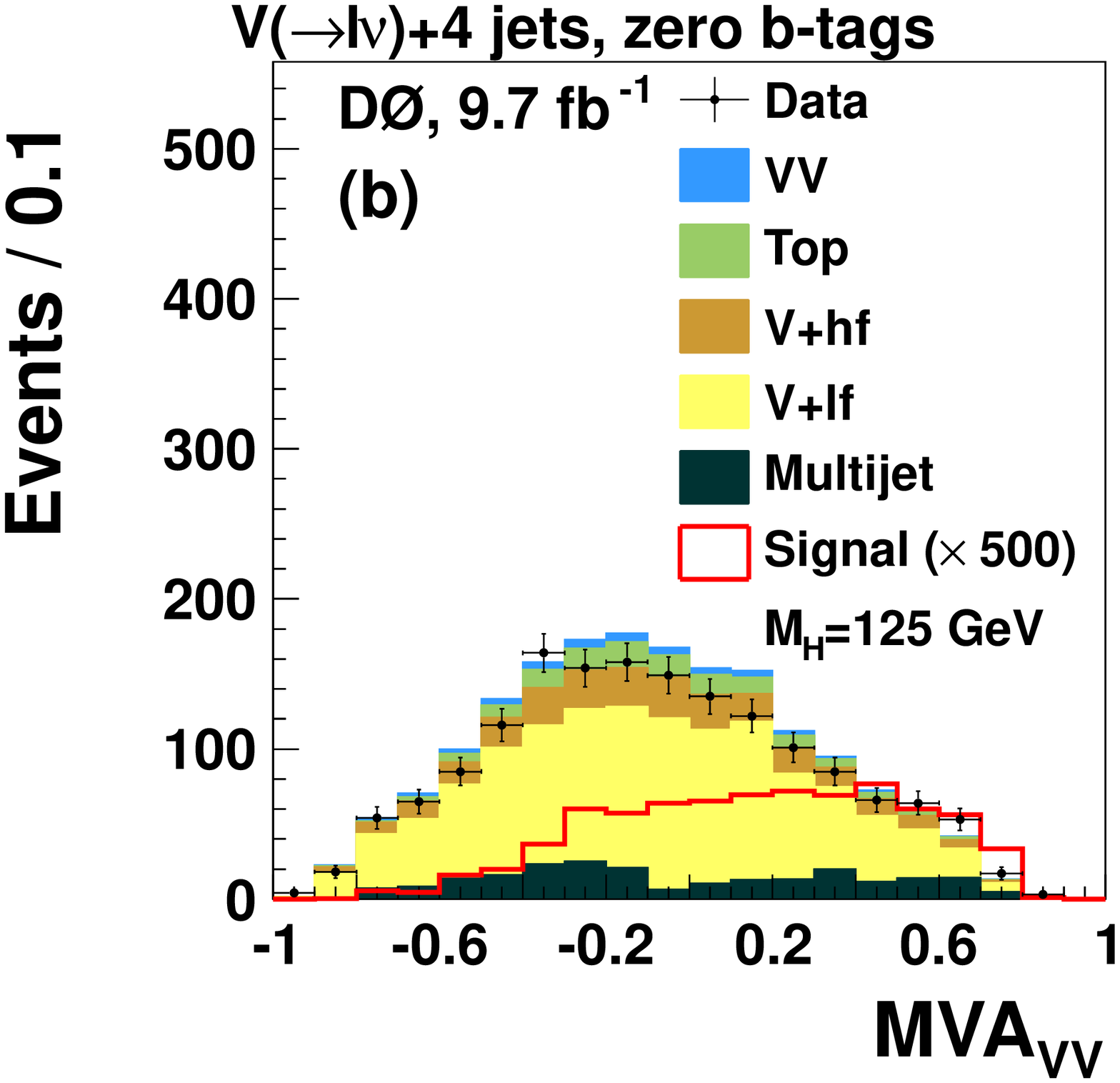}\vspace{-0.15cm}
\includegraphics[width=0.38\textwidth]{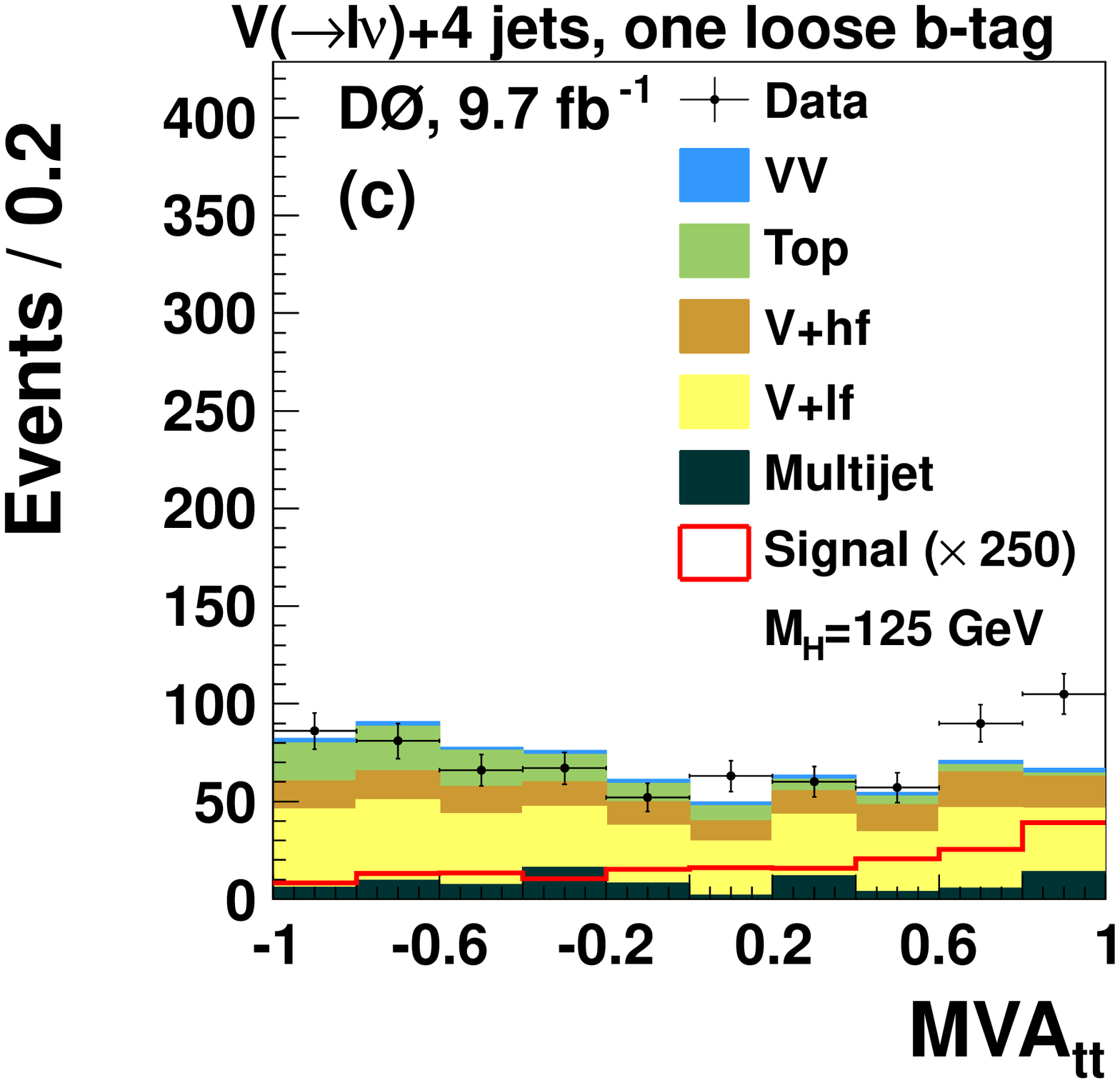}\vspace{-0.15cm}
}\caption{(color online)
Distributions of the output of background-specific multivariate discriminants for the
$\geq 4$-jet subchannels:
(a) discriminant against $V$+jets backgrounds, shown for events with zero $b$-tags;
(b) discriminant against diboson backgrounds, shown for events with zero $b$-tags;
(c) discriminant against top quark backgrounds, shown for events with one loose $b$-tag.
The $M_H=125$~GeV signal is multiplied by 250 in (c) and by 500 in (a) and (b). 
}
\label{fig:lv4jBSMVA-2}
\end{figure}

Diverse signal processes contribute to the inclusive four-jet channel with relative contributions varying with $M_H$.
To help mitigate the effect of having many signal and background contributions to this search channel, 
we use two layers of multivariate discriminants to improve the 
separation of signal from background.  The first layer of training focuses on separating 
the sum of all signal processes from specific sets of backgrounds.  Input variables for each background-specific 
discriminant are selected based on the separation power between the total signal and the backgrounds being 
considered.
Background-specific discriminants are trained to separate the sum of all Higgs boson signal processes from 
three specific background categories: $t\bar{t}$ and single top quark production, $V$+jets production, and diboson production.
The input variables and their descriptions are listed in Appendix~\ref{app:input_var_def}.
Separate background-specific discriminants are trained for each Higgs boson mass point considered.  
Sample inputs and output responses of the background-specific discriminants are shown in Figs.~\ref{fig:lv4jBSMVA} and \ref{fig:lv4jBSMVA-2}, respectively, for $M_H=125$~GeV.


\begin{figure}[htbp]
\centering{
\includegraphics[width=0.38\textwidth]{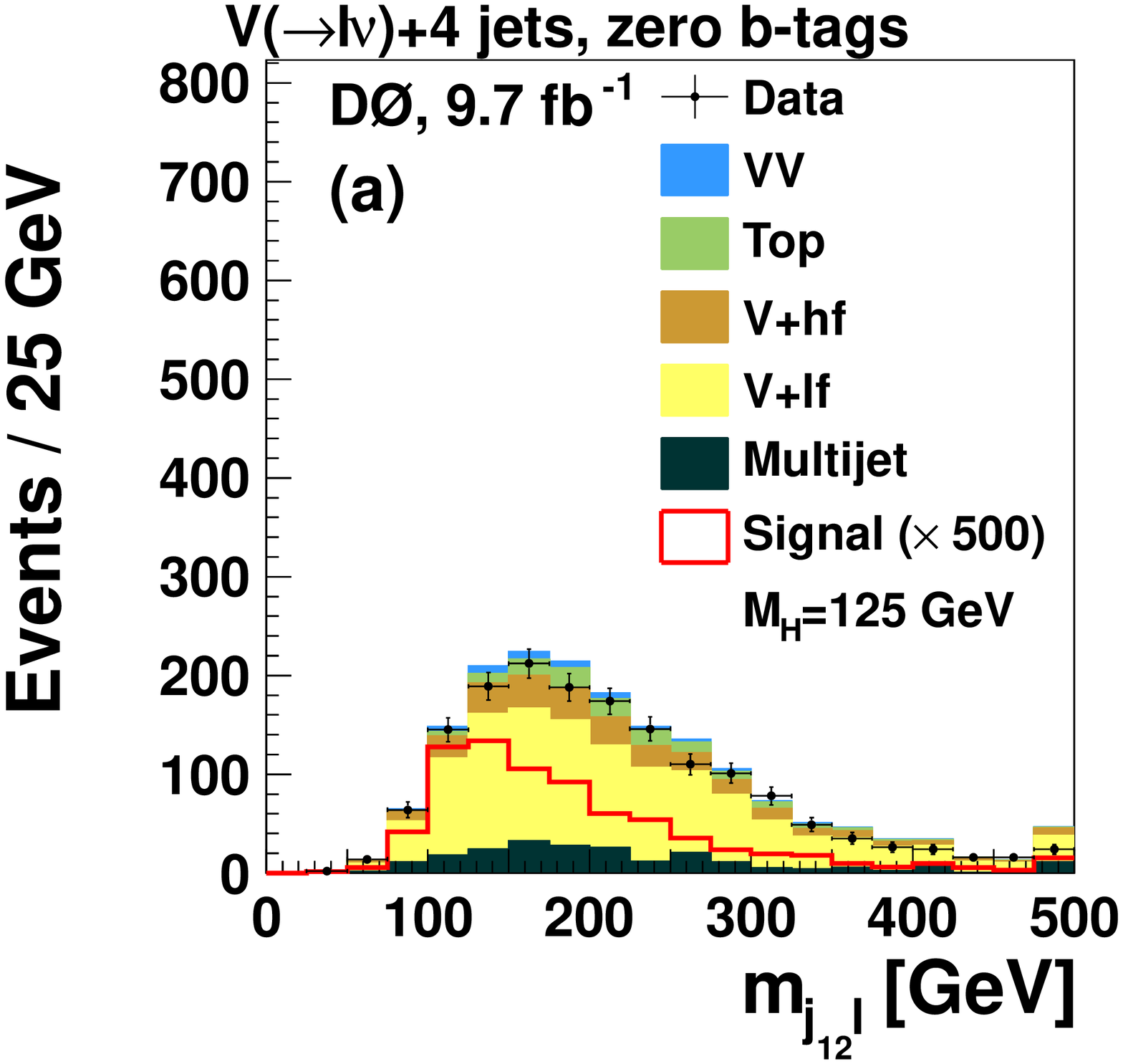}
\includegraphics[width=0.38\textwidth]{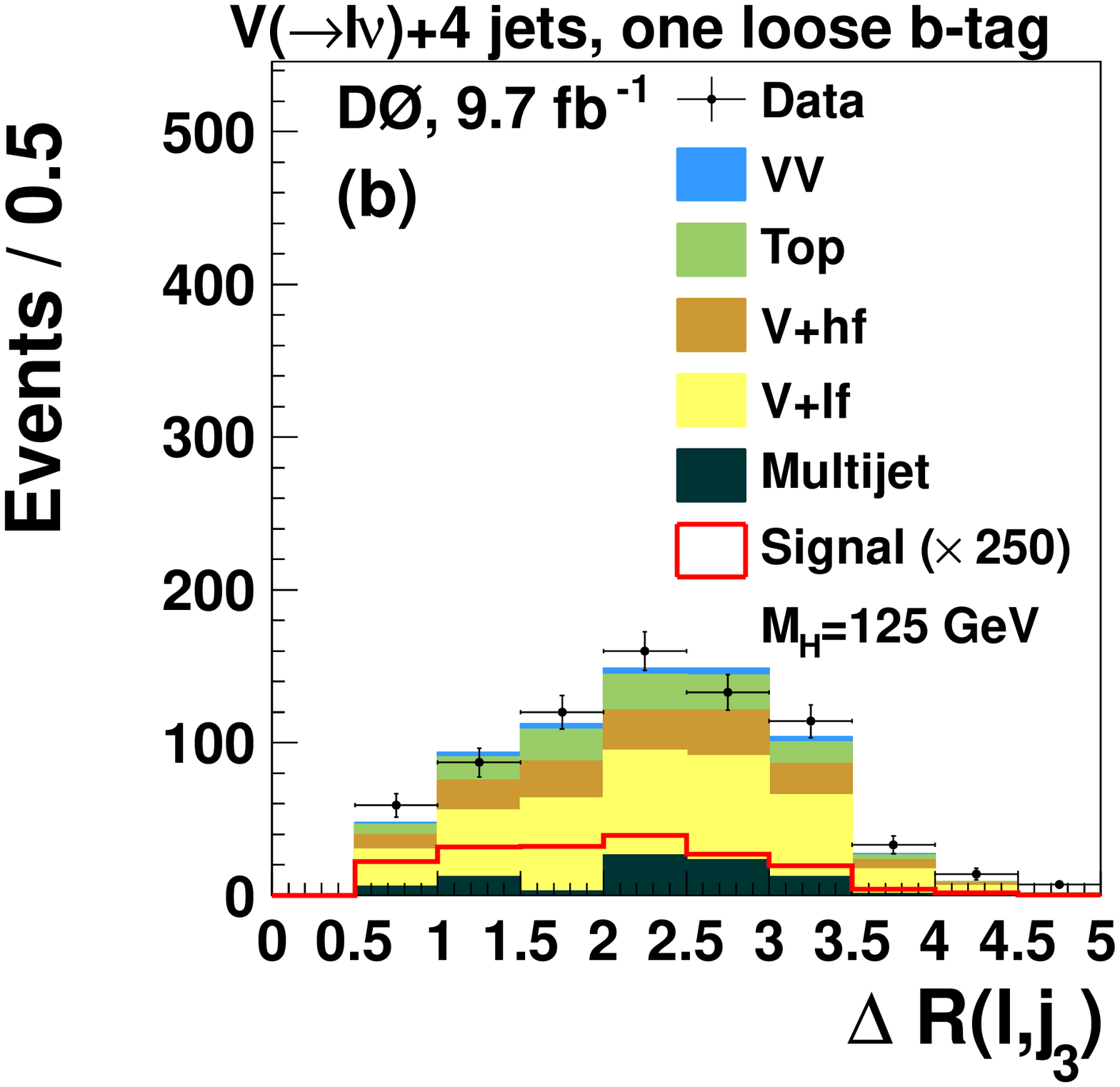}
}\caption{(color online)
Distributions of the most significant inputs, other than background-specific multivariate discriminants,
to the final multivariate discriminants for the
$\geq 4$-jet subchannels:
(a) \jABlepm, shown for events with zero $b$-tags;
(b) \jetlepdrC, shown for events with one loose $b$-tag.
The $M_H=125$~GeV signal is multiplied by 500 in (a) and 250 in (b). 
Overflow events are added to the last bin.
}
\label{fig:lv4jFinalMVA}
\end{figure}

\begin{figure}[htbp]
\centering{
\includegraphics[width=0.38\textwidth]{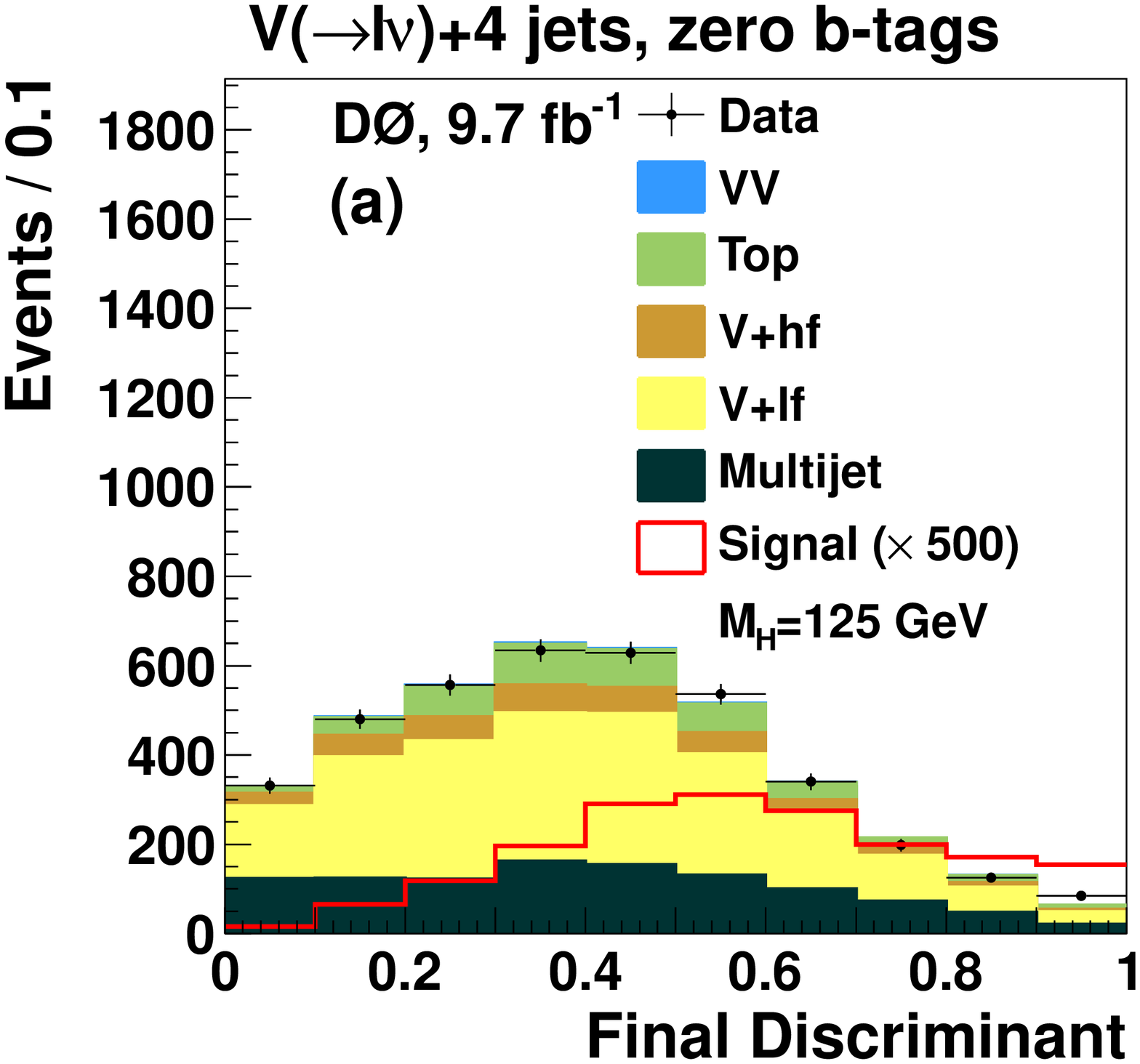}
\includegraphics[width=0.38\textwidth]{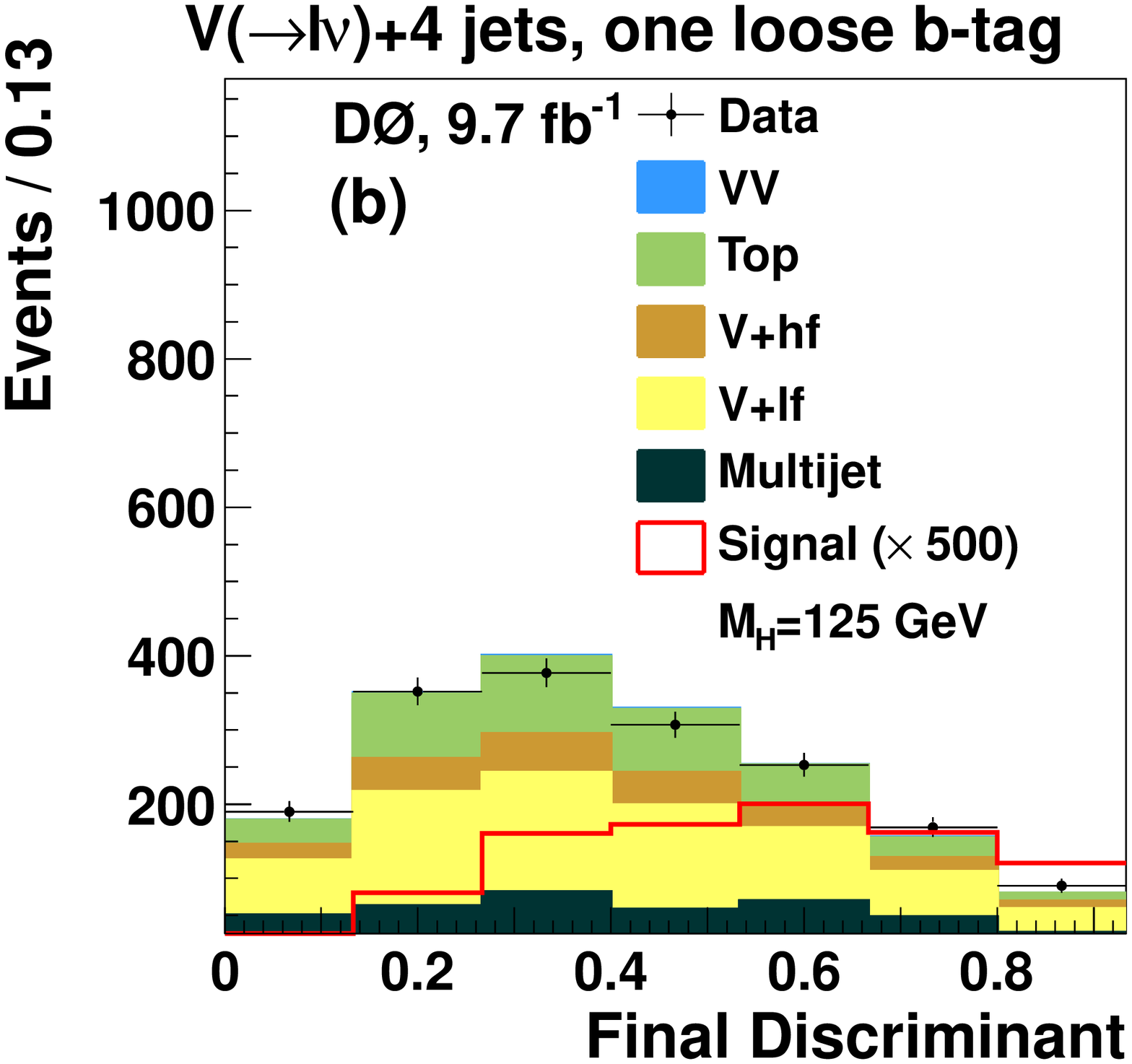}
}\caption{(color online)
Distributions of the final discriminant output, after the maximum likelihood fit (described in Sec.~\ref{sec:limits}), at $M_H=125$~GeV for the four or more jets channels with:
(a) zero $b$-tags and (b) one loose $b$-tag.
The $M_H=125$~GeV signal is multiplied by 500.
}
\label{fig:lv4jFinalMVAout}
\end{figure}

The background-specific discriminants are used as inputs to the final RF discriminant 
that is trained to discriminate all signal processes from the total background contributions.  
Additional input variables for the final discriminant are selected based on their separation power between the total 
signal and the total background, and are required to be well modeled.  The input variables for each lepton and $b$-tag category are listed in Appendix~\ref{app:input_var_def}. 
Sample inputs and output responses of the final discriminants are shown in Figs.~\ref{fig:lv4jFinalMVA} and \ref{fig:lv4jFinalMVAout}, respectively, for $M_H=125$~GeV.

\clearpage

\section{Systematic Uncertainties \label{sec:syst}}

We assess systematic uncertainties on signals and backgrounds for each of the jet multiplicity and $b$-tag
channels by repeating the full analysis after varying each source of uncertainty by $\pm 1$ s.d.
We consider uncertainties that affect both the normalizations and the shapes
 of our MVA outputs.

We include theoretical uncertainties on the $t\bar{t}$ and single top quark production cross sections 
(7\%~\cite{Langenfeld:2009wd,Kidonakis:2006bu}), diboson production cross section (6\%~\cite{Campbell:1999ah}), 
$V+lf$ production (6\%), and $V+hf$ production (20\%, estimated from 
{\sc mcfm}~\cite{Campbell:2001ik,mcfm_code}). Since the $V$+jets experimental scaling factors for the three- and four-jet channels 
are different from unity, we apply an additional systematic uncertainty on the $V$+jets samples that is uncorrelated across 
jet multiplicity and lepton flavor bins. The size of this uncertainty is taken as the uncertainty from the $V$+jets fit to data,
described in Sec.~\ref{sec:MJ}.

An uncertainty on the integrated luminosity (6.1\%~\cite{Andeen:2007zc}) affects the normalization of the expected signal and simulated backgrounds.
Uncertainties that affect the 
final MVA distribution shapes include jet taggability (3\% per jet), $b$-tagging efficiency (2.5\%--3\%\ per heavy-quark jet),
the light-quark jet misidentification rate (10\% per jet), jet identification efficiency (5\%), 
and jet energy calibration and resolution (varying between 5\%\ and 15\%, depending on the process 
and channel) as described in Ref.~\cite{Abazov:2012wh}. 
We also include uncertainties from modeling that affect both the shapes and normalizations of the final MVA distributions. 
These include an uncertainty on the trigger efficiency in the muon channel as derived from the data (3\%--5\%),
lepton identification and reconstruction efficiency (5\%--6\%), the MLM matching~\cite{Mangano:2002ea} applied 
to $V$+light-flavor events ($\approx0.5$\%), the {\sc alpgen} renormalization and factorization 
scales, and the choice of parton distribution functions (2\%) as described in Ref.~\cite{Abazov:2012wh}. 
The trigger uncertainty in the muon channel is calculated as the difference between applying a trigger correction calculated 
using the {\sc alpgen} reweightings derived on the $T_{\mu\text{OR}}$ trigger sample and applying the nominal trigger correction.
Since we reweight our {\sc alpgen} samples, we include separate uncertainties on each of the five functions used to apply
the reweighting. 
The adjusted functions are calculated by shifting the parameter 
responsible for the largest shape variation of 
 the fit by $\pm 1$ s.d.\ then calculating the remaining parameters 
for the function using the covariance matrix obtained from the functional fit.

We determine the uncertainty on the MJ background shape by relaxing the requirement from Sec.~\ref{sec:eventsel} on $\lnumt$ to 
 $\lnumt >30\ \mathrm{GeV} -0.5~\times\MET$ and repeating the analysis with this selection in place. The positive and negative variations 
are taken to be symmetric. The uncertainty on the MJ rate is $15\%$ ($20\%$) for the electron
(muon) channel. Since our MJ sample is statistically limited, we do not correlate the uncertainties on 
the rate and shape across the subchannels. 
Since we simultaneously fit MJ and $V$+jets to match data, we apply a normalization uncertainty to the $V$+jets samples 
that is anticorrelated with the MJ normalization systematics and scales as the relative MJ to $V$+jets normalization.

\section{$\bm{WZ}$ and $\bm{ZZ}$ Production with $\bm{Z\to b\bar{b}}$ \label{sec:diboson}}

\begin{figure}[htbp]
\centering{
\includegraphics[width=0.45\textwidth]{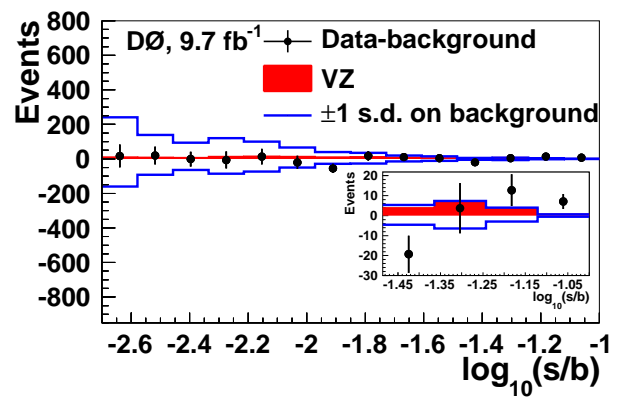}
}\caption{
(color online) Final MVA discriminant output shown for the expected diboson signal and background-subtracted data rebinned as a function of $
\log(S/B)$, after the maximum likelihood fit, summed over $b$-tag channels. 
The error bars on data points represent the statistical uncertainty only.
The post-fit systematic uncertainties are represented by the solid lines. The signal
expectation is shown scaled to the best fit value. The inset gives an expanded view of the high $\log(S/B)$ region.
}
\label{fig:diboson}
\end{figure}

The SM processes $W(\rightarrow \ell \nu)Z(\rightarrow b \bar{b})$ and $Z(\rightarrow \ell \ell)Z(\rightarrow b \bar{b})$ 
where one of the leptons from the $Z\to\ell\ell$ decay is not reconstructed,
result in the same final state signature as the Higgs boson in this search. 
Therefore, we search for these processes to validate our analysis methodology. 
The only change in the analysis is in the training of the final discriminant in events with two or three jets with one tight $b$-tag or two $b$-tags.
We train using the $WZ$ and $ZZ$ diboson processes as signal while leaving the $WW$ process as a background. 
The output of this discriminant is used to measure the combined $WZ$ and $ZZ$ cross section by performing
a maximum likelihood fit to data using signal plus background models, with maximization over the systematic uncertainties as described in detail in Sec.~\ref{sec:limits}.
The expected significance of the measurement using the MVA output is \diboexpsigMVA\ s.d.
We measure a cross section of \diboxsecMVA\ times the expected SM cross section of \diboSMxsec.
Figure~\ref{fig:diboson} shows the MVA discriminant output for the diboson cross section ($WZ + ZZ$) with background-subtracted data and signal scaled to the best fit value.

\section{Upper Limits on the Higgs Boson Production Cross Section \label{sec:limits}}

We derive upper limits on the Higgs boson production cross section multiplied by the 
corresponding branching fraction in units of the SM prediction. 
The limits are
calculated using the modified frequentist
$CL_{s}$ approach~\cite{Junk:1999kv,Read:2002hq,wade_tm}, and the procedure is
repeated for each assumed value of $M_H$. 

Two hypotheses are considered: the background-only hypothesis 
(B), in which only background contributions are present, and the signal-plus-background (S+B) 
hypothesis in which both signal and background contributions are present. 

The limits are determined using the MVA output distributions, 
together with their associated uncertainties, as inputs 
to the limit setting procedure. To preserve the stability of the limit derivation procedure in regions of small 
background statistics in the one tight $b$-tag and all two $b$-tags categories, the width of the 
bin at the largest MVA output value is adjusted by comparing the total background and 
signal+background expectations until the statistical significances for B and S+B are, respectively, 
greater than 3.6 and 5.0 s.d.\ from zero. The remaining part of the 
distribution is then divided into equally sized bins. 
In the zero $b$-tags and one loose $b$-tag categories, the width of the bin at largest MVA output is set such 
that the relative statistical uncertainty on the signals plus background entries is less than 0.15. The remaining 
bins are distributed uniformly.
The rebinning procedure is checked for potential biases in the
determination of the final limits, and no such bias is observed.

We evaluate the compatibility of the data with the background-only and signal+background hypotheses. This is done using the log likelihood ratio ($LLR$), which is twice the negative logarithm of the ratio of the Poisson likelihoods, $L$, of the signal+background hypothesis to the background only hypothesis, $LLR=-2\ln(L_{S+B}/L_{B})$.

Systematic uncertainties are included through nuisance parameters that
are assigned Gaussian probability distributions (priors). 
The signal and background predictions are functions of the nuisance parameters. Each common source of systematic uncertainty (such as the uncertainties on predicted SM cross sections, identification efficiencies, 
and energy calibration, as described in Sec.~\ref{sec:syst}) is taken to be correlated across all 
channels except as otherwise noted in Sec.~\ref{sec:syst}.

The inclusion of systematic uncertainties in the generation of
pseudoexperiments has the effect of broadening the expected $LLR$
distributions and, thus, reducing the ability to resolve
signal-like excesses.  This degradation can be partially reduced
by performing a maximum likelihood fit to each
pseudoexperiment (and data), once for the
B hypothesis and once for the S+B hypothesis.  The
maximization is performed over the systematic uncertainties.  The
$LLR$ is evaluated for each outcome using the ratio of maximum
likelihoods for the fit to each hypothesis.
The resulting degradation of the limits due to systematic uncertainties
is $\sim30\%$ for searches in the vicinity of $M_H=125$~GeV.

\begin{figure*}[htbp]
\includegraphics[width=0.4\textwidth]{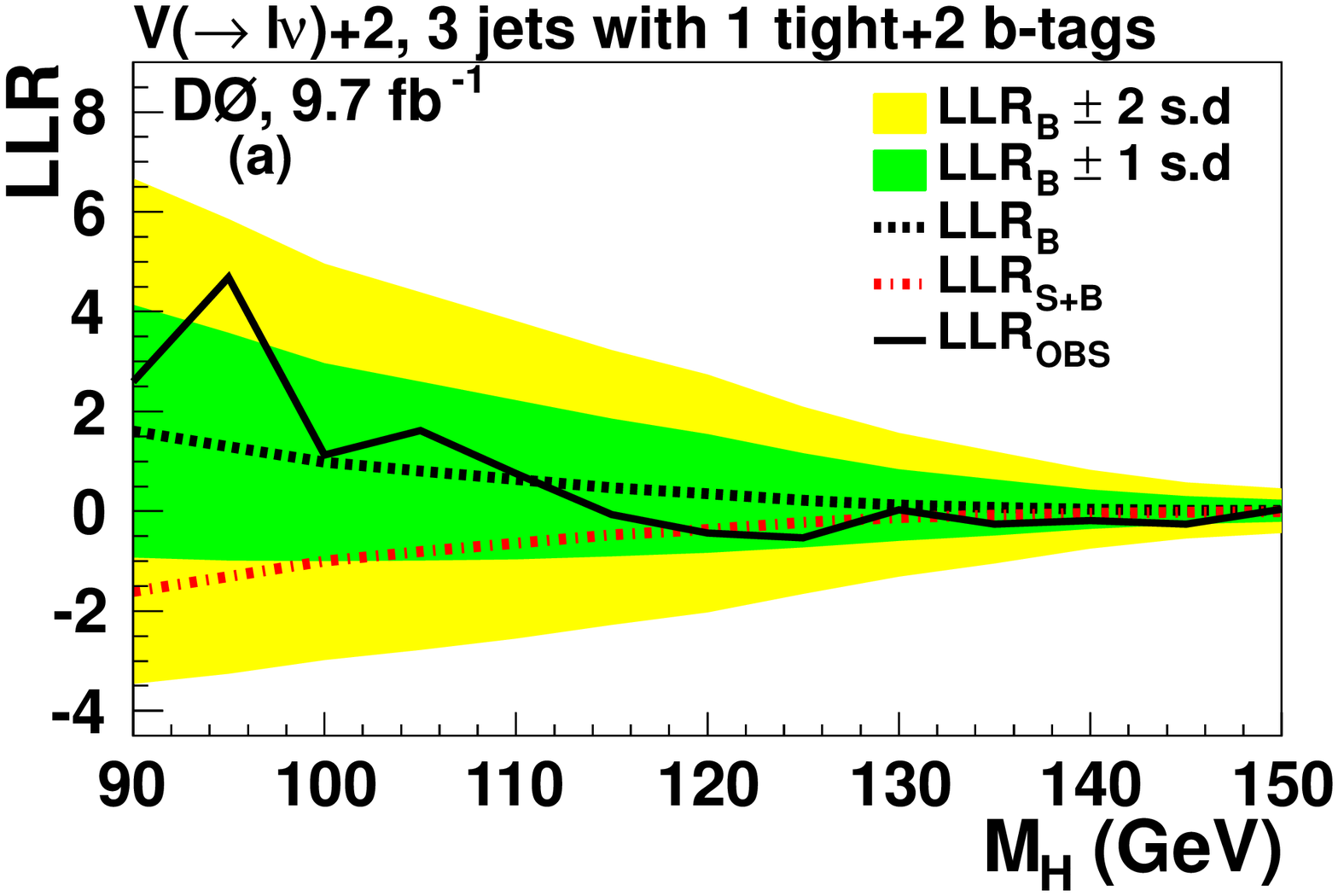}
\includegraphics[width=0.4\textwidth]{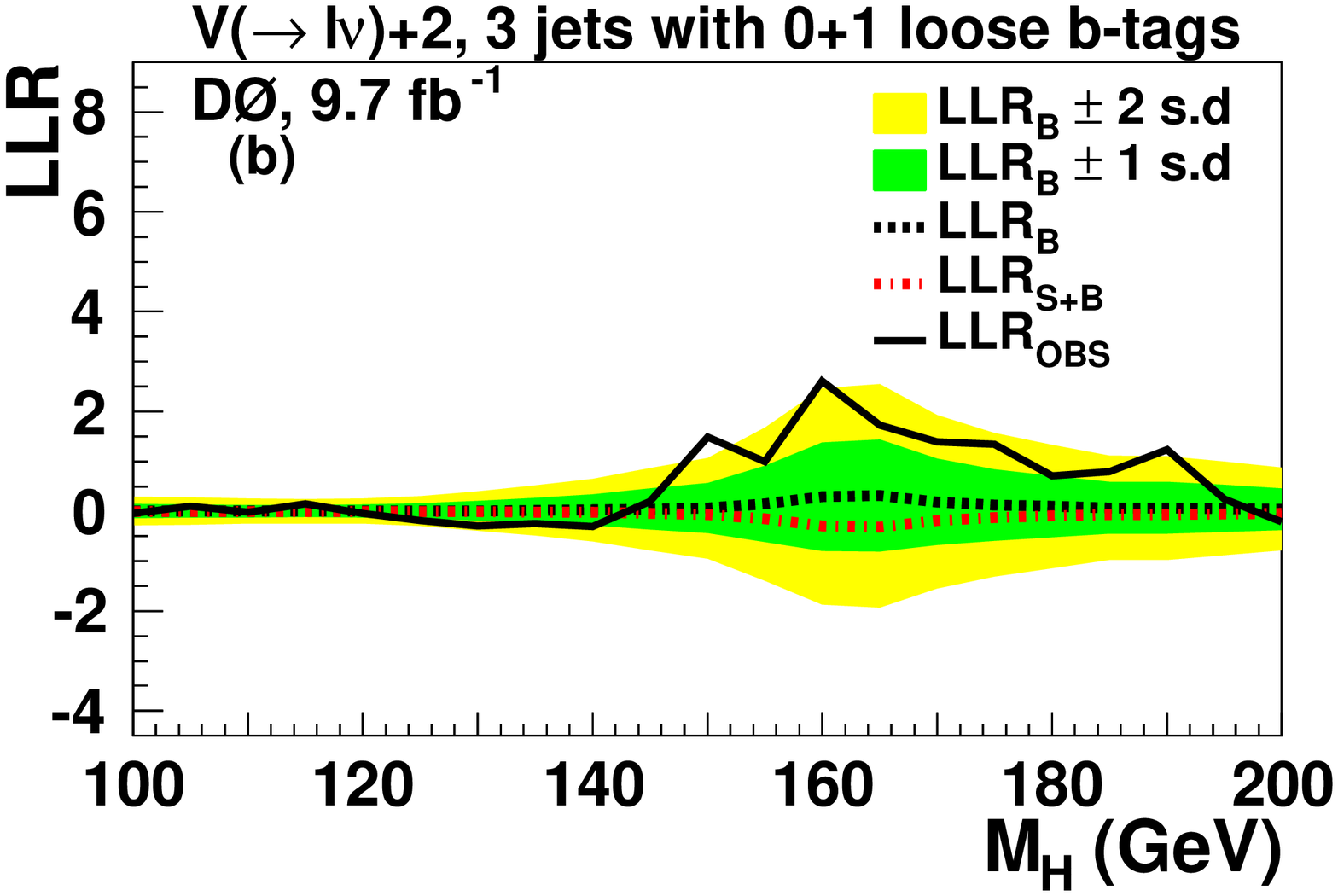}
\includegraphics[width=0.4\textwidth]{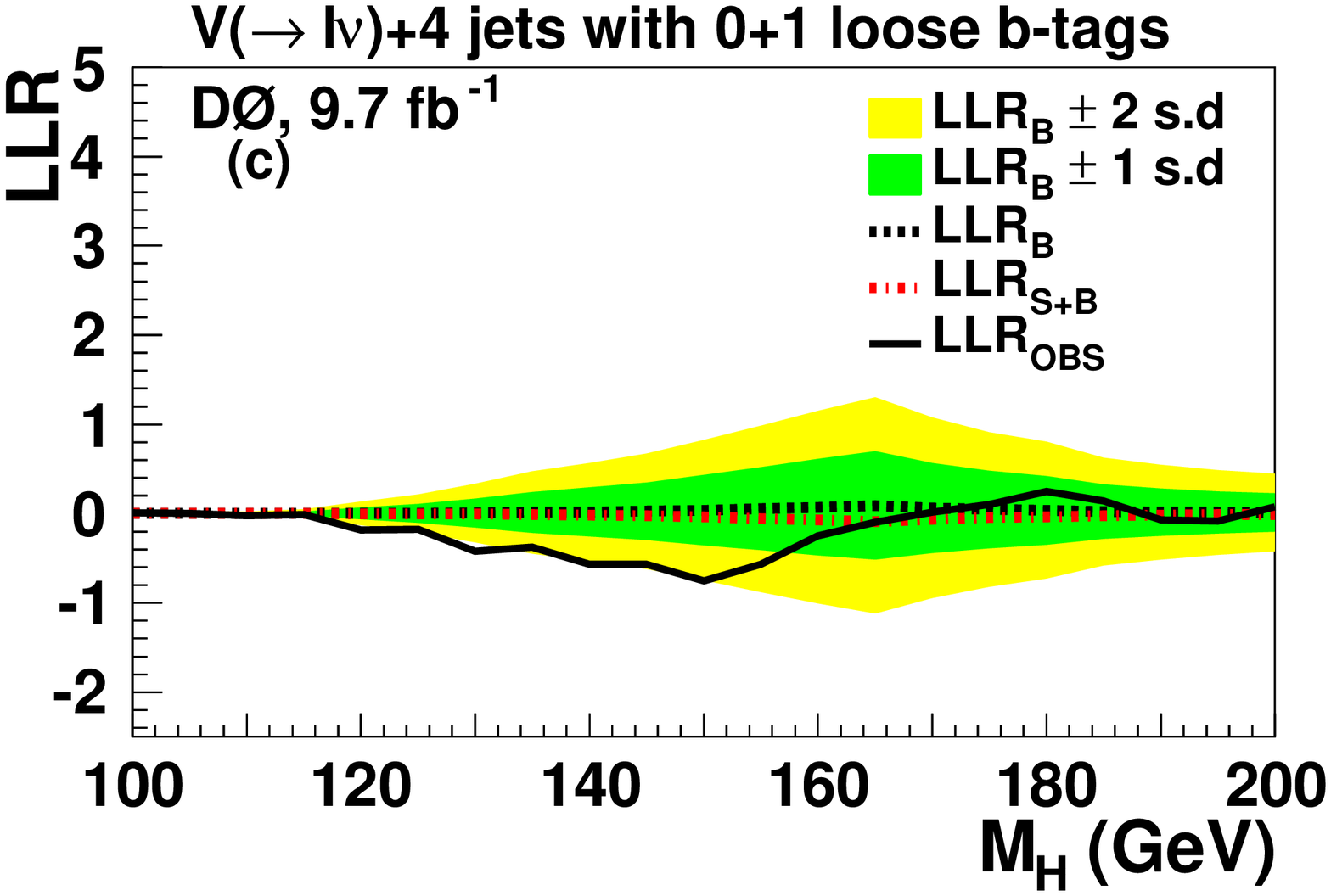}
\includegraphics[width=0.4\textwidth]{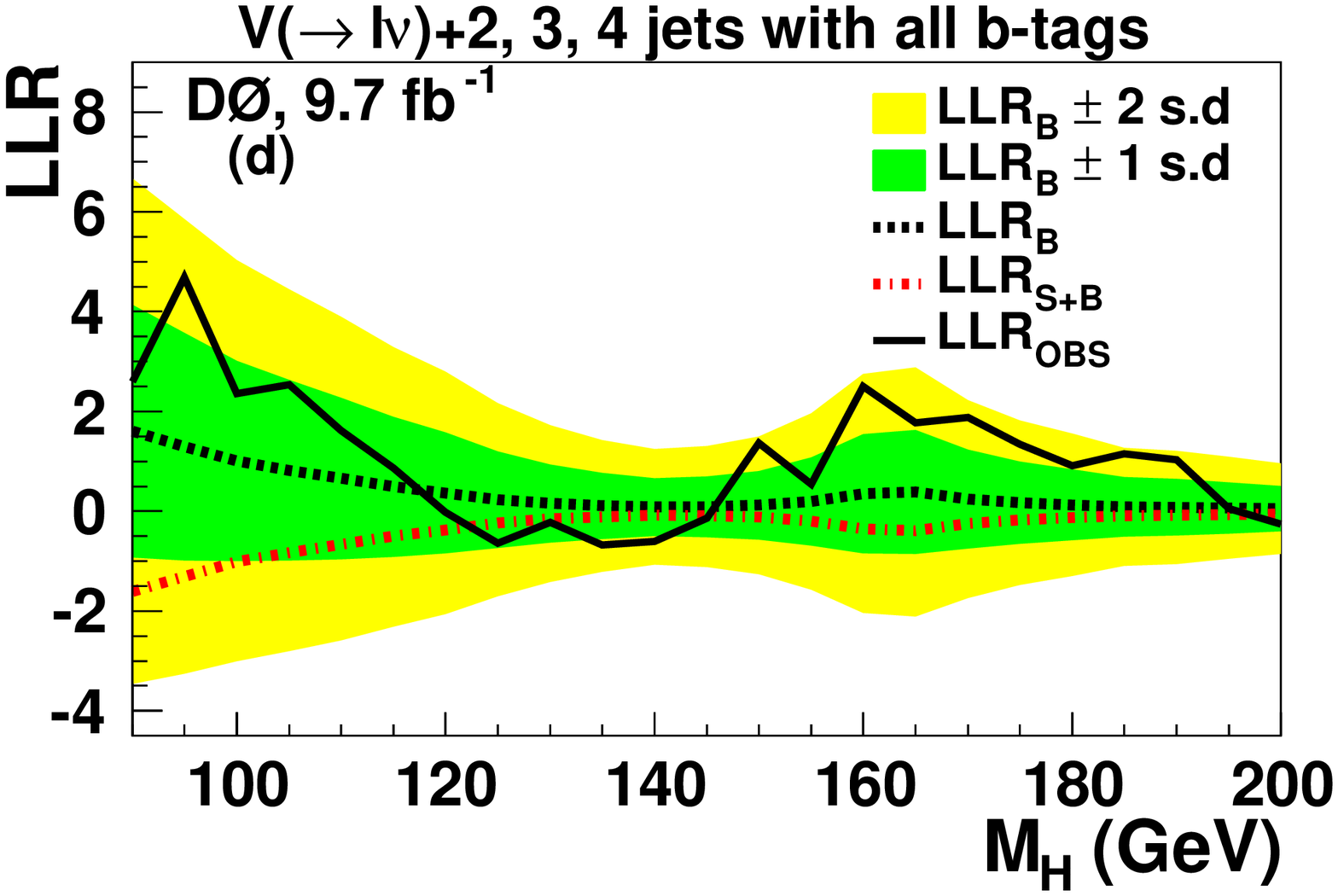}
\caption{(color online) The expected and observed log likelihood ratios as  functions of the hypothesized Higgs boson 
mass $M_H$ for the (a) electron and muon, two and three jets, one tight and two $b$-tag channels; (b) electron and muon, two and three jets, 
zero and one loose $b$-tag channels; (c) electron and muon, four or more jets, zero and one loose $b$-tag channels; (d) combination of all channels. The dashed red and 
black lines correspond to the median $LLR$ of the signal+background and background-only hypotheses, respectively. 
The solid line corresponds to the $LLR$ obtained from the data, and the shaded regions are the $\pm 1$~s.d.\ and $\pm 2$~s.d.\ values for the background-only hypothesis. }
\label{fig:llr} 
\end{figure*}

The medians of the obtained $LLR$ distributions for the B and S+B hypotheses  for each tested mass
are presented in Fig.~\ref{fig:llr}. 
The corresponding $\pm 1$~s.d.\ and 
$\pm 2$~s.d.\ values for the background-only hypothesis at each mass point are represented
by the shaded regions in the figure. The $LLR$ values obtained from the data are also presented in the figure.  

\begin{figure}[htbp]
\includegraphics[width=0.45\textwidth]{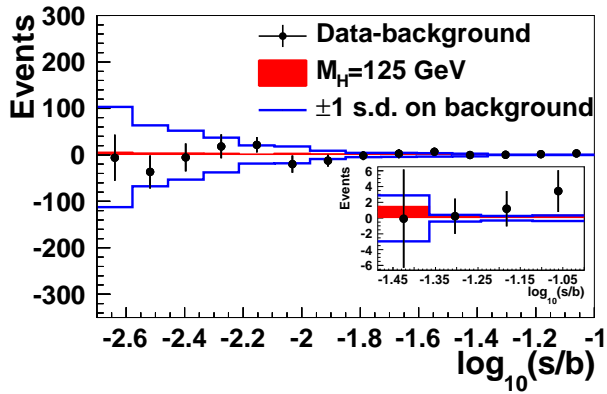}
\caption{(color online) The MVA discriminant output distribution minus the total background expectation for $M_H=125$~GeV rebinned as a function of $
\log(S/B)$. 
The post-fit uncertainties are represented by the solid lines. The signal
expectation is shown scaled to the best fit value.  The inset gives an expanded view of the high $\log(S/B)$ region.}
\label{fig:syst_plot}
\end{figure}

\begin{figure*}[htbp]
\includegraphics[width=0.4\textwidth]{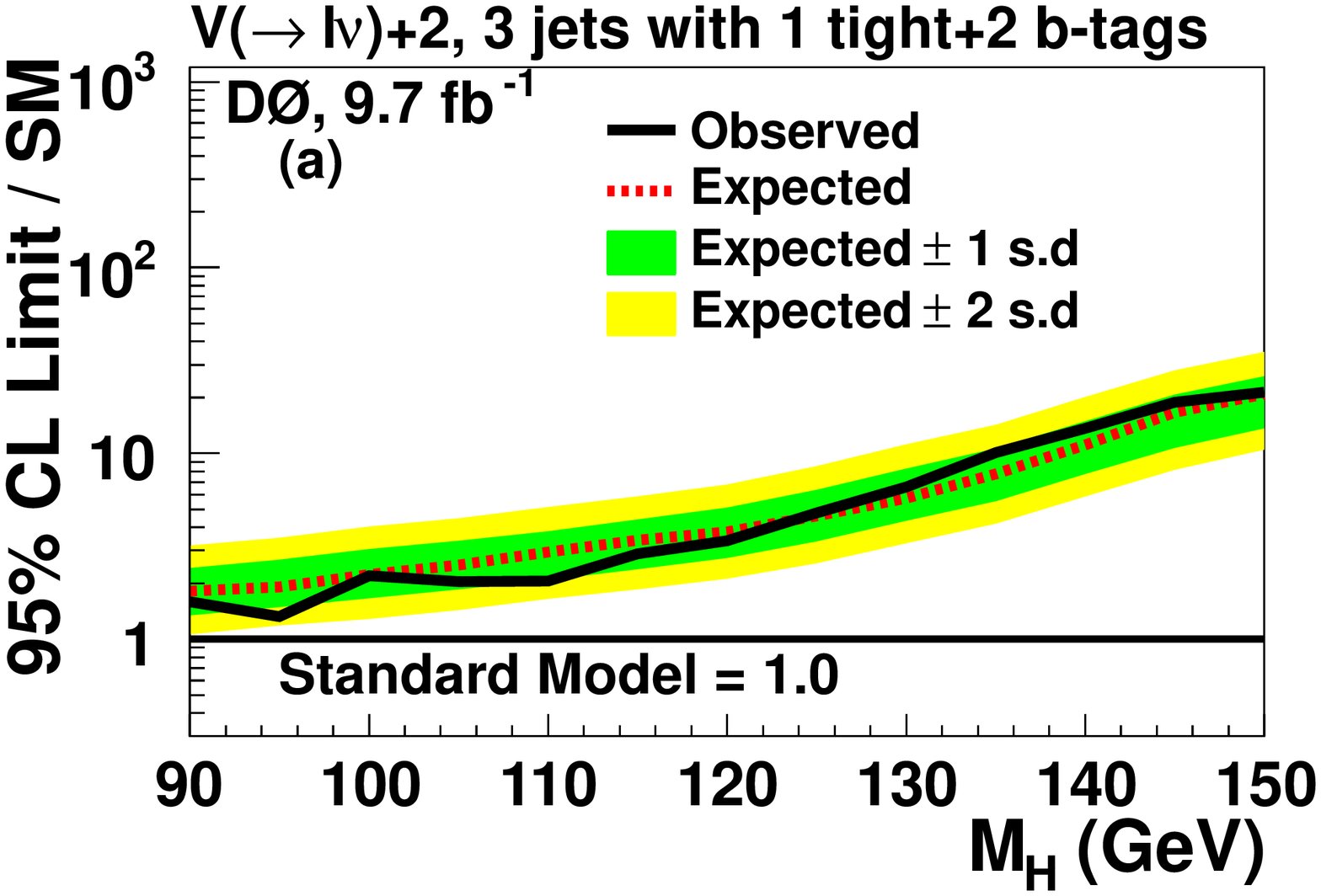}
\includegraphics[width=0.4\textwidth]{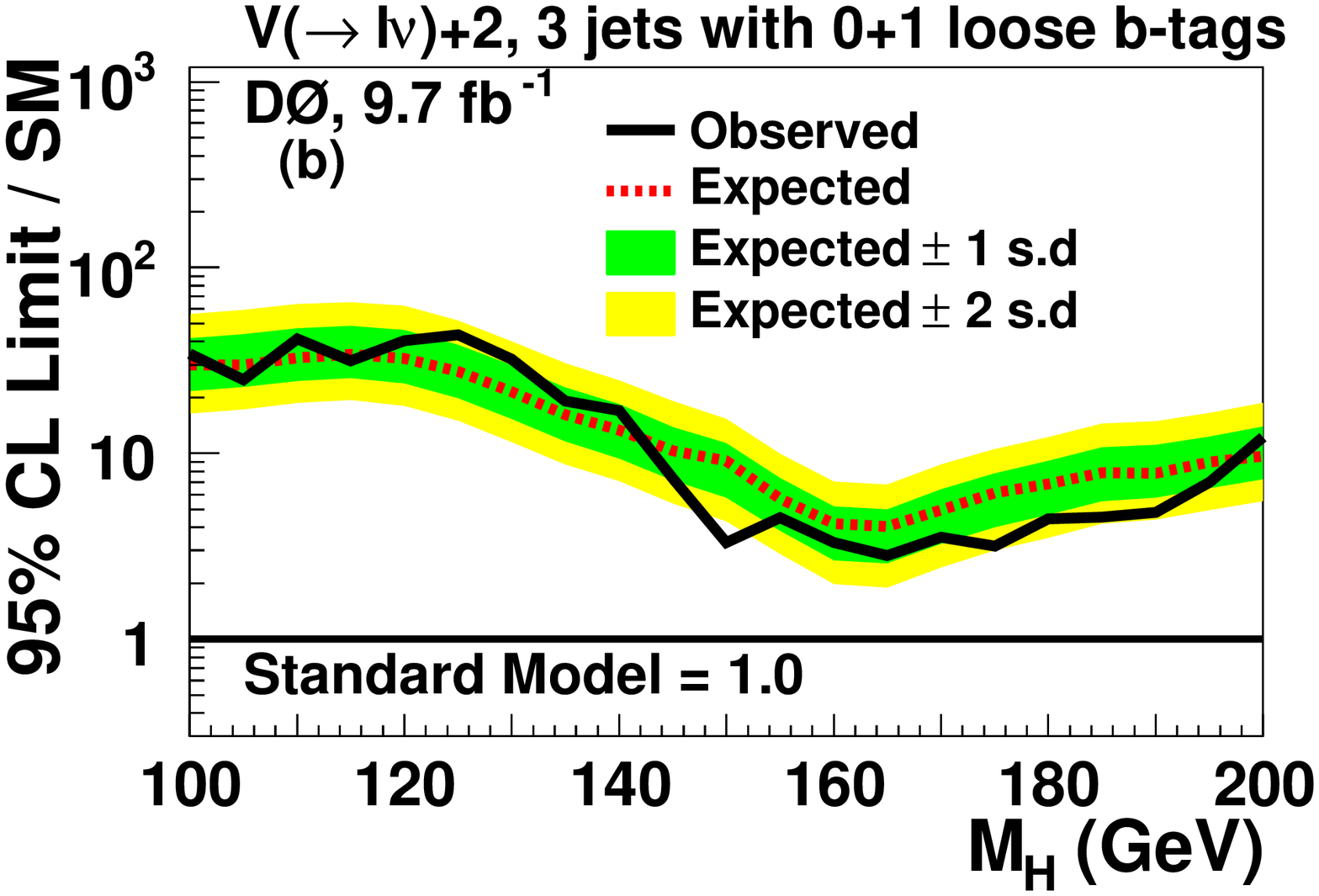}
\includegraphics[width=0.4\textwidth]{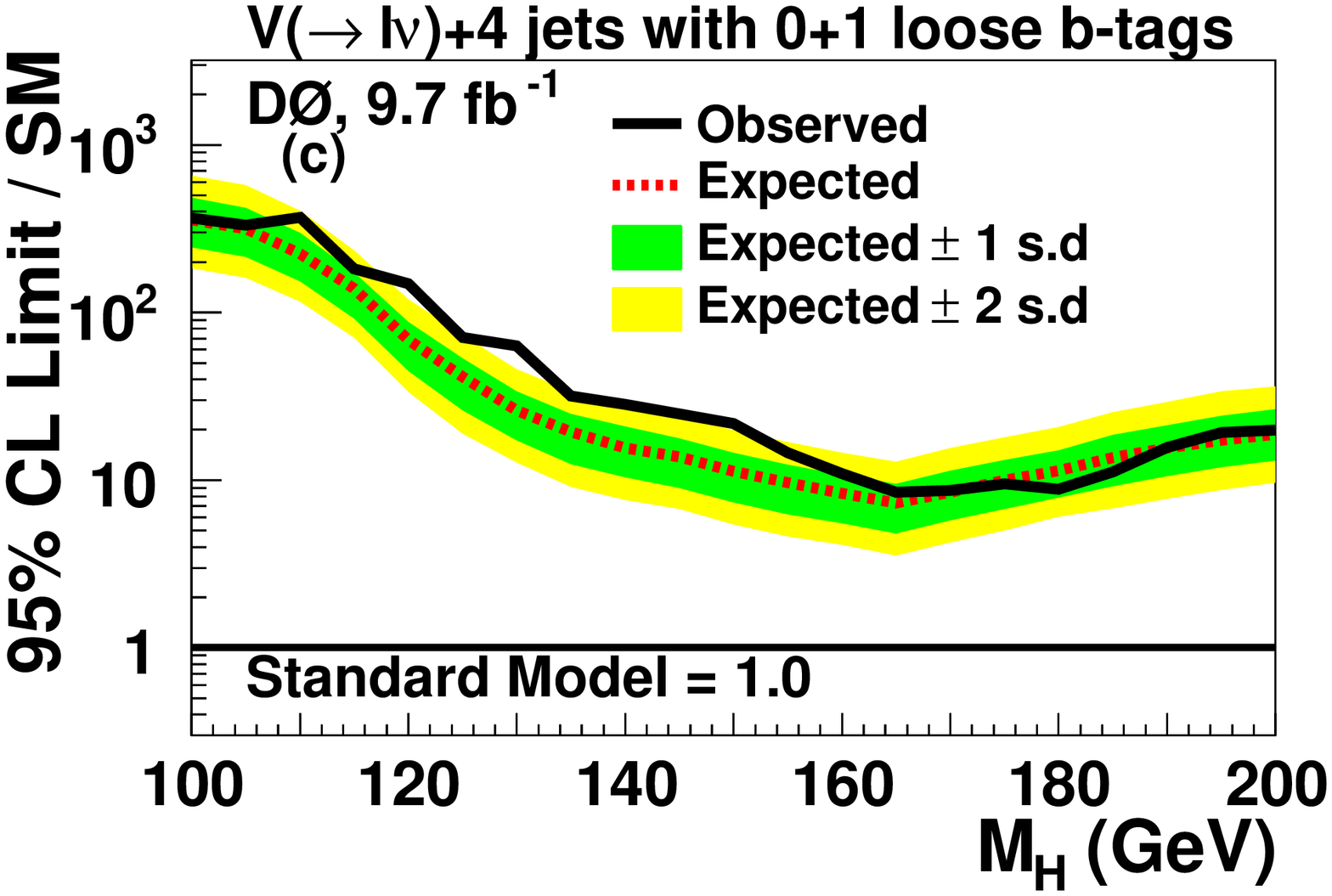}
\includegraphics[width=0.4\textwidth]{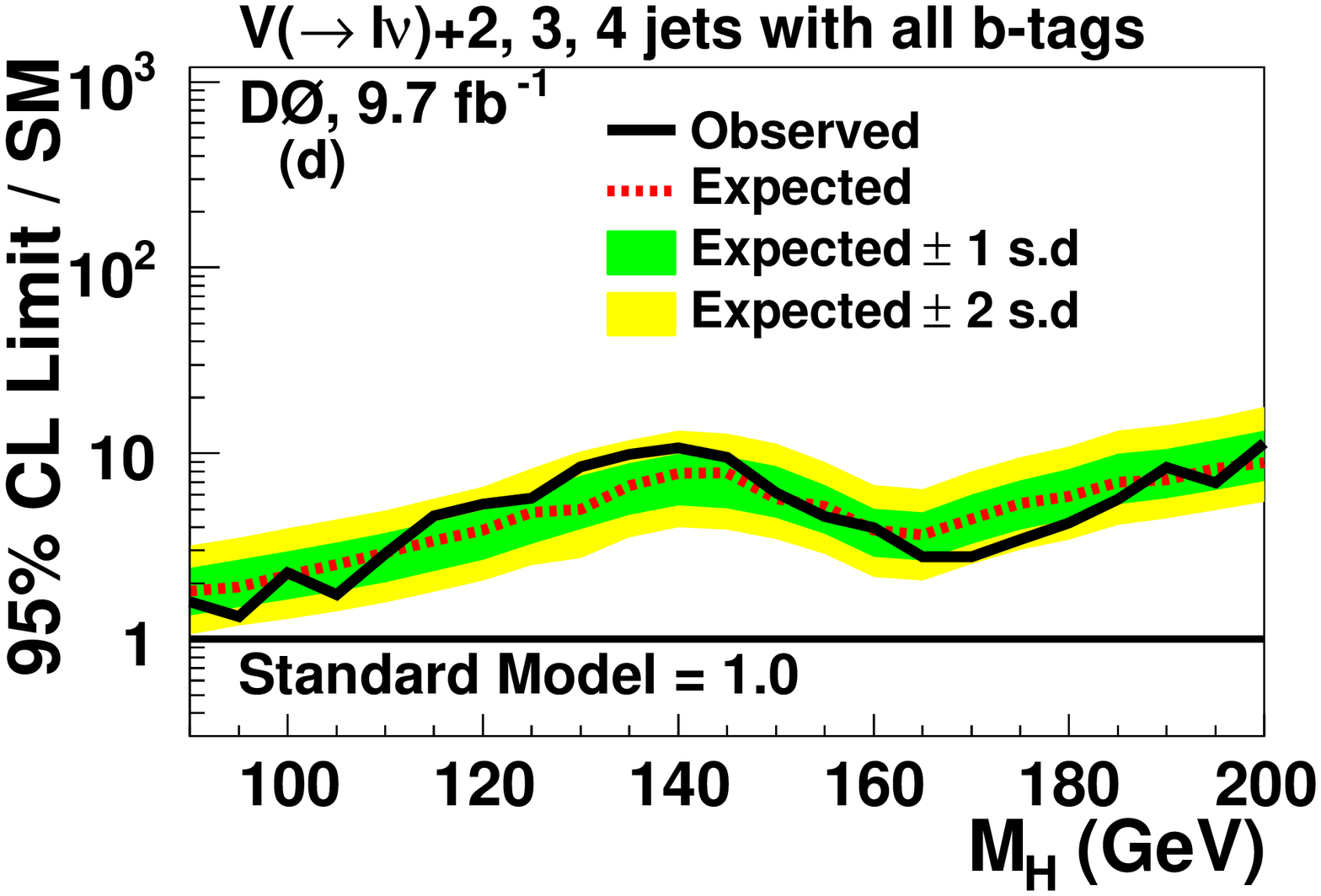}
\caption{(color online) 
The expected and observed 95\% C.L.\ upper limits on SM Higgs boson production for the
(a) electron and muon, two and three jets, one tight and two $b$-tag channels;
(b) electron and muon, two and three jets, zero and one loose $b$-tag channels ($M_H \leq 150$~GeV) and pretag channels ($M_H \geq 155$~GeV);
(c) electron and muon, four or more jets, zero and one loose $b$-tag channels;
(d) combination of all channels.
The limits are presented as ratios 
to the expected SM prediction. The dashed line corresponds to the expected limit, and the solid 
line corresponds to the limit observed in data. The shaded regions are the $\pm 1$~s.d.\ and $\pm 2$~s.d.\ values for the expected limit.
}
\label{fig:limits} 
\end{figure*}

The MVA discriminant distributions, for the Higgs boson mass 
point $M_H=125$~GeV, after 
subtracting the total posterior background expectation are shown in Fig.~\ref{fig:syst_plot}. The signal
expectation is shown scaled to the observed upper limit (described later) and the uncertainties in the background after the constrained fit are shown by the 
solid lines.

\begin{table*}[htbp]
\vspace{-0.3cm}
\caption{The expected and observed 95\% C.L.\ limits, as a function of the Higgs boson mass $M_H$, presented as ratios of
production cross section times branching fraction to the expected SM prediction.\\ }
\label{limitvalues}
\begin{tabular}{cccccccccccccccccccccccc}
\hline
 &    \multicolumn{23}{c}{Combined 95\% C.L. Limit  $ /  \sigma_{SM}$ }\\
 $M_H$ (GeV) & 90 & 95 & 100 & 105 & 110 & 115 & 120 & 125 & 130 & 135 & 140 & 145 & 150 & 155 & 160 & 165 & 170 & 175 & 180 & 185 & 190 & 195 & 200 \\
\hline
 &    \multicolumn{23}{c}{2 or 3 jets with one tight $b$-tag or two $b$-tags }\\
Expected  & 1.8 & 1.9 & 2.2 & 2.5 & 2.9 & 3.4 & 3.8 & 4.7 & 5.8 & 7.9 & 11.1 & 16.7 & 20.8 & -- & -- & -- & -- & -- & -- & -- & -- & -- & -- \\
Observed  & 1.6 & 1.3 & 2.2 & 2.0 & 2.1 & 2.9 & 3.4 & 4.8 & 6.6 & 10.1 & 13.6 & 18.8 & 18.5 & -- & -- & -- & -- & -- & -- & -- & -- & -- & -- \\
\hline
 &    \multicolumn{23}{c}{2 or 3 jets with zero $b$-tags or one loose $b$-tag }\\
Expected & -- & -- & 29.8 & 30.0 & 32.6 & 34.0 & 32.5 & 27.5 & 21.6 & 16.2 & 13.3 & 10.3 & 9.1 & 5.7 & 4.2 & 4.0 & 5.0 & 6.1 & 6.8 & 7.9  & 7.8  & 9.0 & 9.7 \\
Observed & -- & -- & 34.4 & 24.9 & 41.4 & 31.4 & 40.3 & 43.5 & 32.3 & 19.1 & 17.0 & 7.3  & 3.3 & 4.5 & 3.3 & 2.8 & 3.5 & 3.2 & 4.4 & 4.5  & 4.8  & 7.0 & 12.2 \\
\hline
 &    \multicolumn{23}{c}{4 or more jets with zero $b$-tags or one loose $b$-tag }\\
Expected & -- & -- & 357 & 316 & 224 & 139 & 68.6& 41.2 & 26.2 & 19.4 & 15.5 & 13.7 & 11.3 & 9.7  & 8.3  & 7.3 & 8.5 & 10.0 & 11.4 & 13.7 & 15.6 & 17.3 & 18.8 \\
Observed & -- & -- & 365 & 331 & 369 & 182 & 149 & 71.2 & 63.4 & 31.8 & 28.3 & 24.9 & 21.9 & 14.6 & 10.9 & 8.5 & 8.7 & 9.5 & 8.8  & 11.2 & 15.7 & 19.2 & 19.8 \\
\hline
 &    \multicolumn{23}{c}{All channels combined }\\
Expected & 1.8 & 1.9 & 2.2 & 2.5 & 2.9 & 3.4 & 3.8 & 4.7 & 5.0 & 6.7 & 7.8 & 7.9 & 5.7 & 5.2 & 3.8 & 3.7 & 4.4 & 5.4 & 5.9 & 7.0 & 7.2 & 8.3 & 8.9 \\
Observed & 1.6 & 1.3 & 2.3 & 1.7 & 2.9 & 4.6 & 5.3 & 5.8 & 8.5 & 9.9 & 10.7 & 9.6 & 6.1 & 4.6 & 4.0 & 2.8 & 2.8 & 3.4 & 4.2 & 5.7 & 8.4 & 6.9 & 11.4 \\
\hline
\end{tabular}
\end{table*}

Upper limits are calculated at 23 discrete values of the Higgs boson mass, spanning the range 90--200~GeV and spaced in increments of 
5~GeV, by scaling the expected 
signal contribution to the value at which it can be excluded at the 95\% 
C.L. The expected limits are calculated from the background-only $LLR$ distribution whereas the observed limits are quoted with respect 
to the $LLR$ values measured in data. The expected and observed 95\% C.L. upper limits 
results for the  Higgs boson production cross section multiplied by the decay branching fraction are shown, as a function of the Higgs boson mass $M_H$, 
in units of the SM prediction in Fig.~\ref{fig:limits}. The values obtained for the expected and observed limit to SM ratios at each 
mass point are listed in Table~\ref{limitvalues} for all one-tight, two-loose, two-medium, and two-tight $b$-tag subchannels together, 
for the two-jet and three-jet, zero and one loose $b$-tag subchannels (all $b$-tag categories for $M_{H} > 150$~GeV) together, the 
$\geq 4$-jet subchannels, and 
the combination of all subchannels.

\section{Interpretations in fourth generation and fermiophobic Higgs models \label{sec:BSM}}

Extensions of the minimal electroweak symmetry breaking mechanism of the SM may be allowed,
including models with a fourth generation of fermions or with a Higgs boson that has modified couplings to fermions,
as in fermiophobic Higgs models (FHM).
We interpret our results in these scenarios using the subchannels that are sensitive to $H\to WW$ decays: events with two or more jets
and zero or one loose $b$-tag for $M_H \leq 150$~GeV, extended to include pretag two- and three-jet events for $M_H \geq 155$~GeV. These are the first
results for these models in the $\ell \nu+$jets final state.

Previous results from the Tevatron Collider experiments in the context of a fourth generation of fermions
set a limit on the $M_H$ of $131 < M_{H} < 207$~GeV~\cite{Aaltonen:2010sv}.  The ATLAS~\cite{Aad:2011qi} and CMS~\cite{Chatrchyan:2011tz} collaborations
exclude $140 < M_{H} < 185$~GeV and $144 < M_{H} < 207$~GeV, respectively.
Previous searches for the fermiophobic Higgs boson in $H\to\gamma\gamma$ 
and $H\to VV$ channels, with two leptons in the final state, were carried out at the LEP $e^+e^-$
Collider~\cite{Heister:2002ub, Abdallah:2003xf,Achard:2003jb,Abbiendi:2002vu}, by the CDF~\cite{Collaboration:2012pa} and
D0~\cite{Abazov:2011ix} Collaborations, and by the ATLAS~\cite{Aad:2012yq} and CMS~\cite{CMS:2012bd} Collaborations, with the most stringent limits being set by the CMS experiment
where the excluded range is $110 < M_H < 194$~GeV.

\begin{figure}[htbp]
\includegraphics[width=0.4\textwidth]{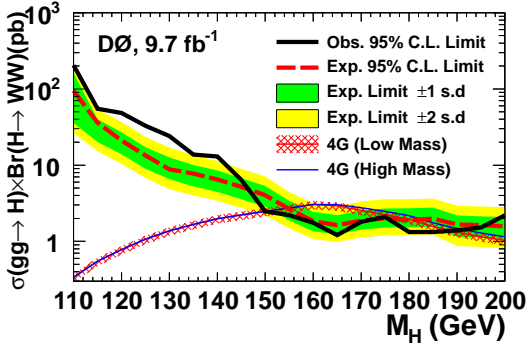}
\caption{(color online) 
The expected and observed 95\% C.L.\ upper limits on $\sigma(gg \to H) \times \mathcal{B}(H \to WW)$
compared to the prediction from the fourth-generation fermion model.
}
\label{fig:bsm4G} 
\end{figure}

The $Hgg$ coupling is enhanced in fourth-generation models, which leads to a higher rate of $gg\to H$ production
and a larger decay width of $H\to gg$ than in the SM~\cite{Holdom:2009rf,Kribs:2007nz,Arik:2005ed,Anastasiou:2010bt}.
However, since $H\to gg$ is loop-mediated, the $H\to WW^*$ decay mode dominates for $M_H > 135$~GeV, as in the SM.
We consider two scenarios for the presence of a fourth generation.
In the ``low-mass'' scenario, we assume a fourth-generation neutrino mass of $m_{\nu 4}=80$~GeV and a value for the fourth-generation
charged lepton mass of $m_{\ell 4}=100$~GeV, while in the ``high-mass'' scenario, we assume values for the fourth-generation neutrino and lepton masses of $m_{\nu 4}=m_{\ell 4}=1$~TeV.
Both scenarios set the fourth-generation quark masses to the values in Ref.~\cite{Anastasiou:2010bt}.
After applying our selection criteria, the total expected signal for $gg \to H$ production in the low-mass (high-mass) fourth-generation model is enhanced by a factor of 7.2 (7.5) over the SM production rate for $M_H=125$~GeV.
We only consider gluon fusion Higgs boson production, and we set limits
on $\sigma(gg \to H) \times \mathcal{B}(H \to WW^*)$.
These limits are compared with the predicted
$gg \to H$ production cross section results from {\sc hdecay}~\cite{Djouadi:1997yw}, as shown in Fig.~\ref{fig:bsm4G}. We exclude the ``low-mass'' scenario for $150<M_H<188$~GeV, and the ``high-mass'' scenario for $150<M_H<190$~GeV.

\begin{figure}[htbp]
\includegraphics[width=0.4\textwidth]{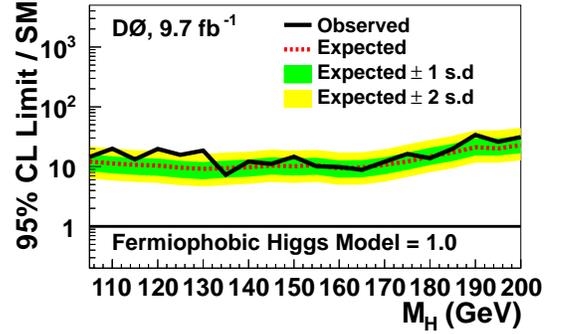}
\caption{(color online) 
The expected and observed 95\% C.L.\ upper limits on fermiophobic Higgs boson production. 
}
\label{fig:bsmFHM} 
\end{figure}

In the FHM, the Higgs boson does not couple to fermions at tree level but is otherwise SM-like.
This suppresses production via gluon fusion to a negligible rate and forbids direct decay to fermions.
Production in association with a vector boson or via vector boson fusion is allowed.
For this interpretation, we set the contribution from $gg\to H$ production to zero and scale the contributions
from other production and decay mechanisms to reflect the predicted rate in the FHM.
After applying our selection criteria, the total expected signal for vector boson fusion and $VH\to VWW$ production in the FHM is enhanced by a factor of 4.2 over the SM production rate for $M_H=125$~GeV.
The expected and observed cross section times branching fraction limits are compared to the FHM predictions in Fig.~\ref{fig:bsmFHM}. 

\section{Summary}

We have presented a search for SM Higgs boson production in lepton + \MET\ + jets final states with a dataset corresponding to 9.7 fb$^{-1}$
of integrated luminosity collected with the D0 detector. The search is sensitive to $VH\rightarrow Vb\bar{b}$, 
$H\rightarrow WW^{*}\rightarrow \ell \nu jj$, and $WH \rightarrow WWW^{*} \rightarrow \ell\nu jjjj$ production and decay, and supersedes previous $VH\rightarrow Vb\bar{b}$ and $H\rightarrow WW^{*}\rightarrow \ell \nu jj$ searches published by D0.
To maximize our signal sensitivity, we subdivide the dataset into 36 independent subchannels according to lepton flavor, jet multiplicity, and the number
and quality of $b$-tagged jets and apply multivariate analysis techniques to further discriminate between signal and background. 
We test our method by examining SM $WZ$ and $ZZ$ production with 
$Z\to b\bar{b}$ decay and find production rates consistent 
with the SM prediction.
We observe no 
significant excess over the background prediction as expected from the amplitude of a 125~GeV SM Higgs boson signal,
given the sensitivity of this single channel. Significance is achieved
by combining this channel with the other low mass channels analyzed
at the Tevatron~\cite{tevatron-bbbar}, while here we set 95\%\ C.L. upper limits on the Higgs boson production cross section for masses between 90 and 
200~GeV. For $M_H=125$~GeV, the observed (expected) upper limit is \obslimA\ (\explimA) times the SM prediction. We interpret the data also in models with fourth generation fermions, or a fermiophobic Higgs boson. In these interpretations, we exclude $150<M_H<188(190)$~GeV in the ``low-mass'' (``high-mass'') fourth generation fermion scenario, and provide 95\% C.L\ limits on the production cross section in the fermiophobic model. 

\section{Acknowledgments}
\input acknowledgement.tex
\FloatBarrier

\bibliography{higgs}
\bibliographystyle{h-physrev3}

\clearpage
\appendix
\input{appendixA}

\end{document}

%% file: dictionary.tex
\newcommand{\MPT}{\mbox{\ensuremath{\not\!p_T}}}
\newcommand{\Apla}{\ensuremath{\cal{A}}}
\newcommand{\Sphe}{\ensuremath{\cal{S}}}
\newcommand{\Centrality}{\ensuremath{\cal{C}}}
\newcommand{\Simila}{\ensuremath{\cal{SIM}}}
\newcommand{\Sigm}{\ensuremath{\cal{SIG}}}
\newcommand{\SigmJ}{\ensuremath{{\cal{SIG}}_{\text{jets}}}}
\newcommand{\Hel}{\ensuremath{\cal{H}}}
\newcommand{\Vel}{\ensuremath{\cal{V}}}
\newcommand{\Twist}{\ensuremath{\cal{T}}}
\newcommand{\Width}{\ensuremath{\cal{W}}}
\newcommand{\jA}{\ensuremath{j_1}}
\newcommand{\jB}{\ensuremath{j_2}}
\newcommand{\jC}{\ensuremath{j_3}}
\newcommand{\jD}{\ensuremath{j_4}}
\newcommand{\jAB}{\ensuremath{j_{12}}}
\newcommand{\jAC}{\ensuremath{j_{13}}}
\newcommand{\jBC}{\ensuremath{j_{23}}}
\newcommand{\jABC}{\ensuremath{j_{123}}}
\newcommand{\jABCD}{\ensuremath{j_{1234}}}
\newcommand{\jABangle}{\ensuremath{\angle(\jA,\jB)}}
\newcommand{\jABcmcostheta}{\ensuremath{\cos\theta(\jA,\jB)_{\text{CM}}}}
\newcommand{\jABdeta}{\ensuremath{\Delta\eta(\jA,\jB)}}
\newcommand{\jABdetamax}{\ensuremath{\max|\Delta\eta(\jAB,\{\jA \text{~or~} \jB\})|}}
\newcommand{\jABdphi}{\ensuremath{\Delta\phi(\jA,\jB)}}
\newcommand{\jABdphiA}{\ensuremath{\Delta\phi(\jAB,\jA)}}
\newcommand{\jABdpt}{\ensuremath{\Delta p_T(\jA,\jB)}}
\newcommand{\jABdr}{\ensuremath{\deltaR(\jA,\jB)}}
\newcommand{\jABdrmin}{\ensuremath{\min\deltaR(\jAB,\{\jA \text{~or~} \jB\})}}
\newcommand{\jABhelicityA}{\ensuremath{{\Hel}(\jAB,\jA)}}
\newcommand{\jABktmin}{\ensuremath{K_T^{\min}}}
\newcommand{\jABm}{\ensuremath{m_{\jAB}}}
\newcommand{\jABmbb}{\ensuremath{m_{b1b2}}}
\newcommand{\jABmt}{\ensuremath{m_T^{\jAB}}}
\newcommand{\jABptoversumpt}{\ensuremath{p_T^{\jAB}/\jABsumpt}}
\newcommand{\jABrecoilpt}{Recoil\ensuremath{(p_T^{\jAB})}}
\newcommand{\jABsigmamax}{\ensuremath{\max[{\Sigm} (\jAB,\{\jA \text{~or~} \jB\})]}}
\newcommand{\jABsigmamin}{\ensuremath{\min[{\Sigm} (\jAB,\{\jA \text{~or~} \jB\})]}}
\newcommand{\jABsigmaA}{\ensuremath{{{\Sigm}} (\jAB,\jA)}}
\newcommand{\jABsumpt}{\ensuremath{\sum\limits_{i=1}^{2} p_T^{j_i}}}
\newcommand{\jABtwist}{\ensuremath{{\Twist}(\jAB)}}
\newcommand{\jABvelj}{\ensuremath{{\Vel}(\jAB)}}
\newcommand{\jABCm}{\ensuremath{m_{\jABC}}}
\newcommand{\jABCpt}{\ensuremath{p_T^{\jABC}}}
\newcommand{\jABCsigma}{\ensuremath{{\SigmJ} (\jABC)}}
\newcommand{\jABCDm}{\ensuremath{m_{\jABCD}}}
\newcommand{\jABCDsigma}{\ensuremath{{\SigmJ} (\jABCD)}}
\newcommand{\jABCDsumpt}{\ensuremath{\sum\limits_{i=1}^{4} p_T^{j_i}}}
\newcommand{\jABbislepangle}{\ensuremath{\angle[\ell,\text{bis}(\jA,\jB)]}}
\newcommand{\jABbislnudphi}{\ensuremath{\Delta\phi[W,\text{bis}(\jA,\jB)]}}
\newcommand{\jABbisnu}{\ensuremath{\text{bis}(\jAB,\nu)}}
\newcommand{\jABlepangle}{\ensuremath{\angle(\ell,\jAB)}}
\newcommand{\jABlepm}{\ensuremath{m_{\jAB\ell}}}
\newcommand{\jABlepsumpt}{\ensuremath{\sum p_T(\jA,\jB,\ell)}}
\newcommand{\jACdphi}{\ensuremath{\Delta\phi(\jA,\jC)}}
\newcommand{\jACdrB}{\ensuremath{\deltaR(\jC,\jAC)}}
\newcommand{\jACvelj}{\ensuremath{{\Vel}_{j13}}}
\newcommand{\jBCdphiB}{\ensuremath{\Delta\phi(\jC,\jBC)}}
\newcommand{\jBCdptA}{\ensuremath{\Delta p_T(\jB,\jBC)}}
\newcommand{\jBCptoversumpt}{\ensuremath{p_T^{\jBC}/\sum\limits_{i=2}^{3} p_T^{j_i}}}
\newcommand{\jBCtwist}{\ensuremath{{\Twist}_{j23}}}
\newcommand{\jeteA}{\ensuremath{E(\jA)}}
\newcommand{\jetptA}{\ensuremath{p_T^{\jA}}}
\newcommand{\jetptB}{\ensuremath{p_T^{\jB}}}
\newcommand{\jetptC}{\ensuremath{p_T^{\jC}}}
\newcommand{\jetwidthC}{\ensuremath{{\Width}_{j3}}}
\newcommand{\jetlepdetaA}{\ensuremath{|\Delta\eta(\ell,\jA)|}}
\newcommand{\jetlepdetaB}{\ensuremath{|\Delta\eta(\ell,\jB)|}}
\newcommand{\jetlepdetaC}{\ensuremath{|\Delta\eta(\ell,\jC)|}}
\newcommand{\jetlepdetamax}{\ensuremath{\max|\Delta\eta(\ell,\{\jA \text{~or~} \jB\})|}}
\newcommand{\jetlepdrA}{\ensuremath{\deltaR(\ell,\jA)}}
\newcommand{\jetlepdrB}{\ensuremath{\deltaR(\ell,\jB)}}
\newcommand{\jetlepdrC}{\ensuremath{\deltaR(\ell,\jC)}}
\newcommand{\jetlepqetaA}{\ensuremath{q^{\ell}\times\eta^{\jA}}}
\newcommand{\jetlepqetaB}{\ensuremath{q^{\ell}\times\eta^{\jB}}}
\newcommand{\jetlepqetaC}{\ensuremath{q^{\ell}\times\eta^{\jC}}}
\newcommand{\jetnudphiA}{\ensuremath{\Delta\phi(\MET,\jA)}}
\newcommand{\jetnudrA}{\ensuremath{\deltaR(\nu,\jA)}}
\newcommand{\jetnudrminjAB}{\ensuremath{\min[\deltaR(\nu,\lbrace \jA~\text{or}~\jB\rbrace)]}}
\newcommand{\bidbl}{\ensuremath{b_{\text{ID}}^{\jAB}}}
\newcommand{\bidblA}{\ensuremath{b_{\text{ID}}^{j_1}}}
\newcommand{\bidblB}{\ensuremath{b_{\text{ID}}^{j_2}}}
\newcommand{\bidblI}{\ensuremath{b_{\text{ID}}^{j_i}}}
\newcommand{\lepe}{\ensuremath{E(\ell)}}
\newcommand{\lepqeta}{\ensuremath{q^{\ell}\times\eta^{\ell}}}
\newcommand{\lepsigma}{\ensuremath{{\SigmJ} (\ell)}}
\newcommand{\lepljetdr}{\ensuremath{\deltaR(\ell, \text{light~jet})}}
\newcommand{\lnuangle}{\ensuremath{\angle(\ell,\nu)}}
\newcommand{\lnucmcostheta}{\ensuremath{\cos\theta(\ell)_{\ell\nu \text{CM}}}}
\newcommand{\lnucosthetaA}{\ensuremath{\cos\theta(\ell)}}
\newcommand{\lnudeta}{\ensuremath{\Delta\eta(\ell,\nu)}}
\newcommand{\lnudetamax}{\ensuremath{\max|\Delta\eta(W,\{\ell \text{~or~} \nu\})|}}
\newcommand{\lnudphi}{\ensuremath{\Delta\phi(\ell,\MET)}}
\newcommand{\lnudphimax}{\ensuremath{\max|\Delta\phi(W,\{\ell \text{~or~} \MET\})|}}
\newcommand{\lnudphimin}{\ensuremath{\min|\Delta\phi(W,\{\ell \text{~or~} \MET\})|}}
\newcommand{\lnudpt}{\ensuremath{\Delta p_T(\ell,\MET)}}
\newcommand{\lnudptoverpt}{\ensuremath{\Delta p_T(\ell,\MET)/\lnupt}}
\newcommand{\lnudr}{\ensuremath{\deltaR(\ell,\nu)}}
\newcommand{\lnudrmax}{\ensuremath{\max[\deltaR(W,\{\ell \text{~or~} \nu\})]}}
\newcommand{\lnudrmin}{\ensuremath{\min[\deltaR(W,\{\ell \text{~or~} \nu\})]}}
\newcommand{\lnulepdeta}{\ensuremath{|\Delta\eta(W,\ell)|}}
\newcommand{\lnulepdpt}{\ensuremath{\Delta p_T(W,\ell)}}
\newcommand{\lnulepdr}{\ensuremath{\deltaR(W,\ell)}}
\newcommand{\lnulepsigma}{\ensuremath{{\Sigm} (W,\ell)}}
\newcommand{\lnumt}{\ensuremath{M_T^W}}
\newcommand{\lnunudpt}{\ensuremath{\Delta p_T(W,\MET)}}
\newcommand{\lnunudr}{\ensuremath{\deltaR(W,\nu)}}
\newcommand{\lnupt}{\ensuremath{p_T^W}}
\newcommand{\lnuptfracmax}{\ensuremath{\max(p_T^{\ell},\MET)/\lnupt}}
\newcommand{\lnuptfracmin}{\ensuremath{\min(p_T^{\ell},\MET)/\lnupt}}
\newcommand{\lnuptoversumpt}{\ensuremath{\lnupt/(\lnusumpt)}}
\newcommand{\lnurecoilpt}{Recoil\ensuremath{(p_T^W)}}
\newcommand{\lnusigmamax}{\ensuremath{\max[{\Sigm}(W,\{\ell \text{~or~} \nu\})] }}
\newcommand{\lnusigmamin}{\ensuremath{\min[{\Sigm}(W,\{\ell \text{~or~} \nu\})] }}
\newcommand{\lnusimilarity}{\ensuremath{\Simila(\ell,\nu)}}
\newcommand{\lnusumpt}{\ensuremath{p_T^{\ell}+\MET}}
\newcommand{\lnutwist}{\ensuremath{{\Twist}_{W\to\ell\nu}}}
\newcommand{\lnujdetaB}{\ensuremath{|\Delta\eta(W,\jB)|}}
\newcommand{\lnujdphiB}{\ensuremath{\Delta\phi(W,\jB)}}
\newcommand{\lnujdrB}{\ensuremath{\deltaR(W,\jB)}}
\newcommand{\lnujmA}{\ensuremath{m_{\ell\nu\jA}}}
\newcommand{\lnujmB}{\ensuremath{m_{\ell\nu\jB}}}
\newcommand{\lnujptB}{\ensuremath{p_T^{\ell\nu\jB}}}
\newcommand{\lnujsumptB}{\ensuremath{p_T^{\ell}+\MET+p_T^{\jB}}}
\newcommand{\lnujjchistar}{\ensuremath{\cos(\chi^* )}}
\newcommand{\lnujjcmjABangle}{\ensuremath{\angle(\jA,\jB)_{\text{HCM}}}}
\newcommand{\lnujjcmjAlepcosangle}{\ensuremath{\cos[\angle(\jA,\ell \nu)]_{\text{HCM}}}}
\newcommand{\lnujjcmlnujAangle}{\ensuremath{\cos[\angle(\jA,\ell)]_{\text{HCM}}}}
\newcommand{\lnujjdeta}{\ensuremath{|\Delta\eta(W,\jAB)|}}
\newcommand{\lnujjjABframejZcosangle}{\ensuremath{\cos[\angle(\jA^{\jAB\text{CM}},(W\to\ell\nu)^{\text{HCM}})]}}
\newcommand{\lnujjlnuframelepZcosangle}{\ensuremath{\cos[\angle(\ell_{\ell\nu CM},(W\to\ell\nu)_{\text{HCM}})]}}
\newcommand{\lnujjm}{\ensuremath{m_{\ell\nu\jAB}}}
\newcommand{\lnujjmasym}{\ensuremath{m^{\text{Asym}}}}
\newcommand{\lnujjmt}{\ensuremath{m_T^{\ell\nu\jAB}}}
\newcommand{\lnujjptoversumpt}{\ensuremath{p_T^{\ell\nu\jAB}/\lnujjsumpt}}
\newcommand{\lnujjptasym}{\ensuremath{(\lnupt-p_T^{\jAB})/(\lnupt+p_T^{\jAB})}}
\newcommand{\lnujjptratio}{\ensuremath{\lnupt/p_T^{\jAB}}}
\newcommand{\lnujjsumpt}{\ensuremath{\sum{p_T(\ell,\MET,\jA,\jB)}}}
\newcommand{\lnujjsumptasym}{\ensuremath{(\lnusumpt-\jABsumpt)/\lnujjsumpt}}
\newcommand{\lnujjsumptratio}{\ensuremath{(p_T^{\ell}+\MET)/\jABsumpt}}
\newcommand{\lnujjthetastar}{\ensuremath{\cos(\theta^* )}}
\newcommand{\lnujjzeronupzm}{\ensuremath{m_{\ell\nu\jAB}(p_Z(\nu)=0)}}
\newcommand{\lnujjjjlnuframelepZcosangle}{\ensuremath{\cos[\angle(\ell_{\ell\nu CM;4j},(W\to\ell\nu)_{\ell\nu jj CM;4j})]}}
\newcommand{\lnujjjjlnujjframelnujAangle}{\ensuremath{\cos(\angle(\jA,\ell\nu))_{\text{HCM};4j}}}
\newcommand{\lnujjjjsumpt}{\ensuremath{\sum p_T(\ell,\MET,\jA,\jB,\jC,\jD)}}
\newcommand{\metscaled}{\ensuremath{\MET^{\text{SC}}}}
\newcommand{\metsig}{\ensuremath{\mbox{\ensuremath{\not\!\!E_T^{\text{Sig}}}}}}
\newcommand{\mptsig}{\ensuremath{\MPT^{\text{Sig}}}}
\newcommand{\mvahiggsvstt}{\ensuremath{\mathrm{MVA_{tt}}}}
\newcommand{\mvahiggsvsvj}{\ensuremath{\mathrm{MVA_{VJ}}}}
\newcommand{\mvahiggsvsvv}{\ensuremath{\mathrm{MVA_{VV}}}}
\newcommand{\mvamjvshvv}{\ensuremath{\mathrm{MVA_{MJ}}(H\to VV)}}
\newcommand{\mvamjvsvh}{\ensuremath{\mathrm{MVA_{MJ}}(VH)}}
\newcommand{\nueta}{\ensuremath{\eta^{\nu}}}
\newcommand{\nupt}{\ensuremath{\MET}}
\newcommand{\nupzA}{\ensuremath{p_Z^{\nu_1}}}
\newcommand{\nupzB}{\ensuremath{p_Z^{\nu_2}}}
\newcommand{\topoaplanarity}{\ensuremath{\Apla}}
\newcommand{\topoaplanarityljAB}{\ensuremath{{\Apla}(\ell\jA\jB)}}
\newcommand{\topoaplanaritynuA}{\ensuremath{{\Apla}(\nu_1)}}
\newcommand{\topocentrality}{\ensuremath{\Centrality}}
\newcommand{\toposphericity}{\ensuremath{\Sphe}}
\newcommand{\toposphericityljAB}{\ensuremath{{\Sphe}(\ell\jA\jB)}}
\newcommand{\toposphericitylnuBjAB}{\ensuremath{{\Sphe}(\ell\nu_{2}\jA\jB)}}
\newcommand{\toposphericitynuA}{\ensuremath{{\Sphe}(\nu_1)}}
\newcommand{\topovispt}{\ensuremath{p_T^{\text{VIS}}}}
\newcommand{\topovissumpt}{\ensuremath{\sum(p_T)^{\text{VIS}}}}

%% file: author_list.tex
\affiliation{LAFEX, Centro Brasileiro de Pesquisas F\'{i}sicas, Rio de Janeiro, Brazil}
\affiliation{Universidade do Estado do Rio de Janeiro, Rio de Janeiro, Brazil}
\affiliation{Universidade Federal do ABC, Santo Andr\'e, Brazil}
\affiliation{University of Science and Technology of China, Hefei, People's Republic of China}
\affiliation{Universidad de los Andes, Bogot\'a, Colombia}
\affiliation{Charles University, Faculty of Mathematics and Physics, Center for Particle Physics, Prague, Czech Republic}
\affiliation{Czech Technical University in Prague, Prague, Czech Republic}
\affiliation{Center for Particle Physics, Institute of Physics, Academy of Sciences of the Czech Republic, Prague, Czech Republic}
\affiliation{Universidad San Francisco de Quito, Quito, Ecuador}
\affiliation{LPC, Universit\'e Blaise Pascal, CNRS/IN2P3, Clermont, France}
\affiliation{LPSC, Universit\'e Joseph Fourier Grenoble 1, CNRS/IN2P3, Institut National Polytechnique de Grenoble, Grenoble, France}
\affiliation{CPPM, Aix-Marseille Universit\'e, CNRS/IN2P3, Marseille, France}
\affiliation{LAL, Universit\'e Paris-Sud, CNRS/IN2P3, Orsay, France}
\affiliation{LPNHE, Universit\'es Paris VI and VII, CNRS/IN2P3, Paris, France}
\affiliation{CEA, Irfu, SPP, Saclay, France}
\affiliation{IPHC, Universit\'e de Strasbourg, CNRS/IN2P3, Strasbourg, France}
\affiliation{IPNL, Universit\'e Lyon 1, CNRS/IN2P3, Villeurbanne, France and Universit\'e de Lyon, Lyon, France}
\affiliation{III. Physikalisches Institut A, RWTH Aachen University, Aachen, Germany}
\affiliation{Physikalisches Institut, Universit\"at Freiburg, Freiburg, Germany}
\affiliation{II. Physikalisches Institut, Georg-August-Universit\"at G\"ottingen, G\"ottingen, Germany}
\affiliation{Institut f\"ur Physik, Universit\"at Mainz, Mainz, Germany}
\affiliation{Ludwig-Maximilians-Universit\"at M\"unchen, M\"unchen, Germany}
\affiliation{Fachbereich Physik, Bergische Universit\"at Wuppertal, Wuppertal, Germany}
\affiliation{Panjab University, Chandigarh, India}
\affiliation{Delhi University, Delhi, India}
\affiliation{Tata Institute of Fundamental Research, Mumbai, India}
\affiliation{University College Dublin, Dublin, Ireland}
\affiliation{Korea Detector Laboratory, Korea University, Seoul, Korea}
\affiliation{CINVESTAV, Mexico City, Mexico}
\affiliation{Nikhef, Science Park, Amsterdam, the Netherlands}
\affiliation{Radboud University Nijmegen, Nijmegen, the Netherlands}
\affiliation{Joint Institute for Nuclear Research, Dubna, Russia}
\affiliation{Institute for Theoretical and Experimental Physics, Moscow, Russia}
\affiliation{Moscow State University, Moscow, Russia}
\affiliation{Institute for High Energy Physics, Protvino, Russia}
\affiliation{Petersburg Nuclear Physics Institute, St. Petersburg, Russia}
\affiliation{Instituci\'{o} Catalana de Recerca i Estudis Avan\c{c}ats (ICREA) and Institut de F\'{i}sica d'Altes Energies (IFAE), Barcelona, Spain}
\affiliation{Uppsala University, Uppsala, Sweden}
\affiliation{Lancaster University, Lancaster LA1 4YB, United Kingdom}
\affiliation{Imperial College London, London SW7 2AZ, United Kingdom}
\affiliation{The University of Manchester, Manchester M13 9PL, United Kingdom}
\affiliation{University of Arizona, Tucson, Arizona 85721, USA}
\affiliation{University of California Riverside, Riverside, California 92521, USA}
\affiliation{Florida State University, Tallahassee, Florida 32306, USA}
\affiliation{Fermi National Accelerator Laboratory, Batavia, Illinois 60510, USA}
\affiliation{University of Illinois at Chicago, Chicago, Illinois 60607, USA}
\affiliation{Northern Illinois University, DeKalb, Illinois 60115, USA}
\affiliation{Northwestern University, Evanston, Illinois 60208, USA}
\affiliation{Indiana University, Bloomington, Indiana 47405, USA}
\affiliation{Purdue University Calumet, Hammond, Indiana 46323, USA}
\affiliation{University of Notre Dame, Notre Dame, Indiana 46556, USA}
\affiliation{Iowa State University, Ames, Iowa 50011, USA}
\affiliation{University of Kansas, Lawrence, Kansas 66045, USA}
\affiliation{Louisiana Tech University, Ruston, Louisiana 71272, USA}
\affiliation{Northeastern University, Boston, Massachusetts 02115, USA}
\affiliation{University of Michigan, Ann Arbor, Michigan 48109, USA}
\affiliation{Michigan State University, East Lansing, Michigan 48824, USA}
\affiliation{University of Mississippi, University, Mississippi 38677, USA}
\affiliation{University of Nebraska, Lincoln, Nebraska 68588, USA}
\affiliation{Rutgers University, Piscataway, New Jersey 08855, USA}
\affiliation{Princeton University, Princeton, New Jersey 08544, USA}
\affiliation{State University of New York, Buffalo, New York 14260, USA}
\affiliation{University of Rochester, Rochester, New York 14627, USA}
\affiliation{State University of New York, Stony Brook, New York 11794, USA}
\affiliation{Brookhaven National Laboratory, Upton, New York 11973, USA}
\affiliation{Langston University, Langston, Oklahoma 73050, USA}
\affiliation{University of Oklahoma, Norman, Oklahoma 73019, USA}
\affiliation{Oklahoma State University, Stillwater, Oklahoma 74078, USA}
\affiliation{Brown University, Providence, Rhode Island 02912, USA}
\affiliation{University of Texas, Arlington, Texas 76019, USA}
\affiliation{Southern Methodist University, Dallas, Texas 75275, USA}
\affiliation{Rice University, Houston, Texas 77005, USA}
\affiliation{University of Virginia, Charlottesville, Virginia 22904, USA}
\affiliation{University of Washington, Seattle, Washington 98195, USA}
\author{V.M.~Abazov} \affiliation{Joint Institute for Nuclear Research, Dubna, Russia}
\author{A.~Abbinante} \affiliation{Fermi National Accelerator Laboratory, Batavia, Illinois 60510, USA}
\author{B.~Abbott} \affiliation{University of Oklahoma, Norman, Oklahoma 73019, USA}
\author{B.S.~Acharya} \affiliation{Tata Institute of Fundamental Research, Mumbai, India}
\author{M.~Adams} \affiliation{University of Illinois at Chicago, Chicago, Illinois 60607, USA}
\author{T.~Adams} \affiliation{Florida State University, Tallahassee, Florida 32306, USA}
\author{G.D.~Alexeev} \affiliation{Joint Institute for Nuclear Research, Dubna, Russia}
\author{G.~Alkhazov} \affiliation{Petersburg Nuclear Physics Institute, St. Petersburg, Russia}
\author{A.~Alton$^{a}$} \affiliation{University of Michigan, Ann Arbor, Michigan 48109, USA}
\author{A.~Askew} \affiliation{Florida State University, Tallahassee, Florida 32306, USA}
% ...
%
%
\collaboration{The D0 Collaboration\footnote{with visitors from
$^{a}$Augustana College, Sioux Falls, SD, USA,
$^{b}$The University of Liverpool, Liverpool, UK,
$^{c}$UPIITA-IPN, Mexico City, Mexico,
$^{d}$DESY, Hamburg, Germany,
$^{e}$SLAC, Menlo Park, CA, USA,
$^{f}$University College London, London, UK,
$^{g}$Centro de Investigacion en Computacion - IPN, Mexico City, Mexico,
$^{h}$ECFM, Universidad Autonoma de Sinaloa, Culiac\'an, Mexico,
$^{i}$Universidade Estadual Paulista, S\~ao Paulo, Brazil,
$^{j}$Karlsruher Institut f\"ur Technologie (KIT) - Steinbuch Centre for Computing (SCC)
and
$^{k}$Office of Science, U.S. Department of Energy, Washington, D.C. 20585, USA.
$^{l}$Visitor from Bradley University, Peoria, IL, USA.
}} \noaffiliation
\vskip 0.25cm

%% file: acknowledgement.tex
%
We thank the staffs at Fermilab and collaborating institutions,
and acknowledge support from the
DOE and NSF (USA);
CEA and CNRS/IN2P3 (France);
MON, NRC KI and RFBR (Russia);
CNPq, FAPERJ, FAPESP and FUNDUNESP (Brazil);
DAE and DST (India);
Colciencias (Colombia);
CONACyT (Mexico);
NRF (Korea);
FOM (The Netherlands);
STFC and the Royal Society (United Kingdom);
MSMT and GACR (Czech Republic);
BMBF and DFG (Germany);
SFI (Ireland);
The Swedish Research Council (Sweden);
and
CAS and CNSF (China).

%% file: appendixA.tex
\section{Multivariate Discriminator Input Variables \label{app:input_var_def}}

The multivariate discriminators used in this search use input variables from five general categories: final state particle information, 
as measured in the D0 detector; kinematics of reconstructed objects, such as $W$ boson candidates reconstructed from the leptonic 
or hadronic decay products; angular distributions between final state particles and reconstructed objects; topological variables that 
examine the net properties of all final state particles in an event; and special variables focused on discriminating Higgs boson 
candidate events from specific backgrounds.  Certain multivariate discriminants trained to separate a Higgs boson signal from a 
specific background are also used as inputs for a final discriminant that is trained to separate the Higgs boson signal from all backgrounds.  

Individual input variables are described in detail below.  In the descriptions, $\ell$ refers to the electron or muon in a selected event, $\nu$ refers to the neutrino candidate,
and $j_n$ refers to jets as ordered by $p_T$ where $j_1$ is the jet with highest $p_T$. The $p_Z$ of the neutrino candidate is estimated by constraining the charged lepton
and the neutrino system to the mass of the $W$ boson and choosing the lowest magnitude solution.

Input variable lists for each multivariate discriminant appear in Tables~\ref{tab:mva-mj}--\ref{tab:mva-4j}. The ranking by importance of the variables is determined 
in the BDT by counting how often the variables are used to
split decision tree nodes, and by weighting each split occurrence by the separation gain-squared it
has achieved and by the number of events in the node~\cite{Hocker:2007ht}.
In the RF, the importance of variables are estimated after training in an independent sample of validation events.
These events are run through the RF, once for each variable used.  On each pass the 
class of each event is randomized whenever the variable under test
is encountered and the change in the quadratic loss figure of merit
is estimated:
\begin{equation}
FOM = \frac{
\displaystyle \sum_{i=1}^{\#events }{wgt_i (event^{class}_i-RF(event_i))^2} }
{\displaystyle \sum_{i=1}^{\#events } wgt_i}
\end{equation}
where $wgt_i$ is an event weight, $event_{i}^{class}$=1 for signal, 0 for 
background, and $RF(event_i)$ is the output of the RF classifier for a 
given event.  Whenever the RF makes an incorrect assignment for an event
the FOM increases in value.  In this test the assignments are randomized for one variable 
at a time, effectively removing the predictive power of that variable, and the FOM
will increase more when more powerful variables are removed in this manner.

\input{var_def}

\input{MVA_tables_v3.tex}

\FloatBarrier

%% file: var_def.tex
The input variable distributions are defined as follows:

\subsection{Final State Particle Information}

\begin{itemize}
\item \jeteA : Energy of the leading jet
\item \jetptA : $p_T$ of the leading jet
\item \jetptB : $p_T$ of the second leading jet
\item \jetptC : $p_T$ of the third leading jet
\item \lepqeta : Product of the lepton charge and its pseudorapidity
\item \nupzA : Smaller absolute value solution for $p_Z$ of the reconstructed neutrino, reconstructed with the assumption all $\MET$ is originating from $W$ boson
\item \nupzB : Larger absolute value solution for $p_Z$ of the reconstructed neutrino
\item \MPT : Missing $p_T$ as determined from charged particle tracks in central tracking detector
\item \metscaled : Scaled $\MET$ is defined as \\
$\sum^{\text{jets}}\{E(j_i)\times\{\overrightarrow{\MET}\cdot\overrightarrow{p(j_i)}/[\MET\times|\overrightarrow{p(j_i)}|]\}^2 \}$
\item \metsig : $\MET$\ significance, a measure of the consistency of the observed \MET\ with respect to zero \MET, accounting for the uncertainty in the 
calorimeter objects that contribute to \MET
\item \mptsig : $\MPT$\ significance, a measure of the consistency of the observed \MPT\ with respect to zero \MPT, accounting for the uncertainty in the 
charged particle tracks that contribute to \MPT
\item \jetlepqetaB : Product of the the lepton charge and pseudorapidity of the second leading jet
\item \jetlepqetaC :  Product of the the lepton charge and pseudorapidity of the third leading jet
\item \bidbl : Averaged $b$-jet identification output for the highest energy $b$-tagged jets
\end{itemize}

\subsection{Kinematics of Reconstructed Objects}

\begin{itemize}
\item \jABm : Invariant mass of the leading and second leading jets
\item \jABmt : Transverse mass of the leading and second leading jets
\item \jABCm : Invariant mass of the leading, second leading, and third leading jets
\item \jABCDm :  Invariant mass of the leading, second leading, third leading, and fourth leading jets
\item \lnudpt : Scalar difference: $|p_T^{\ell}-\MET|$
\item \jABdpt : scalar difference, $p_T^{\jA}-p_T^{\jB}$
\item \jABlepsumpt : Scalar sum of the $p_{T}$ of the two leading jets and the lepton
\item \lnudptoverpt : Ratio of the scalar difference between $p_T^{\ell}$ and the $\MET$, to $\lnupt$
\item \lnuptfracmax : Ratio of the $\max(p_T^{\ell},\MET)$ to $\lnupt$
\item \lnuptfracmin : Ratio of the $\min(p_T^{\ell},\MET)$ to $\lnupt$
\item \lnuptoversumpt : Ratio of the $\lnupt$ to $\lnusumpt$
\item \lnulepdpt : $|\lnupt-p_T^{\ell}|$
\item \lnunudpt : $|\lnupt-\MET|$
\item \jABCpt : $p_T$ of the system consisting of the leading, second leading, and third leading jets
\item \jBCdptA : scalar $\Delta p_T$ between the second leading jet and the system consisting of the second leading and third leading jets
\item \jABsumpt : Scalar sum of the $p_T$ of the two leading jets, $p_T^{\jA}+p_T^{\jB}$
\item \jABCDsumpt : Scalar sum of the $p_T$ of the leading, second leading, third leading and fourth leading jets
\item \jABptoversumpt : Ratio of the $p_T$ of the leading and second leading jet system to the scalar sum of the $p_T$ of the two leading jets,
\item \jBCptoversumpt : Ratio of the $p_T$ of the system consisting of the second leading and third leading jets to the scalar sum 
of the $p_T$ of the second leading and third leading jets
\item \jABrecoilpt : Recoil $p_T$ of the first and second leading jet system
\item \jABlepm : Invariant mass of the dijet system and the lepton
\item \lnumt : Transverse mass of the $\ell\nu$ system
\item \lnupt : $p_T$ of the $\ell\nu$ system
\item \lnusumpt : Scalar sum of $p_T^{\ell}$ and $\MET$
\item \lnurecoilpt : $p_T$ of the $\ell\nu$ system with respect to the thrust vector, $\vec{\ell} - \vec{\nu}$
\item \lnujmA : Invariant mass of the system consisting of the charged lepton, reconstructed neutrino (assuming \nupzA), and leading jet
\item \lnujmB : Invariant mass of the system consisting of the charged lepton, reconstructed neutrino (assuming \nupzA), and second leading jet
\item \lnujptB : $p_T$ of the system consisting of the charged lepton, reconstructed neutrino (assuming \nupzA), and second leading jet
\item \lnujsumptB : Scalar sum of the $p_T$ of the charged lepton, $\MET$, and second leading jet
\item \lnujjm : Invariant mass of the charged lepton, reconstructed neutrino (assuming \nupzA), and two leading jets
\item \lnujjmt : Transverse mass of the charged lepton, $\MET$, and two leading jets
\item \lnujjzeronupzm : Invariant mass of the charged lepton, reconstructed neutrino (assuming \nupzA), and two leading jets, with the assumption that $p_Z^{\nu}=0$
\item \lnujjsumpt : Scalar sum, $p_T^{\ell}+\MET+\jABsumpt$
\item \lnujjjjsumpt : Scalar sum, $p_T^{\ell}+\MET+\jABCDsumpt$
\item \Hel : Helicity is defined for an object $A$, coming from the decay of object $C$ via $C\to AB$, as $\arccos\theta_{AC}[(\overrightarrow{C} \cdot \overrightarrow{A})/(|C|\times |A|)]$
\item \jABhelicityA : Helicity of the leading jet in the dijet system, calculated in the laboratory frame
\item \Vel : Velocity is defined for an object $C\to AB$ as $-\ln\lbrace 1-\lbrace 1-4\times [(m_A^{2}+m_B^{2})/m_{C}^{2}]^{1/2}\rbrace ^{1/2}\rbrace$
\item \jABvelj : Velocity of the dijet system
\item \jACvelj : Velocity of the system consisting of the leading and third leading jets
\item \Twist : Twist is $\arctan(\Delta\phi/\Delta\eta)$
\item \jABtwist : Twist of the dijet system
\item \jBCtwist : Twist of the system consisting of the second leading and third leading jets
\item \lnutwist : Twist of the $\ell\nu$ system
\item \Width : Width of a jet in $(\eta,\phi)$ space defined as $\sqrt{\eta _w^{2} + \phi _w^{2}} $, where $\eta _w$ and $\phi _w$ are the $p_{T}$ weighted RMS $\eta$ and $\phi$ of energy deposits around the jet centroid.
\item \jetwidthC : Width of the third leading jet
\end{itemize}

\subsection{Angular Distributions}

\begin{itemize}
\item \jABdeta : Separation in $\eta$ between the two leading jets, $|\eta_{\jA}-\eta_{\jB}|$
\item \jABdetamax : Maximum $\Delta\eta$ between the dijet system and the leading or second leading jet
\item \jABdphi : Separation in $\phi$ between the two leading jets, $|\phi_{\jA}-\phi_{\jB}|$
\item \jABdr : Angular separation in $(\eta,\phi)$ space between the two leading jets
\item \jABdrmin : Minimum angular separation in $(\eta,\phi)$ space between the dijet system and the leading or second leading jet
\item \jABdphiA : $|\phi_{\jAB}-\phi_{\jA}|$, where $\phi_{\jAB}$ is the $\phi$ of the dijet system
\item \jACdphi : Separation in $\phi$ between the first and third leading jets, $|\phi_{\jA}-\phi_{\jC}|$
\item \jACdrB : $\deltaR$ between the third leading jet and the system consisting of the leading and third leading jets
\item \jBCdphiB : $\Delta\phi$ between the third leading jet and the system consisting of the second leading and third leading jets
\item \jABangle : 3D angle between the two leading jets
\item \jABcmcostheta : Cosine of the angle between the two leading jets in the center of mass (CM) of the dijet system
\item \jABbislepangle : 3D angle between the charged lepton and the bisector of the dijet system
\item \jABbislnudphi : Signed $\Delta\phi$ between the $\ell\nu$ system and the bisector of the dijet system
\item \jetlepdetaA : Separation in $\eta$ between the lepton and the leading jet, $|\eta^{\ell}-\eta^{\jA}|$ 
\item \jetlepdetaB : Separation in $\eta$ between the lepton and the second leading jet, $|\eta^{\ell}-\eta^{\jB}|$ 
\item \jetlepdetaC : Separation in $\eta$ between the lepton and the third leading jet, $|\eta^{\ell}-\eta^{\jC}|$ 
\item \jetlepdetamax : Maximum $\Delta\eta$ between the charged lepton and the leading or second leading jet
\item \jetlepdrA : $\deltaR$ between the charged lepton and the leading jet
\item \jetlepdrB : $\deltaR$ between the charged lepton and the second leading jet
\item \jetlepdrC : $\deltaR$ between the charged lepton and the third leading jet
\item \jetnudphiA : $\Delta\phi$ between the $\MET$ and the leading jet
\item \jetnudrA : $\deltaR$ between the reconstructed neutrino (assuming \nupzA), and the leading jet
\item \jetnudrminjAB :  Minimum $\deltaR$ between the reconstructed neutrino (assuming \nupzA), and the leading or second leading jet
\item \lepljetdr : $\deltaR$ between the charged lepton and the leading non-$b$-tagged jet
\item \jABlepangle : 3D angle between the charged lepton and the dijet system
\item \lnudeta : Separation in $\eta$ between the lepton and the reconstructed neutrino (assuming \nupzA), $|\eta^{\ell}-\nueta|$
\item \lnudetamax : Maximum $\Delta\eta$ between the $\ell\nu$ system and charged lepton or reconstructed neutrino (assuming \nupzA)
\item \lnudphi : $\phi$ angle between the lepton and $\MET$.
\item \lnudphimax : Maximum $\Delta\phi$ between the $\ell\nu$ system and the charged lepton or $\MET$
\item \lnudphimin : Minimum $\Delta\phi$ between the $\ell\nu$ system and the charged lepton or $\MET$
\item \lnudr : $\deltaR$ between the charged lepton and the reconstructed neutrino (assuming \nupzA)
\item \lnudrmax : Maximum $\deltaR$ between the $\ell\nu$ system and the charged lepton or reconstructed neutrino (assuming \nupzA)
\item \lnudrmin : Minimum $\deltaR$ between the $\ell\nu$ system and the charged lepton or reconstructed neutrino (assuming \nupzA)
\item \lnulepdeta : $\Delta\eta$ between the $\ell\nu$ system and the charged lepton 
\item \lnulepdr : $\deltaR$ between the $\ell\nu$ system and the charged lepton 
\item \lnunudr : $\deltaR$ between the $\ell\nu$ system and the reconstructed neutrino (assuming \nupzA)
\item \lnuangle : 3D angle between the charged lepton and the reconstructed neutrino (assuming \nupzA)
\item \lnucmcostheta : Cosine of the angle between the charged lepton and the proton beam axis in the CM of $\ell\nu$ system
\item \lnucosthetaA : Cosine of the angle between the charged lepton and the proton beam axis in the detector
\item \lnujdetaB : $\Delta\eta$ between $\ell\nu$ system and the second leading jet
\item \lnujdphiB : $\Delta\phi$ between $\ell\nu$ system and the second leading jet
\item \lnujdrB : $\deltaR$ between $\ell\nu$ system and the second leading jet
\item \lnujjdeta : $\Delta\eta$ between $\ell\nu$ system and the dijet system
\item \lnujjcmjABangle :  3D angle between the two leading jets in the $H\to WW\to\ell\nu jj$ CM frame (HCM)
\item \lnujjcmjAlepcosangle : Cosine of the 3D angle between the leading jet and $\ell\nu$ system in the HCM
\item \lnujjcmlnujAangle : Cosine of the 3D angle between the charged lepton and the leading jet in the HCM
\item \lnujjjABframejZcosangle : Cosine of the 3D angle between the leading jet in energy in the CM of the dijet system and $\ell\nu$ system in the $H\to WW\to\ell\nu jj$ CM frame; jet energy is calculated in the $H\to WW\to\ell\nu jj$ CM frame
\item \lnujjlnuframelepZcosangle : Cosine of the 3D angle between the charged lepton in the $\ell\nu$ system CM and $\ell\nu$ system in the $H\to WW\to\ell\nu jj$ CM frame
\item \lnujjjjlnuframelepZcosangle : Cosine of the 3D angle between the charged lepton in the $\ell\nu$ system CM and $\ell\nu$ system in the $H\to WW\to\ell\nu jj$ CM frame for $V(\to jj)H(\to WW \to \ell\nu jj)$ candidate events; jet energy is calculated in the $H\to WW\to\ell\nu jj$ CM frame
\item \lnujjjjlnujjframelnujAangle : Cosine of the 3D angle between the leading jet and $\ell\nu$ system in the $H\to WW\to\ell\nu jj$ CM frame frame for $V(\to jj)H(\to WW \to \ell\nu jj)$ candidate events
\item \lnujjthetastar: $\theta^* = \angle(W,\text{incoming}~u\text{-type quark})$ in HCM frame~\cite{Parke:1999gx} 
\item \lnujjchistar : $\chi^* = \angle(\ell, \text{spin}_W)$ in $\ell\nu$ system CM frame~\cite{Parke:1999gx} 
\end{itemize}

\subsection{Topological Variables}

\begin{itemize}
\item \topoaplanarity : Aplanarity is $3\lambda_{3}/2$ where $\lambda_{3}$ is the smallest eigenvalue of the normalized momentum tensor 
$S^{\alpha \beta}=(\sum_{i}p_{i}^{\alpha}p_{i}^{\beta})/(\sum_{i} |\vec{p_{i}}|^{2})$ , where $\alpha,\beta=1,2,3$ correspond to the 
$x,y,z$ momentum components, and $i$ runs over selected objects. Without arguments, it is calculated for all visible objects
\item \topoaplanarityljAB : \Apla\ calculated for the charged lepton,  and leading and second leading jets
\item \topoaplanaritynuA : \Apla\ calculated for the charged lepton, reconstructed neutrino (assuming \nupzA), and all selected jets
\item \topocentrality : Centrality is $(\sum_{i} p_{T}^{i})/(\sum_{i} |\vec{p_{i}}|)$, where $i$ runs over $\ell$ and all jets
\item \toposphericity : Sphericity is $3(\lambda_{2}+\lambda_{3})/2$ where $\lambda_{3}$ ($\lambda_{2}$) is the smallest (second-smallest) eigenvalue of the normalized momentum tensor described under \Apla. Without arguments, it is calculated for all visible objects
\item \toposphericityljAB : \Sphe\ calculated for the charged lepton, and leading and second leading jets
\item \toposphericitylnuBjAB : \Sphe\ calculated for the charged lepton, reconstructed neutrino (assuming \nupzB), and leading and second leading jets
\item \toposphericitynuA : \Sphe\ calculated for the charged lepton, reconstructed neutrino (assuming \nupzA), and all selected jets
\item \topovispt : Magnitude of the vector sum  of the $\vec{p}_T$ of the visible particles
\item \topovissumpt : Scalar sum of the $p_T$ of the visible particles
\item \jABktmin : $\jABdr\times E_T^{\jB}/(\MET+E_T^{\ell})$, where $E_T$ is the transverse energy
\item \jABbisnu : Scalar product of the bisector of the dijet system and the $\MET$ vector, 
i.e.\ $\overrightarrow{\text{bis}( \jA,\jB ) } \cdot \overrightarrow{ \MET } $
\item \lnujjmasym : Mass asymmetry between $\ell\nu$ system and the dijet system: $(m_{\ell\nu}-\jABm)/(m_{\ell\nu}+\jABm)$
\item \lnujjptasym : $p_T$ asymmetry between $\ell\nu$ system and the dijet system
\item \lnujjptratio : Ratio of $\lnupt$ to $p_T^{\jAB}$
\item \lnujjsumptasym : $\sum p_T$ asymmetry between $\ell\nu$ system and the dijet system
\item \lnujjsumptratio : Ratio of $\lnusumpt$ to $\jABsumpt$
\item \lnujjptoversumpt : Ratio of the $p_T^{\ell\nu\jAB}$ to $\lnujjsumpt$
\item \jABsigmaA : Based on the pull variables described in Ref.~\cite{Black:2011}. Sigma, $\Sigm$, of the dijet system with respect to the leading jet defined as\ $p_T^{\jB} \deltaR(\jA,\jB)/\jABsumpt$ 
\item \jABsigmamax : Maximum $\Sigm$ of  the leading or second leading jet defined as\ $p_T^{\max}(\jA,\jB) \deltaR(\jA,\jB)/\jABsumpt$ 
with respect to the dijet system
\item \jABsigmamin : Minimum $\Sigm$ of the leading or second leading jet defined as\ $p_T^{\min}(\jA,\jB) \deltaR(\jA,\jB)/\jABsumpt$
with respect to the dijet system
\item \lnulepsigma: $\Sigm$ of the $\ell\nu$ system with respect to the lepton, defined as $(\deltaR(\ell,\nu)\times \MET)/(\lnusumpt)$ 
\item \lnusigmamax :  Maximum $\Sigm$ of the lepton or $\MET$ defined as $\max(p_T^{\ell},\MET)\times\deltaR(\ell,\nu)/(\lnusumpt)$ 
with respect to the $\ell\nu$ system
\item \lnusigmamin : Minimum $\Sigm$ of the lepton or $\MET$ defined as
$\min(p_T^{\ell},\MET)\times\deltaR(\ell,\nu)/(\lnusumpt)$ with respect to the $\ell\nu$ system
\item ${\SigmJ}(A)$ : $Sigm$ defined for an object $A$, with respect to all jets in an event, as 
\begin{itemize}
\item[] $\sum_{\text{jets}}[p_T^{\text{jet}}\times\deltaR (A,{\text{jet}})]/\sum_{\text{jets}}(p_T^{\text{jet}})$
\end{itemize}
\item \jABCsigma : ${\SigmJ}$ of the system consisting of the leading, second leading and third leading jets
\item \jABCDsigma : ${\SigmJ}$ of the system consisting of the leading, second leading, third leading and fourth leading jets
\item \lepsigma : ${\SigmJ}$ of the charged lepton
\item \Simila : Similarity is defined for two objects, $A$ and $B$, as $\min(p_T)^2\times \deltaR^2/(\sum p_T)^2$, where $\min(p_T)$ is the 
minimum $p_T$ of the objects $A$ and $B$, $\deltaR$ is the angular separation in $(\eta,\phi)$ space between objects $A$ and $B$, and 
$\sum p_T$ is the scalar sum of the $p_T$s of $A$ and $B$
\item \lnusimilarity : Similarity of the charged lepton and the reconstructed neutrino (assuming \nupzA)
\end{itemize}

\subsection{Discriminants to Separate Higgs Boson Events from a Specific Background}

\begin{itemize}
\item \mvahiggsvstt : Output of the multivariate discriminant trained against $t\bar{t}$ and single top quark backgrounds
\item \mvahiggsvsvj : Output of the multivariate discriminant trained against $V+$jets backgrounds
\item \mvahiggsvsvv : Output of the multivariate discriminant trained against diboson backgrounds
\item \mvamjvshvv : Output of the multivariate discriminant trained to distinguish $H\to WW\to\ell\nu jj$ from the MJ background
\item \mvamjvsvh : Output of the multivariate discriminant trained to distinguish $WH\to\ell\nu b\bar{b}$ from the MJ background

\end{itemize}

%% file: MVA_tables_v3.tex
\FloatBarrier
\newpage

\begin{table}[htbp]
\caption{Input variables for the  $\mvamjvsvh$ and $\mvamjvshvv$ discriminants.
Two discriminants were trained to reject MJ events, one trained using $VH\to\ell\nu b\bar{b}$
events as a signal and the second using $H\to VV\to\ell\nu jj$ events as signal.
Both discriminants use the same list of inputs.
Variables are listed by their importance in the MVA.
\label{tab:mva-mj}}
{ \renewcommand{\arraystretch}{1.0}
\begin{tabular}{l}
\hline
\hline
$\mathrm{MVA_{MJ}}$ Input Variables \\
\hline                  
\nueta                  \\
\metsig                 \\
\lnudeta                \\
\lnutwist               \\
\lnucmcostheta          \\
\jABvelj                \\
\lnujjmasym             \\
\topocentrality         \\
\nupt                   \\
\topovispt              \\
\jetlepdetamax          \\  
\hline
\hline
\end{tabular}
}
\end{table}

\FloatBarrier
\newpage

\begin{table}[htbp]
\caption{Table of input variables for the final signal discriminant for the $WH\to\ell\nu b\bar{b}$ channel.
Variables are listed by their rank of importance when used in the two tight $b$-tagged (2T),
two medium $b$-tagged (2M), two loose $b$-tagged (2L), and one tight $b$-tagged (1T) categories.
\label{tab:mva-lvbb}}
{ \renewcommand{\arraystretch}{1.0}
\begin{tabular}{p{4.5cm}cccc}
\hline
\hline
Variable & 2T & 2M & 2L & 1T \\
\hline
\mvamjvsvh			& 1	& 1	& 	&    \\
\jABmbb				& 2	& 4	& 3	& 1  \\
\lnuptoversumpt			& 3	& 6	& 4	& 2  \\
\bidbl				& 4	& 13	& 1	& 4  \\
\lnujjchistar			& 5	& 3	& 	&    \\
\jetlepdetamax			& 6	& 11	& 2	& 3  \\
\lepqeta			& 7	& 2	& 6	& 6  \\
\jetlepdrA			& 8	& 5	& 	&    \\
\jABsigmamin			& 9	& 15	& 9	& 5  \\
\jetlepqetaA			& 10	& 7	& 11	& 9  \\
\jABvelj			& 11	& 12	& 7	& 11 \\
\lnujjthetastar			& 12	& 10	& 	&    \\
\lnujmB				& 13	& 16	& 12	& 13 \\
\jABmt				& 14	& 14	& 	&    \\
\topocentrality      		& 15	& 8	& 8	& 10 \\
\topovissumpt			& 16	& 9	& 	&    \\
\lnujjmasym			& 	& 	& 5	& 8  \\
\topoaplanarity			& 	& 	& 10	& 12 \\
\jetptB				& 	& 	& 13	& 7  \\
\hline
\hline
\end{tabular}}
\end{table}

\FloatBarrier
\newpage

\begin{table}[htbp]
\caption{Table of input variables for the final signal discriminant for the $H\to WW\to e\nu jj$ channel for two jets events.
Variables are by their rank of importance when used in the zero $b$-tags (0T) and one loose $b$-tag (1T) categories for $M_H \leq 150$~GeV.
\label{tab:mva-ev2jlm}}
{ \renewcommand{\arraystretch}{1.0}
\begin{tabular}{lcc}
\hline
\hline
Variable & 0T & 1T \\
\hline
\lnujjsumptratio     &   1  &     1   \\
\lnulepdeta	     &   2  &         \\
\jABm	             &   3  &     4   \\
\mvamjvsvh	     &   4  &     6   \\
\lnujdetaB	     &   5  &         \\
\jABdetamax	     &   6  &     10  \\
\topoaplanarityljAB  &   7  &         \\
\jetlepdrA	     &   8  &         \\
\topovissumpt	     &   9  &     15  \\
\jABhelicityA	     &   10 &         \\
\lnudphimax	     &   11 &         \\
\lnudphi	     &   12 &         \\
\jABdpt	             &   13 &         \\
\jetptA	             &   14 &     7   \\
\jetlepdetaA	     &      &     2   \\
\mvamjvshvv	     &      &     3   \\
\jetlepdetaB	     &      &     5   \\
\jABvelj	     &      &     8   \\
\lnumt		     &      &     9   \\
\lnucmcostheta	     &      &     11  \\   
\lnujptB	     &      &     12  \\
\lnunudpt	     &      &     13  \\
\lnusumpt	     &      &     14  \\
\jABlepsumpt	     &      &     16  \\
\jABsumpt	     &      &     17  \\
\hline
\hline
\end{tabular}}
\end{table}

\begin{table}[htbp]
\caption{Table of input variables for the final signal discriminant for the $H\to WW\to e\nu jj$ channel for three jets events.
Variables are listed by their rank of importance when used in the zero $b$-tags (0T) and one loose $b$-tag (1T) categories for $M_H \leq 150$~GeV.
\label{tab:mva-ev3jlm}}
{ \renewcommand{\arraystretch}{1.0}
\begin{tabular}{lcc}
\hline
\hline
Variable & 0T & 1T \\
\hline
\mvamjvsvh              & 1   &  6    \\
\jABm	                & 2   &   15  \\
\lnujjsumptratio        & 3   &   2   \\  
\jetlepdrA              & 4   &       \\
\lnujdetaB              & 5   &       \\
\lnulepdeta             & 6   &       \\
\jABdetamax             & 7   &   8   \\
\topoaplanarityljAB     & 8   &       \\
\topovissumpt           & 9   &   16  \\
\lnudphimax             & 10  &       \\
\jABdpt	                & 11  &       \\
\jABhelicityA           & 12  &       \\
\lnudphi	        & 13  &       \\
\jetptA	                & 14  &       \\
\mvamjvshvv             &     &   1   \\
\jetlepdetaB            &     &   3   \\
\jetlepdetaA            &     &   4   \\
\lnucmcostheta          &     &   5   \\
\jetptB	                &     &   7   \\
\jABvelj	        &     &   9   \\
\lnumt	                &     &   10  \\
\lnujptB	        &     &   11  \\
\lnunudpt               &     &   12  \\
\lnusumpt               &     &   13  \\
\jABlepsumpt            &     &   14  \\
\jABsumpt               &     &   17  \\
\hline
\hline
\end{tabular}}
\end{table}

\FloatBarrier
\newpage

\begin{table}[htbp]
\caption{Table of input variables for the final signal discriminant for the $H\to WW\to \mu\nu jj$ channel for two jets events.
Variables are listed by their rank of importance when used in the the zero $b$-tags (0T) and one loose $b$-tag (1T) categories for $M_H \leq 150$~GeV.
\label{tab:mva-mv2jlm}}
{ \renewcommand{\arraystretch}{1.0}
\begin{tabular}{lcc}
\hline
\hline
Variable & 0T & 1T \\
\hline
\jABbislnudphi       &   1  &        \\  
\lnulepdeta          &   2  &        \\
\jABm	             &   3  &     7  \\
\lnujsumptB          &   4  &     16 \\
\lnudrmin            &   5  &        \\
\jABdr	             &   6  &     14 \\
\lnusimilarity       &   7  &        \\
\jABsumpt            &   8  &     4  \\
\lnuptfracmin        &   9  &        \\
\lnudpt	             &   10 &        \\
\jetnudrA            &   11 &        \\
\jetnudphiA          &   12 &        \\
\jABsigmaA           &   13 &        \\
\topovissumpt        &   14 &        \\
\jABdpt	             &   15 &        \\
\jetptA	             &   16 &     6  \\
\jetlepdetaA         &      &     1  \\
\jABhelicityA        &      &     2  \\
\jetlepdetaB         &      &     3  \\
\lnucmcostheta       &      &     5  \\
\lnujdrB	     &      &     6  \\
\lnujjdeta           &      &     8  \\
\lnujjmasym          &      &     9  \\
\jABbisnu            &      &     10 \\
\lnulepdpt           &      &     11 \\
\lnujmB	             &      &     12 \\
\lnutwist            &      &     13 \\
\jABdphi	     &      &     15 \\
\lnujjmt	     &      &     17 \\
\jABlepsumpt         &      &     18 \\
\lnudr	             &      &     19 \\
\lnudphi	     &      &     20 \\
\lnupt	             &      &     21 \\
\hline
\hline
\end{tabular}}
\end{table}

\begin{table}[htbp]
\caption{Table of input variables for the final signal discriminant for the $H\to WW\to \mu\nu jj$ channel for three jets events.
Variables are listed by their rank of importance when used in the zero $b$-tags (0T) and one loose $b$-tag (1T) categories for $M_H \leq 150$~GeV.
\label{tab:mva-mv3jlm}}
{ \renewcommand{\arraystretch}{1.0}
\begin{tabular}{lcc}
\hline
\hline
Variable & 0T & 1T \\
\hline
\jetptC	              &   1 &        \\   		
\jetlepdrB	      &   2 &        \\
\lnulepdeta	      &   3 &        \\
\jetlepdetaA	      &   4 &    2   \\
\jABcmcostheta	      &   5 &    22  \\
\jACdphi	      &   6 &        \\
\jABvelj	      &   7 &    21  \\
\lnurecoilpt	      &   8 &    16  \\
\topoaplanarity	      &   9 &        \\
\lnujmA	              &   10&	     \\
\jABlepm	      &   11&	 18  \\
\jABktmin	      &   12&	     \\
\lnujjcmjAlepcosangle &   13&	     \\
\lnujjm	              &   14&	     \\
\jBCdphiB	      &     &    1   \\
\lepqeta	      &     &    3   \\
\lnujdphiB	      &     &    4   \\
\metscaled	      &     &    5   \\
\toposphericityljAB   &     &    6   \\
\jABCm		      &     &    7   \\
\jetnudrA	      &     &    8   \\
\lepljetdr	      &     &    9   \\
\jABdeta	      &     &    10  \\
\lnujjmasym	      &     &    11  \\
\lnuptoversumpt	      &     &    12  \\
\jABCsigma	      &     &    13  \\
\topocentrality	      &     &    14  \\
\jABsigmaA	      &     &    15  \\
\lnucmcostheta	      &     &    17  \\
\lnumt		      &     &    19  \\
\lnujjzeronupzm	      &     &    20  \\
\jABm		      &     &    23  \\
\lnutwist	      &     &    24  \\
\lnudeta	      &     &    25  \\
\hline
\hline
\end{tabular}}
\end{table}

\FloatBarrier
\newpage

\begin{table}[htbp]
\caption{Table of input variables for the final signal discriminant for the $H\to WW\to e\nu jj$ channel for two jets events
in the pretag category for  $M_H \geq 155$~GeV.  
Variables are listed by their importance in the MVA.
\label{tab:mva-ev2jhm}}
{ \renewcommand{\arraystretch}{1.0}
\begin{tabular}{l}
\hline
\hline
MVA Input Variables      \\
\hline  
\topovissumpt            \\  
\lnurecoilpt             \\
\jABm                    \\
\lnujjmasym              \\
\lnulepsigma             \\
\mvamjvsvh               \\
\lnulepdeta              \\
\jeteA                   \\
\jetlepdrA               \\
\lnudphimax              \\
\jABangle                \\
\lnulepdr                \\
\jABrecoilpt             \\
\lnujjcmjAlepcosangle    \\
\jABhelicityA            \\
\lepljetdr               \\
\hline
\hline 
\end{tabular}}
\end{table}

\FloatBarrier
\newpage

\begin{table}[htbp]
\caption{Table of input variables for the final signal discriminant for the $H\to WW\to e\nu jj$ channel for three jets events
in the pretag category for  $M_H \geq 155$~GeV.  
Variables are listed by their importance in the MVA.
\label{tab:mva-ev3jhm}}
{ \renewcommand{\arraystretch}{1.0}
\begin{tabular}{l}
\hline
\hline
MVA Input Variables      \\
\hline
\topovissumpt            \\
\lnurecoilpt             \\
\mvamjvsvh               \\
\lnudphimax              \\
\lnulepsigma             \\
\lnujjmasym              \\
\lnulepdeta              \\
\jABm                    \\
\jetlepdrA               \\
\jeteA                   \\
\jABangle                \\
\lnujjcmjAlepcosangle    \\
\jABhelicityA            \\
\jABrecoilpt             \\
\lepljetdr               \\ 
\hline
\hline
\end{tabular}}
\end{table}

\FloatBarrier
\newpage

\begin{table}[htbp]
\caption{Table of input variables for the final signal discriminant for the $H\to WW\to \mu\nu jj$ channel for two jets events
in the pretag category for  $M_H \geq 155$~GeV.  
Variables are listed by their importance in the MVA.
\label{tab:mva-mv2jhm}}
{ \renewcommand{\arraystretch}{1.0}
\begin{tabular}{l}
\hline
\hline
MVA Input Variables  \\
\hline
\jABvelj              \\
\lnuangle             \\
\toposphericityljAB   \\
\jABbislnudphi        \\
\jetptA               \\
\jetnudrminjAB        \\
\lnujjcmlnujAangle    \\
\nupt                 \\
\jABdr                \\
\jABptoversumpt       \\
\jABdrmin             \\
\lnupt                \\
\lnujjsumpt           \\
\lnudphimin           \\
\lnuptoversumpt       \\
\lnudrmin             \\
\jABdphiA             \\
\lnusigmamax          \\
\jABdpt               \\
\lnuptfracmax         \\
\jABsigmaA            \\
\lnuptfracmin         \\  
\hline
\hline
\end{tabular}}
\end{table}

\FloatBarrier
\newpage

\begin{table}[htbp]
\caption{Table of input variables for the final signal discriminant for the $H\to WW\to \mu\nu jj$ channel for three jets events
in the pretag category for  $M_H \geq 155$~GeV.  
Variables are listed by their importance in the MVA.
\label{tab:mva-mv3jhm}}
{ \renewcommand{\arraystretch}{1.0}
\begin{tabular}{l}
\hline
\hline
MVA Input Variables \\
\hline
\topovissumpt         \\ 
\lepe                 \\
\jABdr                \\
\lnujjptasym          \\
\jABlepm              \\
\lnujjcmjABangle      \\
\jABmt                \\
\jABbislnudphi        \\
\lnujdphiB            \\
\jABktmin             \\
\lnujmB               \\
\jetlepdrA            \\
\toposphericitynuA    \\
\lnujjzeronupzm       \\
\lnucmcostheta        \\
\jABcmcostheta        \\
\lnusigmamin          \\
\lnuptoversumpt       \\
\jABsigmaA            \\
\jABCsigma            \\
\jABptoversumpt       \\
\jABtwist             \\
\lnudr                \\
\jABCpt               \\
\hline
\hline
\end{tabular}}
\end{table}

\FloatBarrier
\newpage

\begin{table}[htbp]
\caption{Table of input variables for the $\mvahiggsvstt$ discriminant for the $H\to WW\to \ell\nu jjjj$ channel.
Variables are listed by their importance in the MVA.
\label{tab:mva-4jBStt}}
{ \renewcommand{\arraystretch}{1.0}
\begin{tabular}{l}
\hline
\hline
MVA Input Variables \\
\hline
\lnujjjjsumpt                  \\
\jABCDm                        \\
\lnusumpt                      \\
\metscaled                     \\
\jetlepdrC                     \\
\lnujjcmjABangle               \\
\lnujjptasym                   \\
\jACvelj                       \\
\lepe                          \\
\lnumt                         \\
\lnujjmasym                    \\
\lnujjptoversumpt              \\
\jABktmin                      \\
\toposphericity                \\
\lnulepdeta                    \\
\lnujjjjlnujjframelnujAangle   \\
\lnujjjABframejZcosangle       \\
\nupzA                         \\
\jABbislnudphi                 \\
\jBCdptA                       \\
\jACdrB                        \\
\lnulepdr                      \\
\jABlepangle                   \\
\lnuangle                      \\
\nupzB                         \\
\lnurecoilpt                   \\
\mptsig                        \\
\nueta                         \\
\lnujjptratio                  \\
\lnudphimin                    \\  
\hline
\hline
\end{tabular}}
\end{table}

\begin{table}[htbp]
\caption{Table of input variables for the $\mvahiggsvsvj$ discriminant for the $H\to WW\to \ell\nu jjjj$ channel.
Variables are listed by their importance in the MVA.
\label{tab:mva-4jBSvj}}
{ \renewcommand{\arraystretch}{1.0}
\begin{tabular}{l}
\hline
\hline
MVA Input Variables \\
\hline
\jetwidthC                   \\   
\lnucmcostheta               \\
\jetlepdrC                   \\
\jBCtwist                    \\
\lepqeta                     \\
\jetlepdetamax               \\
\jABlepm                     \\
\lnujjsumptasym              \\
\lnudetamax                  \\
\lnudr                       \\
\jABdeta                     \\
\lnulepdr                    \\
\lnujjjjlnuframelepZcosangle \\
\lepsigma                    \\
\topocentrality              \\
\jetlepdrB                   \\
\lnujjptasym                 \\
\topoaplanarity              \\
\lnujjlnuframelepZcosangle   \\
\jABCDsumpt                  \\
\lnusimilarity               \\
\lnulepdeta                  \\
\lnupt                       \\
\topoaplanaritynuA           \\
\lnuangle                    \\
\nueta                       \\
\lepe                        \\
\lnurecoilpt                 \\
\lnujjptratio                \\
\hline
\hline
\end{tabular}}
\end{table}

\begin{table}[htbp]
\caption{Table of input variables for the $\mvahiggsvsvv$ discriminant for the $H\to WW\to \ell\nu jjjj$ channel.
Variables are listed by their importance in the MVA.
\label{tab:mva-4jBSvv}}
{ \renewcommand{\arraystretch}{1.0}
\begin{tabular}{l}
\hline
\hline
MVA Input Variables \\
\hline
\lnucmcostheta                \\  
\lepsigma                     \\
\jetlepqetaB                  \\
\lnujjjABframejZcosangle      \\
\lnujjsumptasym               \\
\jetlepdetamax                \\
\toposphericitylnuBjAB        \\
\lepqeta                      \\
\lnuangle                     \\
\jetlepqetaC                  \\
\nueta                        \\
\lnulepdeta                   \\
\lnujjptasym                  \\
\lnujjcmjABangle              \\
\lnulepdr                     \\
\lnujjptoversumpt             \\
\lnusimilarity                \\
\jetlepdrB                    \\
\jABlepm                      \\
\lnujjjjlnuframelepZcosangle  \\
\jABCsigma                    \\
\jABCDsigma                   \\
\jABbislepangle               \\
\jACdrB                       \\
\jABCDsumpt                   \\
\lnudrmax                     \\
\lnujjptratio                 \\
\hline
\hline
\end{tabular}}
\end{table}

\begin{table}[htbp]
\caption{Table of input variables for the final signal discriminant for the $H\to WW\to \ell\nu jjjj$ channel.
Variables are listed by their rank of importance when used in the zero $b$-tags (0T) and one loose $b$-tag (1L) categories for each charged lepton type.
\label{tab:mva-4j}}
{ \renewcommand{\arraystretch}{1.0}
\begin{tabular}{lcccc}
\hline
\hline
Variable & $e$ 0T & $e$ 1L & $\mu$ 0T & $\mu$ 1L \\
\hline
\mvahiggsvstt	                & 1	& 5    &  5	& 3     \\ 
\mvamjvshvv	                & 2	& 7    &  4	& 9     \\
\mvahiggsvsvj	                & 3	& 1    &  1	& 1     \\
\mvahiggsvsvv	                & 4	& 8    &  7	& 2     \\
\lepsigma	                & 5	& 12   &  14	&       \\
\jABlepm	                & 6	&      &  	&       \\
\jBCptoversumpt	                & 7	&      &  9	& 5     \\
\metsig	                        & 8	& 14   &  	&       \\
\mvamjvsvh	                & 9	& 3    &  12	& 14    \\
\lepe	                        & 10	& 6    &  	& 16    \\
\jABptoversumpt	                & 11	&      &  	&       \\
\lepqeta	                & 12	&      &  3	& 8     \\
\lnujjlnuframelepZcosangle	& 13	&      &  	&       \\
\topoaplanaritynuA	        & 14	&      &  	&       \\
\lnujjptasym	                & 15	&      &  	&       \\
\lnulepdr	                & 16	&      &  	& 13    \\
\lnudptoverpt	                & 17	& 11   &  	&       \\
\lnucosthetaA	                & 18	& 9    &  	&       \\
\topovispt		        &       &   2  &  	&       \\
\jetnudrA		        &       &   4  &  	&       \\
\jABCsigma		        &       &   10 &  	&       \\
\jetlepqetaA		        &       &   13 &  	&       \\
\lnulepdeta		        &       &   15 &  13	&       \\
\lnudetamax		        &       &   16 &  	&       \\
\jABcmcostheta		        & 	&      &    2	& 12    \\
\topocentrality		        & 	&      &    6	&       \\
\jetlepdetaC		        & 	&      &    8	&       \\
\lnujjcmjAlepcosangle	        & 	&      &    10	&       \\
\topoaplanarity		        & 	&      &    11	&       \\
\lnusumpt		        & 	&      &    15	&       \\
\jABbislepangle		        & 	&      &    16	&       \\
\lnucmcostheta		        & 	&      &    17	&       \\
\lnutwist		        & 	&      &    18	&       \\
\lnudeta		        & 	&      &    19	&       \\
\lnumt			        &       &      &      20& 	\\
\jetlepdrC		        & 	&      &        &   4   \\
\lnujjcmlnujAangle	        & 	&      &  	&   6   \\
\lnujjzeronupzm		        & 	&      &        &   7   \\
\toposphericityljAB	        & 	&      &  	&   10  \\
\lnudrmax		        & 	&      &        &   11  \\
\jABsigmamax		        & 	&      &        &   15  \\
\lnurecoilpt		        &       &      &        &   17  \\  
\hline
\hline
\end{tabular}}
\end{table}

%% file: PRD_lvbb_9.7fb.bbl
\begin{thebibliography}{10}
\bibitem{defseta}
The pseudorapidity $\eta = - \ln \left[ \tan\frac{\theta}{2}\right]$, where
  $\theta$ is the polar angle as measured from the proton beam axis.
\bibitem{defphi}
The azimuthal angle, $\phi$, is defined as the opening angle with respect to
  the +$x$ direction in a right-handed coordinate system defined by +$y$ as up
  and +$z$ as the proton beam direction.
\bibitem{defsDR}
$\Delta\mathcal{R} = \sqrt{\left(\Delta \eta\right)^2 + \left(\Delta
  \phi\right)^2}$ is the separation between two objects in ($\eta ,\phi$)
  space, where $\phi$ is the azimuthal angle.
\bibitem{defwtm}
Transverse mass of a leptonically decaying $W$ is defined as $M_T^2 \equiv
  2\,p_T^\ell\,$\mbox{$\not\!\!E_T$}$ [1 - \cos
  \Delta\phi(\ell,$\mbox{$\not\!\!E_T$}$) ]$.
\bibitem{defslepqeta}
Product of the lepton charge and its pseudorapidity.
\bibitem{defstopoaplanarity}
Aplanarity is defined as $3\lambda_{3}/2$, where $\lambda_{3}$ is the smallest
  eigenvalue of the normalized momentum tensor $S^{\alpha
  \beta}=(\sum_{i}p_{i}^{\alpha}p_{i}^{\beta})/(\sum_{i} |\vec{p_{i}}|^{2})$ ,
  where $\alpha,\beta=1,2,3$ correspond to the $x,y,z$ momentum components, and
  $i$ runs over all visible objects.
\bibitem{defsnupz}
The $p_Z$ of the neutrino candidate is estimated by constraining the charged
  lepton and the neutrino system to the mass of the $W$ boson and choosing the
  lowest magnitude solution.
\bibitem{defslepsigma}
${\cal{SIG}}_{\text{jets}}$ for $\ell$ is defined with respect to all jets in
  an event as $\sum_{\text{jets}}[p_T^{\text{jet}}\times\Delta\mathcal{R}
  (\ell,{\text{jet}})]/\sum_{\text{jets}}(p_T^{\text{jet}})$.
\end{thebibliography}
